\documentclass[acmsmall]{acmart}

\AtBeginDocument{%
  \providecommand\BibTeX{{%
    \normalfont B\kern-0.5em{\scshape i\kern-0.25em b}\kern-0.8em\TeX}}}

\setcopyright{acmcopyright}
\copyrightyear{2022}
\acmYear{2022}
\acmDOI{10.1145/1122445.1122456}

\acmConference[CSCW '22]{CSCW '22: The 25th ACM Conference On Computer-
Supported Cooperative Work And Social Computing }{November 12-16, 2022}{Taipei, Taiwan}
\acmBooktitle{CSCW '22: The 25th ACM Conference On Computer-
Supported Cooperative Work And Social Computing ,
  November 12-16, 2022, Taipei, Taiwan}
\acmPrice{15.00}
\acmISBN{978-1-4503-XXXX-X/18/06}
\usepackage{wrapfig}
\usepackage{multirow}
\usepackage{ulem}
\usepackage{comment}
\usepackage{xcolor}
\excludecomment{toexclude}

\begin{document}

%\title{CommunityNews: Visualizing political bias and datavoids on social media}
\title[An AI System for Addressing Political Data Voids on Social Media]{Datavoidant: An AI System for Addressing Political Data Voids on Social Media}
%Addressing Political Data Voids on Social Media
%human-AI interface for collective sensemaking


\author{Claudia Flores-Saviaga}
\affiliation{%
  \institution{Northeastern University}
  %\streetaddress{1 Th{\o}rv{\"a}ld Circle}
  \city{Boston,MA}
  \country{USA}}
\email{floressaviaga.c@northeastern.edu}

\author{Shangbin Feng}
\affiliation{%
  \institution{Xi'an Jiaotong University}
  \city{Xi'an}
  \country{China}
  \email{wind_binteng@stu.xjtu.edu.cn}
}

\author{Saiph Savage}

\affiliation{%
  \institution{Northeastern University}
  \city{Boston,MA}
  \country{USA}}
  \affiliation{%
  \institution{Universidad Nacional Autonoma de Mexico}
  \city{CDMX}
  \country{MX}}
  
\email{s.savage@northeastern.edu}

%\author{Lars Th{\o}rv{\"a}ld}
%\affiliation{%
%  \institution{The Th{\o}rv{\"a}ld Group}
%  \streetaddress{1 Th{\o}rv{\"a}ld Circle}
%\email{larst@affiliation.org}

%\affiliation{%
%  \institution{Inria Paris-Rocquencourt}

%\affiliation{%
% \institution{Rajiv Gandhi University}

%\affiliation{%
%  \institution{Tsinghua University}
%  \streetaddress{30 Shuangqing Rd}

%\affiliation{%
%  \institution{Palmer Research Laboratories}
%  \streetaddress{8600 Datapoint Drive}
%  \country{USA}

%\affiliation{%
%  \institution{The Th{\o}rv{\"a}ld Group}
%  \streetaddress{1 Th{\o}rv{\"a}ld Circle}
%\email{jsmith@affiliation.org}

%\affiliation{%
%  \institution{The Kumquat Consortium}
%  \city{New York}
%\email{jpkumquat@consortium.net}

\renewcommand{\shortauthors}{Flores-Saviaga C., Feng, S. and Savage S.}

\begin{abstract}
The limited information (data voids) on political topics relevant to underrepresented communities has facilitated the spread of disinformation. Independent journalists who combat disinformation in underrepresented communities have reported feeling overwhelmed because they lack the tools necessary to make sense of the information they monitor and address the data voids. In this paper, we present a system to identify and address political data voids within underrepresented communities. Armed with an interview study, indicating that the independent news media has the potential to address them, we designed an intelligent collaborative system, called Datavoidant. Datavoidant uses state-of-the-art machine learning models and introduces a novel design space to provide independent journalists with a collective understanding of data voids to facilitate generating content to cover the voids. We performed a user interface evaluation with independent news media journalists (N=22). These journalists reported that Datavoidant's features allowed them to more rapidly while easily having a sense of what was taking place in the information ecosystem to address the data voids. They also reported feeling more confident about the content they created and the unique perspectives they had proposed to cover the voids. We conclude by discussing how Datavoidant enables a new design space wherein individuals can collaborate to make sense of their information ecosystem and actively devise strategies to prevent disinformation.

\end{abstract}



%\keywords{Human-centered computing → Collaborative and social computing systems and tools; Collaborative and social computing design and evaluation methods;}

\maketitle

\section{Introduction}

Disinformation erodes the integrity of the information circulating on social media and reduces our capacity to make sense of it \cite{starbird2019disinformation}. Together, this has impacted our society in negative ways. For instance, disinformation is negatively impacting our elections \cite{recuero2020hyperpartisanship,woolley2017computational,flores2018mobilizing}, and it is even hurting people’s health by having them follow dangerous health conspiracy theories \cite{dupuis2021misinformation,lu2021positive,lee2021viral}. Consequently, journalists and academics have spent significant time studying and identifying different ways to mitigate and address the problem of disinformation \cite{xu2021unified,wild2020designing,schwartz2020disinformation,stray2019institutional,zhou2019fake,zhang2018structured}. Journalists \cite{haque2020combating,mcclure2020misinformation,shu2019defend}, professional fact-checkers \cite{noain202013,noain202013}, and automatic disinformation detection systems \cite{alam2021survey,castelo2019topic,jiang2021structurizing,zeng2021automated} have contributed to countering the infodemic. However, bad actors are using a disinformation dynamic that journalists and academics have not yet been able to address \cite{donovan2021stop}. This disinformation dynamic has dangerously targeted underrepresented groups to spread political lies and hinder their civic participation \cite{flores2019anti,Factsand79:online}. The dynamic consists of weaponizing the limited information that exists about a political topic to promote disinformation, especially concerning an underrepresented population. In other words, the bad actors are weaponizing ``data voids'' \cite{golebiewski2018data}.  A data void or data deficit, occurs when there is high demand for information about a topic, but credible information is non-existent or in low supply. The low supply can help the bad actors fill the void with their own ideological, economic, or political agendas more easily \cite{Datadefi93:online,TheCovid22:online}. Bad actors can expose their problematic content to wider audiences, by filling the voids with their own information. Especially, because when people search for the topic, search engines and social media platforms will tend to give the problematic content higher visibility (as there is no other content available) \cite{NeimanLab:online,kou2017conspiracy,hagen2019emoji}. Recent studies have highlighted that an effective way to start to address the problem is through collaborations of independent news media \cite{Factsand79:online, ziff2016countering,spangenberg2014news}. Independent news media, different from the mainstream, has the incentives to collaborate and cover important data voids affecting society \cite{anderson2017future,donovan2021stop,bayer2019disinformation}. However, independent journalists currently are understaffed and have limited resources and tools \cite{ismail2018strengthening}; while, mainstream media is more tied to monetary incentives that can limit the type of news that can be covered \cite{entman2010media}.
\begin{wrapfigure}{L}{0.5\textwidth}
\centering
  \includegraphics[width=0.5\textwidth]{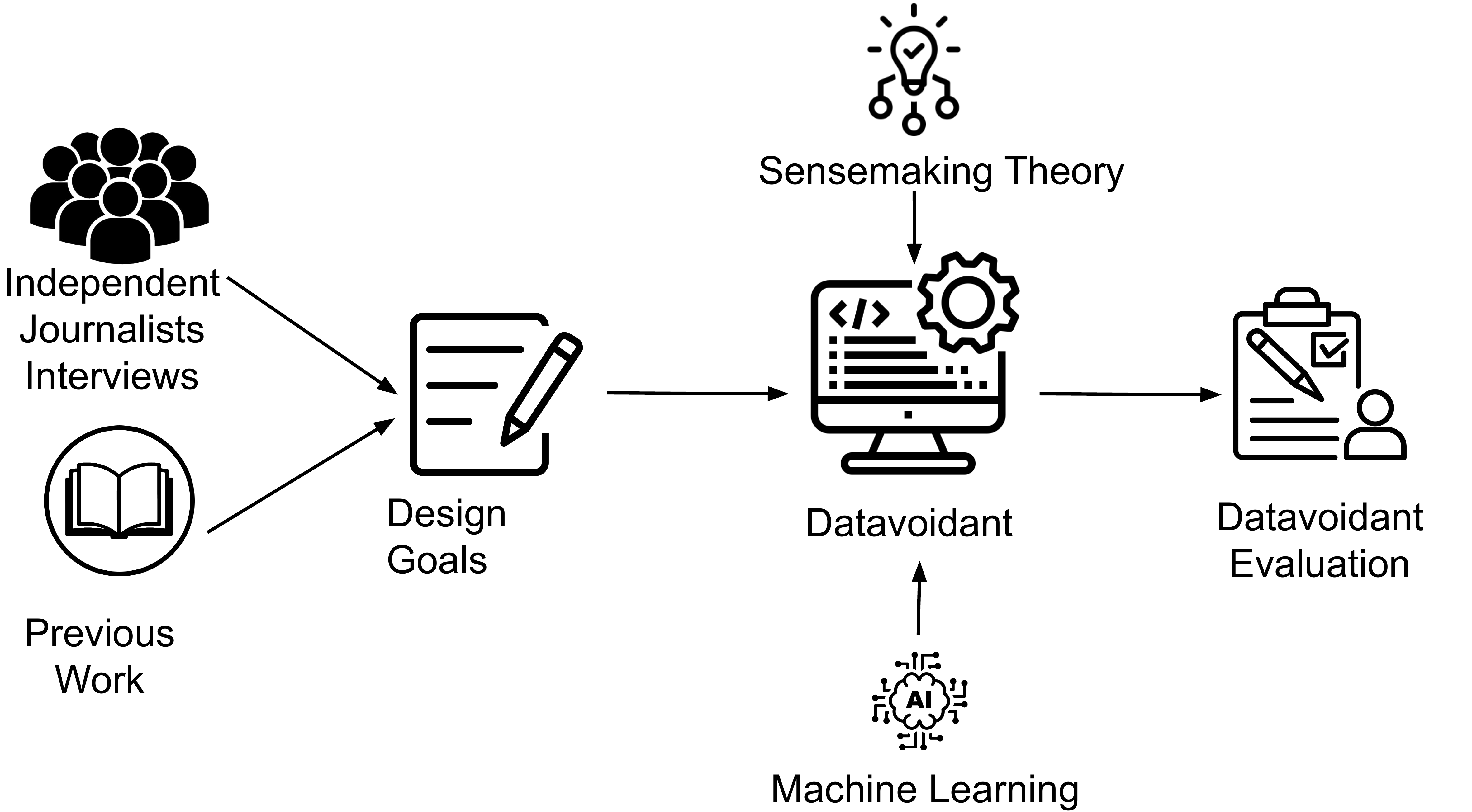}
  \caption{{\small Overview of our process for creating Datavoidant.}}
  \label{fig:process}
  \vspace{-0.5cm}
\end{wrapfigure}

To empower independent journalists to identify and address data voids, we follow a human-centered design approach \cite{norman2013design}  to ground the creation of  
a system that supports journalists in these tasks (See Figure \ref{fig:process}). First, we conduct an interview study to understand the social processes, needs and challenges that independent journalists currently face for addressing data voids. We interview journalists who address political data voids concerning underrepresented populations (a critical type of data void given its implications in elections and civic engagement  \cite{Howtroll78:online, PoliticalDataVoid2:online}). Through interviews, we found that independent journalists work together to monitor social media, particularly Facebook, as part of their daily routine to find data voids. These journalists perceived themselves as uniquely positioned to collectively counter disinformation and data voids targeting underrepresented communities. However, the whole process usually takes them days because they lack systems to support their work effectively. They revealed that a significant amount of manual labor takes place. 

Armed with this knowledge of how independent journalists operate, prior work, and Pirolli's et al. sensemaking theory \cite{pirolli2005sensemaking}, we designed an intelligent collaborative system to support journalists in addressing data voids: Datavoidant. Datavoidant has two primary modules for empowering journalists to identify and cover data voids: {\it``Intelligent Data Void Visualizer''} and {\it``Collaborative Data Void Addresser.''} The Intelligent Visualizer deploys state-of-the-art machine learning models and data visualizations to help journalists collectively identify data voids on multiple levels. The module visualizes categorized social media data with intuitive figures and automated summaries to help journalists conduct collaborative sensemaking and understand where the data voids exist. The {\it ``Collaborative Data Void Addresser''} introduces collaboration features to help journalists combine findings and create strategies on how they will fill the voids. It is important to note that most systems for journalists focus on helping them to fact check disinformation or conduct collaborative storytelling of local news \cite{karmakharm2019journalist, dalgali2020algorithmic,lin2021sync}.  
Instead, Datavoidant, integrates key design features to enable journalists to  specifically work for underrepresented populations and cover data voids present in their information ecosystem. Some of these key design features are that Datavoidant: operates within Facebook, which is the largest social network used for news consumption, especially among underrepresented communities \cite{DigitalN17:online}; facilitates collaborations among journalists on a diverse set of political topics considering that diverse expertise is needed when working with underrepresented populations \cite{Failuret73:online}; and has a ``backstage'' space to enable journalists to strategize what content to produce to cover a data void (having a backstage is important because the journalists need to identify best ways to engage underrepresented populations with content that will be presented to them for the first time). Our evaluation study revealed that journalists found our tool easy to use, and appreciated the intelligent summaries, deep dives, and multiple perspectives that Datavoidant offered for inspecting the data. These features enabled journalists to quickly visualize what was occurring in the information ecosystem; collaborate and create strategies to collectivelly fill the data voids more effectively, while feeling more confident about the content they created and the unique perspectives they were able to offer. In this paper, we contribute: 1) an investigation of independent journalists' practices for covering political data voids targeting underrepresented populations; 2) a system supporting independent journalists to cover data voids; 3) novel mechanisms for collective sensemaking and knowledge production; and 4) an interface evaluation showing that Datavoidant allows journalists to identify data voids on multiple levels. 

\section{Related Work}
In this section, we first summarize the relationship between data voids and disinformation. Next, we discuss the evolution of journalism due to the popularity of social media and how disinformation has affected how they work. Furthermore, we examine how disinformation affects underrepresented communities (e.g. Latinx) and why advocacy groups are urging independent journalists within these groups to assist in tackling the problem. Then, we overview the sensemaking process used to help design our system. Finally, we discuss the current tools used by journalists for disinformation detection \cite{pirolli2005sensemaking} and the challenges they face that motivate our system. 

\subsection{Data Voids, Data Deficits and, Disinformation.} Data voids were first studied in the context of search engines. According to Golebiewski and Boyd  \cite{golebiewski2018data}, a data void, or data deficit, occurs when there is high demand for information about a topic, but credible information is non-existent or in low supply. Fewer conversations regarding topics in different bipartisanship contexts can create deficits with weighted narratives of the predominant groups, leaving malicious actors free to exploit these deficits and instill their political and ideological agenda ~\cite{golebiewski2018data}. As a result, when people are searching for the topic, search engines and social media platforms will show high visibility to the problematic content \cite{NeimanLab:online,golebiewski2018data}, helping to increase the exposure to the political agendas of conspiracy theorists, white nationalists, and other extremist groups. Consequently, data voids on social media actively contribute to the spread of disinformation and cause real-world harm  \cite{NeimanLab:online,TheCovid22:online,Talkinga39:online}.

The terms misinformation and disinformation are frequently used interchangeably; however, they are not synonymous. Although both terms refer to inaccurate, incorrect, or misleading information, the main difference is the intention behind them. Misinformation refers to inaccurate or erroneous information spread without intending to cause harm \cite{jack2017lexicon}, such as in crises when there is a lack of verified information \cite{starbird2014rumors}. In contrast, disinformation refers to the dissemination of false information with the intent of deceiving the public; for instance, for political purposes \cite{bittman1985kgb}. The danger of disinformation is that it is designed to resonate with the existing beliefs of a targeted audience, therefore giving it a greater likelihood of being accepted as fact  \cite{rid2020active}. For years, state actors and partisan groups had spread disinformation through ``disinformation campaigns'' \cite{rid2020active,bittman1985kgb}. In this study, we focus on political information, in which bad actors can spread information to deceive public opinion, therefore, we use the term ``disinformation''. 
Regarding our use of the term ``data voids'', we recognize the term ``data'' has experienced a variety of definitions \cite{ackoff1989data}; however, we use the term  ``data voids'' to remain consistent with how previous research has defined it \cite{Datadefi93:online,PoliticalDataVoid2:online}. 

Recent research has analyzed how malicious actors weaponize data voids \cite{starbird2019disinformation,Factsand79:online,flores2019anti,UndertheSurface:online} and orchestrate disinformation campaigns \cite{pasquetto2022disinformation,wilson2021cross}, some of which target underrepresented communities  \cite{starbird2019disinformation,arif2018acting}. A recent study found that data voids emerged on social media in the early stages of the COVID-19 pandemic as underrepresented communities sought health information ~\cite{UndertheSurface:online}. The study revealed that the data voids were being filled with disinformation, resulting in the development of narratives detrimental to vaccine confidence and trust in governmental institutions by underrepresented communities. For example, Facebook posts claimed that the COVID-19 vaccine altered people's DNA and cause infertility in recipients. Researchers concluded that, since there was no high-quality information to challenge these narratives (i.e., there was a data void), the public, especially underrepresented populations, could not understand the development of the COVID-19 vaccine, and disinformation narratives easily spread without being confronted \cite{UndertheSurface:online}. For this reason, researchers recommend to stop relying on fact-checking efforts and platforms' content moderation, since these approaches are reactive, insufficient, and potentially counterproductive \cite{chou2021covid, lewandowsky2012misinformation}. Researchers have argued that instead, there should be a focus on adopting a proactive stance where data voids are addressed before they are weaponized \cite{TheCovid22:online,UndertheSurface:online,Talkinga39:online}. Next, we discuss more on how journalists work on social media, and their efforts to proactively address disinformation and data voids. We connect how this influenced our system design.

\subsection{The emergence of Social Media Journalism and Online Disinformation.}
The rising popularity of social media has led news organizations and independent journalists to publish their news reports on these platforms \cite{newman2011mainstream}. These platforms have become especially important as people increasingly use them to consume their daily news \cite{shearer2018social}. Similarly, journalists have started to heavily rely daily on social media to discover and share breaking news, connect with sources, and promote their work \cite{hermida2012social}. For instance, during breaking news events, journalists monitor the social media accounts of institutions, people, and public figures to contextualize information and use it as part of their reporting \cite{newman2009rise}. Within the context of journalists focused on underrepresented audiences, most typically go on Facebook and post their news reports within Facebook groups and pages related to the underrepresented populations they target \cite{robe2020reflections}. The journalists also tend to create their own Facebook groups or pages, and post their news stories on those spaces \cite{direito2020social}. Such practices have been adopted by both mainstream media and independent journalists; the latter have tended to use social media more, as it allows them to better connect and expand their audience \cite{hermida2012social}. 
However, due to the increasing popularity of social media, the low entry barriers, and the data voids present \cite{ireton2018journalism, golebiewski2018data, NeimanLab:online}, it has also been possible for disinformation to spread on these platforms. A recent survey of more than one-thousand journalists revealed that the proliferation  of disinformation on social media has negatively affected them \cite{HardNews11:online}. They feel overwhelmed and outmatched \cite{mcclure2020misinformation}, especially because they do not believe they have the skills and tools necessary to \textit{make sense} of the amount of information they encounter to address disinformation properly \cite{mcclure2020misinformation,beers2020examining}. When aiming to counter disinformation, journalists must typically decide if they want to debunk it and how, since debunking the disinformation could also amplify it further \cite{HardNews11:online,mcclure2020misinformation}. Thus, researchers argue for proactive measures to address disinformation. In such setting, social media content that may be  weaponized for disinformation should be proactively addressed before it becomes problematic \cite{TheCovid22:online,UndertheSurface:online,Talkinga39:online}. However, the problem is that we currently lack tools to empower journalists for this task, especially within the context of the information ecosystem of underrepresented communities \cite{Facebook53:online}. This problem inspired our research. Next, we present more about underrepresented populations, disinformation, and the role of independent journalism in this context.

\subsection{Disinformation, Underrepresented Groups, and Independent Journalism.}
Disinformation targeting underrepresented groups can have significant consequences. Such consequences include eroding trust in institutions, suppressing their vote, increasing hatred against them, or even putting their health at risk \cite{judit2021,Factsand79:online}. For example, in the case of the Latinx community, advocacy groups and governments are concerned about the role disinformation targeting the Latinx community could have on democracy \cite{Spanishl94,RepsMuca18:online,Howfaken58:online,Thisisf55:online}. In the run-up to U.S. 2020 election, data voids were filled with disinformation and conspiracy theories regarding political issues in order to influence and divide Latinx voters or spark violence \cite{Spanishl89:online}. Examples of disinformation that emerged from data voids that were circulating on Facebook (the go-to platform for Latinx \cite{Demograp95:online}), included: narratives that connected Biden to socialism (which may have been intended to dissuade Latino voters who fled socialist regimes in Venezuela, Cuba and Nicaragua) \cite{Thisisf55:online};  narratives intended to put Latino and Black voters against one another \cite{Spanishl58:online}; or narratives that questioned in Spanish the reliability of mail-in voting \cite{sbee61:online,Spanishl89:online}. Organizations focused on Native Americans, Black, Afro-Latinx, and Latinx communities found that, during that time, large newsrooms were tackling disinformation based on what they assumed were relevant issues rather than attempting to learn which election-related issues were most important to those underrepresented communities \cite{Failuret73:online,Talkinga39:online}. Part of the problem, is that main stream media is interested
in large profits \cite{herman2010manufacturing} or lacks proper representation of minorities in their staff \cite{Decadeso78:online,HardNews11:online}. As a result, they do not cover news stories tailored to the needs of those communities \cite{Failuret73:online}. Here is where independent journalists play a key role. Without independent journalism that focuses on underrepresented communities and creates content for them, the communities are less likely to be politically informed, become civically engaged, or get access to information shared by authentic sources they trust \cite{HardNews11:online}. As a result, independent journalists have become the ones addressing the information needs of these communities, as well as combating disinformation targeting them \cite{Failuret73:online}. However, this is also time consuming and difficult \cite{Facebook53:online}, especially because most journalists lack tools to help them in the process \cite{HardNews11:online}. In this research, we focus on creating a tool to help journalists address data voids in underrepresented communities. We argue we can accomplish this task by connecting to sensemaking theory and integrating it into our design. Next, we present about sensemaking theory, how it connects to data voids and our design process.%Thus, experts recommend collaborating with independent journalists from these communities to monitor data deficiencies, prioritize them and choose what disinformation to debunk \cite{Failuret73:online}. However, independent journalists, who combat disinformation on Facebook targeting underrepresented communities, acknowledge that it is time-consuming and diverts resources from advocating for and serving those communities \cite{Facebook53:online}.}

\subsection{Disinformation and Sensemaking.} 
\begin{toexclude}
\vspace{-0.5cm}
\begin{wrapfigure}{l}{0.5\textwidth}
\centering
  \includegraphics[width=0.5\textwidth]{figures/sense-making loop-tachado.pdf}
  \vspace{-0.75cm}
  \caption{{\small Pirolli et al.\cite{pirolli2005sensemaking} Sensemaking loop.}}~\label{fig:loop}
\end{wrapfigure}
\end{toexclude}
The process of understanding a data void can be understood as a sensemaking task to gather and analyze a large variety of unstructured data and arrive at a conclusion \cite{pirolli2011introduction, pirolli2005sensemaking}. Pirolli and Card \cite{pirolli2005sensemaking} characterize sensemaking as a bottom-up process that involves a series of iterations: foraging for relevant source data (e.g., searching and filtering for relevant Facebook groups/pages to study data voids and disinformation in them), extracting useful information (e.g., collecting and reading the data from the groups/pages), organizing and re-representing the information (e.g., schematizing the extracted data from the groups/pages), developing hypotheses from different perspectives (e.g., building a case about the different types of data voids present in the Facebook groups/pages), and deciding on the best explanation or outcome (e.g., deciding what story/narrative to write to address a particular void). There are active research efforts in the CSCW community focused on building tools that support collaborative sensemaking \cite{qu2008building}. Some example include tools for collaborative sensemaking in: web search \cite{paul2009cosense}, mystery solving \cite{li2018crowdia}, self-directed learning \cite{butcher2011self}, and knowledge creation  \cite{pirolli2011introduction}. In this paper, we develop a human-AI collaboration solution that automates parts of the sensemaking pipeline to help journalists to more effectively address political data voids together. Next, we discuss general tools that journalists have for addressing disinformation, and we contextualize them with our system. 

\subsection{Journalism Tools to Address Disinformation.}
According to Zubiaga et al. \cite{zubiaga2018detection}, social media have become a critical publishing tool for journalists  \cite{diakopoulos2012finding,tolmie2017supporting}. However, the absence of control and fact-checking of posts makes social media a fertile ground for spreading unverified and/or false information \cite{zubiaga2018detection}. The traditional approach to combating disinformation gaining popularity in recent years is fact-checking. Nevertheless, fact-checking is tedious work that does not keep up with the staggering amount of content posted on social media every day \cite{allen2020scaling}. For this reason, research efforts have been devoted to designing \textit{collaborative systems} to help fact-checkers and journalists address mis- and disinformation. For instance, collaborative tools for fact-checking news \cite{sethi2018extinguishing}, videos  \cite{carneiro2019deb8} and visual disinformation \cite{venkatagiri2019groundtruth,matatov2018dejavu}.  Similar systems were also proposed to combat disinformation with crowdsourcing such as Newstrition \cite{Joinusin5:online}, Checkdesk \cite{Checkche53:online} and Truly Media \cite{caled2021digital}. However, these methods generally underperform professional fact-checkers and rely heavily on politically knowledgeable individuals \cite{godel2021moderating}.  Additionally, none of the tools are tailored to monitor and detect data voids. 
In fact, journalists do not use sophisticated tools to perform their job \cite{beers2020examining,mcclure2020misinformation}. For instance, Brands et al. \cite{brands2018social} found that journalists use TinEye, FotoForensics, and Google’s reverse image search for image verification; InVid for video verification; Google Maps for audiovisual verification and Botometer for verifying unauthentic accounts on Twitter. Additionally, they found that journalists used Excel and Google Sheets to perform their analyses, including  visualizing and aggregating data.They report that some of them used TweetDeck and CrowdTangle. Brandtzaeg et al. \cite{brandtzaeg2016emerging}  found that journalists in Europe also use traditional methods such as Exif, Topsy,
Tungstene to verify images and videos posted on social media. Meanwhile, researchers coincide in the view that many journalists locate stories to cover lurking on social media feeds 
\cite{brands2018social,beers2020examining, brandtzaeg2016emerging}. However, journalists report having a limited understanding of how to use some of these tools because some are not explicitly designed for journalists, are not intuitive to use \cite{HardNews11:online}, and not tailored for them \cite{beers2020examining}. Other researchers also acknowledge the lack of understanding of journalists' needs and values to design tools to support their work \cite{mcclure2020misinformation,komatsu2020ai}. Additionally, researchers have not developed systems that detect data voids  \cite{NeimanLab:online}. Our work complements
the lack of systems for addressing political data voids on social media. We also take a human centered design approach and conduct interviews with journalists to understand their needs and practices, and thus create a tool that will be useful for journalists to tackle data voids. We also connect to sensemaking theory to further help us in our design. 

%In this paper, we address this challenge by incorporating journalists' work and insight towards online disinformation into our scalable data void detection system. Specifically, we resort to previous research \cite{mcclure2020misinformation,beers2020examining} while also conducting our own interviews with journalists to understand their work and gain insights into data voids on social media. With the help of these insights, we designed Datavoidant, an intelligent tool that allows independent media journalists to visualize the political data voids that exist on social media, conduct collective sensemaking, and collaborate to create content that fills those voids.
\section{Interview Study}
Our interview study aimed to understand the practices that independent journalists currently follow to address data voids targeting underrepresented communities. We use the findings from this study to help us explore and understand the design space. Notice that due to the niche nature of combating disinformation narratives that target underrepresented communities, the pool of interviewees was limited. It was therefore important to avoid sharing detailed information about our interviewees as they could be more easily identifiable given the few actors in the space \cite{carlson2021digital,waisbord2020mob}. 
\subsection{Interview Study Participants}
We invited independent news journalists to our study through social media, professional networks, and by attending disinformation workshops for independent journalists working with underrepresented communities. We also used snowball sampling to invite more individuals \cite{goodman1961snowball}. For the purpose of the study, we considered independent news journalists as those who felt they were free to report on issues of public interest given that the organizations where they worked were free of the influence of governments, and other partisan interests \cite{deane2016role}. In total, we recruited 22 individuals, all self-identified as independent news journalists; they created content to address disinformation targeting underrepresented populations (See in our appendix Table \ref{tab:participants} with details of our interviewees). The majority of our participants %(90\%)
(20) specialized exclusively in underrepresented communities; they worked in either niche newspapers, digital first outlets, or non-profit newsrooms; most %(90\%)
(20) self-identified as underrepresented individuals, and worked primarily with either Latinx, Black, and Native American communities in the US. It is important to highlight that two of the authors of this paper are from underrepresented communities and have previously worked with independent journalists who concentrated on underrepresented communities. This helped us identify mailing lists, workshops, and social media spaces where we could connect with these type of journalists. Note, however, that participant recruitment was done entirely separately from our prior direct engagement with journalists. Futthermore, the recruitment was primarily led by students who were unknown to the journalists to ensure that participants saw participation as voluntary. In the rest of the paper we refer to these participants with the identification of ``J''. %In this paper, we have anonymized these participants, but we have provided basic information about each of them in Table \cite{}. 

\subsection{Interview Study Protocol}
%We conducted an interview with 22 independent news journalists. who self-identified as independent news journalists and have addressed disinformation among underrepresented populations in the past. 
We interviewed 22 independent news journalists. These interviews helped us obtain information detailing how independent journalists addressed data voids, their motivations for covering them, and the challenges they faced. Although we could have interviewed more independent journalists to obtain additional insights, the interviews conducted allowed us to achieve sufficient data saturation \cite{guest2006many, weller2018open} for the themes presented here. Our interview focused first on asking questions about the nature of participants' jobs in journalism, their background (e.g., what they studied, other jobs they had in the past), and experiences using social media and related tools for their journalism. Next, we elicited information about how they worked, how they decided what stories to cover, how they used social media for news reporting, and how they tackled disinformation in their work. We also asked them to mention the opportunities and challenges they faced in their general journalism work and when addressing disinformation. We also questioned whether they had witnessed data voids and how they tackled them (pain points, high points, and also a walk-through of the process they adopted to cover the data voids). We were also interested in the type of values they had adopted for conducting their work (to help us understand their priorities). All of our interview questions were checked with our partners in journalism  to ensure they were appropriate. %Note that we do not report details of participants to help keep their anonymity.   

\subsection{Data Analysis of the Interview Study} To analyze the interview responses, we coded them to extract initial concepts \cite{mihas2019qualitative}. To develop a set of codes for the data, two of the authors independently coded the data. They then worked together to create 17 axial codes, which were applied top-down to the responses from journalists. From the top-down axial codes, the authors then organized the interview data into 7 themes and produced a final list of mutually exclusive themes that denoted the main findings from our interviews. The 7 themes were highly agreed upon by the intercoders (Cohen's Kappa coefficient (k) = 0.79). Authors discussed the disagreements during the writing and final synthesis process of the themes.

\subsection{Interview Study Findings}
%\textcolor{red}{Link findings with collaboratively sensemaking theory }\cite{pirolli2005sensemaking}\\
%Through analysis of the interviews a number of themes emerged. We focus on seven main insights.
We now present the 7 main themes (findings) that emerged from our interview analysis.
%First, (1) we discuss how participants considered that independent media journalists were in a unique position to counter disinformation due to their lack of economic and political ties. Next, (2) we explain how they consider important to be aware about the opinions of different actors of society, such as the public, politicians and other news media outlets. (3) We explain how journalists face the challenge of identifying when a data void might be originating; (4) journalists expressed a lack of essential tools to identify data voids that might be tight to political actors and (5) how people interact with certain content;  (6); we also talk about how journalists want to be able to detect fake accounts in order to know which narratives may be manipulated. Finally, (7) we discuss journalist's collaborative practices to understand those voids. \newline

\textbf{Finding 1. Independent media organizations perceive themselves as uniquely positioned to counter disinformation targeting underrepresented populations}. %86\% 
19 of the interviewed journalists thought mainstream news outlets had a difficult time addressing disinformation targeting underrepresented populations.
They believed it was better for independent news media to take on the responsibility: \textit{``...In the fight against disinformation, we [independent media] play a crucial role. This is especially true if we consider that we [independent media] are economically and legally protected from having our editorial line influenced by the interests of our funding agencies...''} J5. Part of the reason why they considered that independent news media were better for this is that they viewed the task as a social justice activity that tried to fix some of the distortions that mainstream media originally produced: \textit{``There's a lot of social justice involved, especially because we know mainstream media is often run by monopolies and has a political agenda; a lot of the Hispanic media distorts reality [...] hence, independent media brings you the truth no one else wants to tell. The independent journalists are like ``social justice warriors'', they are the ones who risk their lives in the middle of a protest, and the ones who sometimes get gassed by the police. They are the ones who dare to be able to say what is happening, both from one side and the other.''} J3. According to %36\% 
8 of the journalists, a major advantage and opportunity that independent media is free from economic ties to any particular organization: \textit{``Media outlets outside the mainstream have a better chance of communicating truthfully and openly with citizens, as mainstream media have to follow agreements with governments, political parties, or companies based on economics. This can lead them to report in a biased way or report only one side of the coin.''} J9. However, %10\% 
2 of the journalists in our study also recognized that this economic freedom acts as a double-edged sword. They believe that it can also put independent media at a disadvantage, as independent media can then have less funding for accessing  the same tools and resources as mainstream media: \textit{``Independent media are limited by funding, personnel, space, and infrastructure, unlike all the mainstream media's machinery.''} J4.  %18\% 
4 of participants considered that community trust in independent media helped these journalists to address disinformation because they were already considered a reliable source: \textit{``When the public reads our news they naturally think whatever is shared is already vetted [without disinformation]. You know, information they can trust. The best we can do as journalists is to take care of the relationship with the public by posting clear, useful, trustworthy information'' J11}. Generally, independent journalists are considered freer of economic or political interests \cite{hyde2002independent}. Consequently, people can trust independent media for specific topics more \cite{deane2016role}, since independent journalists have more freedom about the topics they can cover \cite{price2000enabling}. Together, these dynamics appeared to have helped independent journalists become the most prompt for covering data voids.

\textbf{Finding 2. Independent journalists monitor conversations from multiple stakeholders with different political leanings to find stories to cover.} Journalists reported that they constantly monitored the conversations from various actors in society to better determine the stories they will cover: \textit{``I always watch social media to see what people are saying, because we determine what we are going to verify or cover based on factors of public interest...''} J7. Monitoring the different conversations also meant that journalists had to analyze what political actors from both sides discussed: \textit{``Our last elections were very polarized [...] it was essential to pay attention to what both sides were saying...''} J8. The monitoring of both political sides not only involved political actors but also analyzing what news sites with different political leanings covered: \textit{``There are news media outlets that are obviously blinded by a political side. I always read what other news outlets are covering, what topics they're covering more, what kinds of stuff they want to put on the daily schedule. Then, I check who's behind the editorial line and think: ``why are they so interested in talking so much about this topic?'' That gives me an idea about which mediums [news sites] to trust and which ones to fact-check.''} J13. However, part of the problem was that it was difficult to study and quantify to what extent political groups pushed certain narratives: \textit{``Oftentimes the stories are promoted by groups with political interests, which is why they're dominating the conversation in the media, but it's hard to quantify''}. J22. In order to devise strategies for adequately addressing disinformation, especially disinformation targeting underrepresented populations, it is important to understand who can be behind the disinformation, as well as with what groups the narrative is most resonating with \cite{persily2020social}. However, such analysis is not simple \cite{bradshaw2018global}. Our goal with the design of Datavoidant was to further help journalist in this endeavour.

\textbf{Finding 3. There is a need to identify and understand data voids}. 
%41\% 
9 of the interviewed journalists believed that social media suffered from {data voids}. Interviewees considered that the problem of data voids involves identifying and understanding what information about a given topic is inaccurate or insufficient. %23\% 
5 indicated that they had difficulty finding data voids and identifying what information needed to be better clarified or covered: \textit{``Sometimes there is a lot of interest in a certain topic on social media, but not much supply of quality information; however, we don't have a way to measure that and they can slip by [the existence of data voids].''} J15. Journalists believed there were benefits associated with identifying data voids, especially as it allowed them to bring unique and quality perspectives to certain topics: \textit{``Finding a topic that is not adequately explained and has manipulative information,  is a gem because it gives us  the opportunity to talk about the topic,  investigate, and communicate. If it is something that is already on people's lips at that time, it is a gold mine. But it is not always easy to find those topics.''} J18.  Even though it is challenging for independent journalists to find data voids, %18\% 
4 considered that identifying data voids was crucial, especially during election periods where data voids could adversely impact people's vote: \textit{``In a political campaign, citizens' perception of a politician will help them vote for or against him. In this case, we need to find out what limited information minority populations have [about a politician] and then disrupt it, improve it, and make it more accurate.''} J20. Researchers agree that by identifying data voids early, journalists can more easily fill these gaps \cite{Datadefi93:online}. 

\textbf{Finding 4. To strategy how they will cover data voids, journalists need to understand how people interact with the limited content created around the void.} Journalists indicated that it was important to know how people engage with manipulative content created around a data void. However, this was difficult: \textit{``We must know how people react to those limited stories [data voids] that are being created. I just have no sense of how to find that out. We don't have tools to help us understand which messages are indeed resonating; it would be essential to know.''} J14. Information about engagement mattered because it helped journalists to better plan how they would address certain data voids, especially given their own limited capacity and also not wanting to amplify problematic content:  \textit{``We can't just analyze every topic, and try and address it. We don't have the capacity to do that, and we don't want to draw attention to something that's not getting any attention [engagement] in the first place.''} J7. According to J10, prioritization is important since it is impossible to cover everything: \textit{``There's too much information out there, so no matter  how hard we try, we can't cover everything. So we decide what to verify based on factors of public interest and virality, and the consequences it might have, like health risks.''} J10. This finding is consistent with previous studies showing that journalists struggle to understand the extent and impact of problematic narratives on social media, impacting their decisions as to when to publish and when not to \cite{mcclure2020misinformation}. It is also consistent with previous research that documents journalists' need to understand how people are reacting to certain content to avoid amplifying the problematic content, or give bad actors more exposure than they would have otherwise \cite{phillips2018oxygen}. Within underrepresented communities, understanding what voids are most engaging is crucial because independent journalists are even more under resourced \cite{deane2016role}. Journalists recognized this as an important undertaking, particularly in political situations where the voting decisions of under-represented populations can be affected \cite{sbee61:online} or even suppressed \cite{sbee61:online}. Governments and advocacy groups have been raising alarm on this issue for years \cite{TheCovid22:online}. During the 2018 midterm elections, previous research found a shortage of information about political candidates and voting rights for minority groups, prompting bad actors to spread disinformation \cite{flores2019anti,woolley2017computational}.

\textbf{Finding 5. Independent journalists want to detect automated accounts to know which narratives may be manipulated.} The journalists in our study were aware that automated social media accounts existed and had strategies to identify them: \textit{ ``They usually don't use their full names in the account, or they use random pictures. These accounts don't have a lot of Facebook friends. One becomes more or less aware of them.''} J2. Journalists noted that identifying automated accounts was important in allowing them to determine if a narrative is being manipulated and forced into the public discourse: \textit{``Identifying these accounts [automated accounts] gives us an idea of how legitimate or not a piece of information is''} J15.  This was agreed by J21: \textit{``You can get a better sense of what might be going on, especially when elections are coming up. For instance, I've noticed that politicians did polls on social media during elections to figure out who might win. They always showed the ultra-right candidate winning, but this didn't happen in the end. By knowing that fake accounts [automated accounts] are pushing this, I could have a better understanding of the situation''J21.} For us it was interesting to identify that, in difference to prior work \cite{beers2020examining}, none of the journalists in our study reported using tools to detect automated accounts. This may be due to the fact that tools like Botometer are not adapted for use outside the English language, nor are they tailored for the bots in underrepsented communities \cite{forelle2015political}. Our hope is that our system can help address this gap, by providing information about where there is automation within underrepresented communities.

\textbf{Finding 6. Independent journalists collaborate to identify and address data voids}.  As part of the process of identifying data voids, journalists turn to other colleagues to corroborate information and discuss whether there is a data void in order to create plans for addressing the void together. %22\% 
5 of journalists mentioned that they consulted their colleagues for assistance in identifying data voids and created plans for how they would handle them: \textit{``I do collaborate to verify notes, even share sources of information. In the absence of sufficient data about a subject, this is crucial. We also do joint efforts to coordinate what information each person will look up. This can be quite time consuming when we have little information. You don't know whether what little exists is problematic.''} J5. Similarly, J14 stated the need to collaborate and brainstorm ideas about what stories around data voids to publish: \textit{``...We need to analyze the information [information about the data void] and brainstorm about what we should and shouldn't publish because we as journalists have a big responsibility. We can't just publish whatever, we have to be a filter [...] we either brainstorm via chat (WhatsApp) or I call him and say: Let's talk about our upcoming publications! [...] Sometimes, how it works is that I create a document and the other person will say:  ``Look, I'll go through it and see if I can add anything else, or we leave it like that'' [...] And it [the content to address the data void] is being worked on collaboratively.''} J14. Overall, we saw that independent journalists reported working together to address data voids, at times even creating alliances to address data voids targeting underrepresented populations: \textit{``We now maintain an alliance with other Hispanic media outlets at the national level, especially during election seasons. In the last national elections, we had an alliance with 15 media outlets from all over the country to verify information and locate the items that are missing. Right now, we are collaborating to create content about COVID as it is a topic we all care about, and unfortunately, our audience isn't always provided with good information [about COVID-19].''} J16. 
According to our interviewees, collaboration had become an integral part of how they extended their capability and reach. J16 considered that by collaborating with others, they were able to produce more accurate and comprehensive stories with less budget. Working in collaboration also allowed journalists to reach broader networks (as they could reach the audiences of each involved journalist): \textit{``It's the beauty of independent media, you can publish on one [news outlet] and then share it on others, we help each other reach more people''} J11. Collaborations were likely even more important in this setting given the lower resources of these journalists \cite{deane2016role}, as well as the diverse knowledge that is needed to understand data voids within underrepresented populations \cite{mesquita2021collaborative}. %This helped also journalists to enhance their reputation as more people were reading their timely and trustworthy stories.} 
 
\textbf{Finding 7. Independent journalists address data voids to help their audience make better decisions}.
Our participants expressed that they typically aimed to create content around data voids that could educate people and help them make more informed decisions. %36\% 
8 of the interviewed journalists  considered that educating the public about data voids is important to prevent the spread of disinformation: \textit{``Part of our job is to be "information translators". That's why we're creating content [content around a data void] to explain why some information is fake and how to spot it. By creating these articles [articles around data voids], we hope to increase the public's media literacy.''} J19. This was something that other journalists also echoed. For instance, J12 expressed: \textit{``We're interested in having an educational role, so we create content about topics that might not be newsworthy right now. We do, however, consider it a good idea for the public to educate themselves on the issue, as their decisions may be impacted by it [the topic].''} J12. Similarly, J15 expressed that they were interested in covering data voids that would help educate communities to make better decisions : \textit{``In our work, we try to address gaps related to community needs and will help community decisions be more educated, more informed.}

\subsection{Connecting Journalists' Interviews to System Design}
Based on our interview study, we identified 4 design goals (DG) and 6 subgoals (SG) to guide our system design:

\textit{\textbf{Design Goal 1. Design for Independent Journalists Targeting Minorities}}. Through our interviews we identified that independent journalists were who could address data voids because, unlike mainstream media, they had less restrictions on the content to create (Finding 1). We therefore tailored our tool to independent journalists (SG1). Our interviews and prior work \cite{Failuret73:online,flores2019anti},  also helped us to understand that these journalists focused primarily on underrepresented populations where data voids were present (Finding 1). We consequently focused on designing the interface for independent journalists working with underrepresented groups (SG2). Based on prior work \cite{Mappingd79:online, HowMisin70:online}, we can also expect that journalists working with underrepresented populations will have to work in multiple languages, especially as the underrepresented populations could be immigrants for whom English is a second language, and consequently, will likely consume information in different languages \cite{Mappingd79:online, HowMisin70:online}. We considered that having to navigate between multiple languages can make it hard for journalists to understand  data voids. We thus set out to create an interface that would help journalists navigate the different languages easily (SG3).

\textit{\textbf{Design Goal 2. Collective Sensemaking to Understand Data Voids on Multiple Levels}}. 
Journalists reported constantly browsing social media and performing manual multi-level analyses to understand the different types of data voids (Findings 2,3,4,5). They were interested in finding topics with limited coverage (Finding 3). Thus, we argued for visualizations to allow for a topical analysis on multiple levels (SG4). We focus on visualizations that highlight the specific multi-level analysis that our interviewees mentioned was important: topical, political leanings, and bot analysis. We also provide visualization-friendly summaries of different variables that journalists reported they analyzed (e.g., amount of posts per topic, how different political leanings are discussing different topics).

\textit{\textbf{Design Goal 3. Facilitate ``Backstage Space'' to Discuss Data Voids.}}
Journalists collaborated to develop plans for addressing data voids together (Finding 6). Such planning was important as the journalists were often the first to create content for the underrepresented population. They had to strategize what to cover to best engage and educate their audience. Based on prior work \cite{flores2021fighting, lampinen2011we}, we considered that a way to address this need was via a ``backstage'' space that facilitated such discussions. To that end, we introduced into our interface a chat with voice and video capabilities(SG5).

\textit{\textbf{Design Goal 4. Collaborative Spaces for Creating Content To Address the Data Voids.}} Journalists explained that they typically attempted to produce content collaboratively to help address the data voids they had identified previously (Finding 7). To that end, we enabled in our interface a shared document through which journalists could create articles with their colleagues to address these voids together (SG6).

\section{Datavoidant}
Guided by our design goals, we created: Datavoidant, a collaborative online interactive system with state-of-the-art machine learning models and a dashboard to categorize social media content and help journalists visualize data voids on multiple levels.  
Next, we provide a scenario where Datavoidant can be employed, followed by the system description.
\begin{wrapfigure}{R}{0.55\textwidth}

\centering
\includegraphics[width=0.55\textwidth]{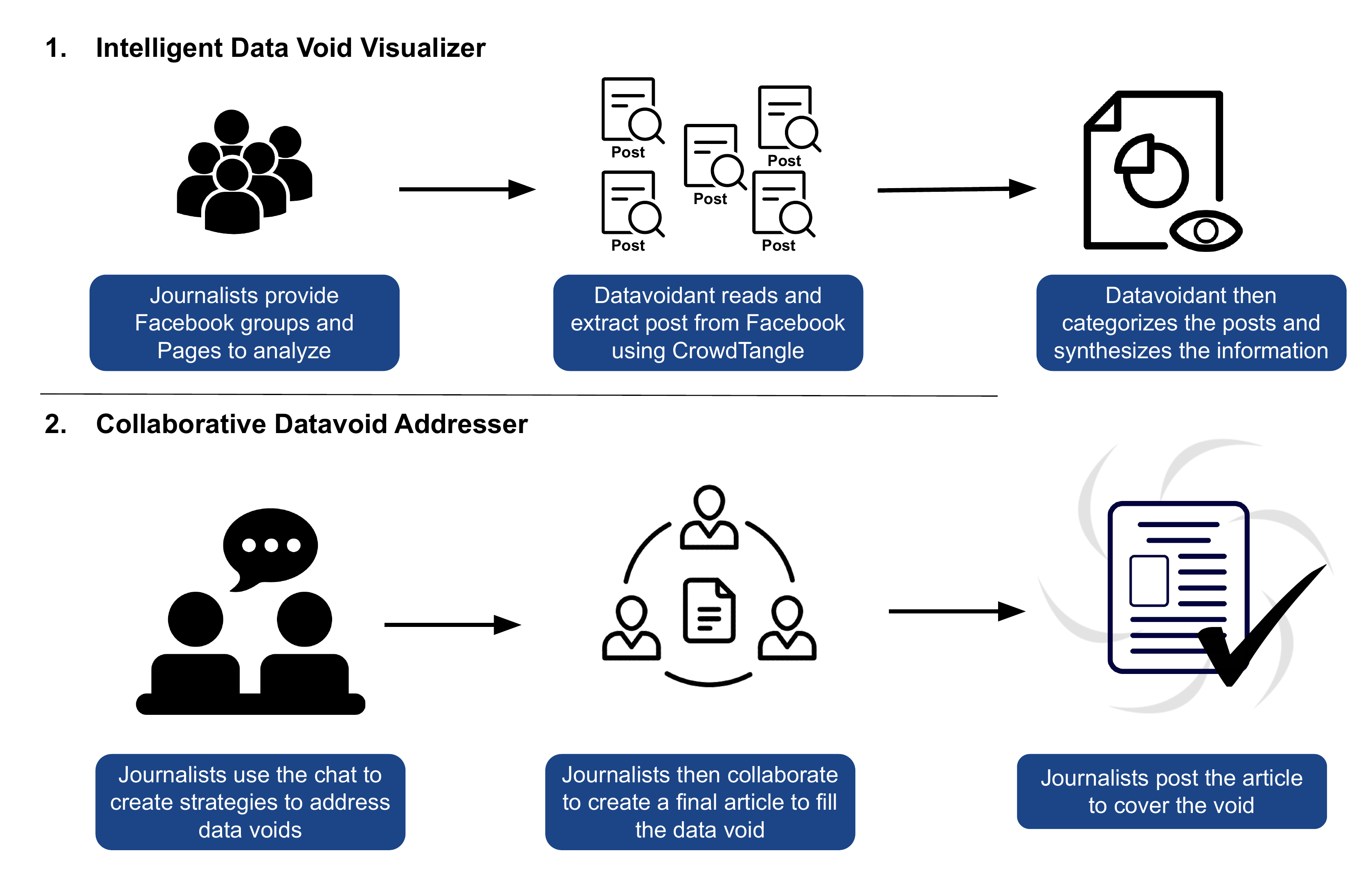}
\caption{\label{fig:datavoidOverview} Overview of Datavoidant’s functionality.}
\end{wrapfigure}
\subsection{USER SCENARIO.} Laura is an independent journalist from the NGO ``Voto Latino'', aiming to help the Latinx community access quality information for the upcoming election. Laura logs into Datavoidant and noticed by looking at the \textit{Post per Topic} graph that immigration is among the topics most discussed on Facebook by the Latinx community. Based on her examination of the \textit{Political Leaning} graph, she realized that the topic of ``immigration'' is also highly politicized. This topic has a crucial data void: it receives almost NO neutral coverage. Conservative news outlets predominantly cover the topic. Looking at the  \textit{Type of Groups/Pages Generating Content} graph, she discovered that a partisan citizen group, ``Latinos Conservadores'',  is among the top groups posting content about immigration. Laura asked Juan, a colleague from a neutral Latinx news media outlet, ``Latino Justice'',  to take a look. She hopes, she and Juan can devise a strategy to confront the data void. Juan examined the \textit{Percentage Bots per Topic} graph and realized that around 20\% of the posts on immigration are automated; in addition, over 60\% of such posts are being commented on and shared. It worries him to see the lack of neutral content and how much people engage with non-neutral immigration content. The analysis of individual posts also reveals a false claim that Democrats were planning to send a caravan of Cuban immigrants to storm the U.S. border to disrupt the election \cite{Howfaken58:online}. Juan and Laura decided to use the built-in chat function to formulate a strategy on how to fill the void and limit the spread of disinformation. The authors wrote a neutral article to discredit the disinformation and explain what is actually happening. The authors posted the story on the Facebook group of Latinos Conservadores and the Facebook page of Latino Justice. They also plan to organize a press conference to give visibility to their article and inform Latin voters about it through more neutral Latin media. Laura and Juan have been able to address data voids targeting Latinx communities within hours by using Datavoidant, instead of taking days.

\subsection{SYSTEM DESCRIPTION.} Datavoidant modularizes the sensemaking process to allow journalists to visualize existing data voids and devise strategies for covering the voids across different types of Facebook groups and pages (citizen, political, and news media). Fig. \ref{fig:datavoidOverview} presents an overview of our system. In the following section, we describe how Datavoidant is designed to be tailored for journalists working with underrepresented communities and the two major components of Datavoidant: {\it ``Intelligent Data Void Visualizer''}; and {\it ``Collaborative Data Void Addresser''}. 

\begin{wrapfigure}{r}{0.65\textwidth}
\centering
\includegraphics[width=0.65\textwidth]{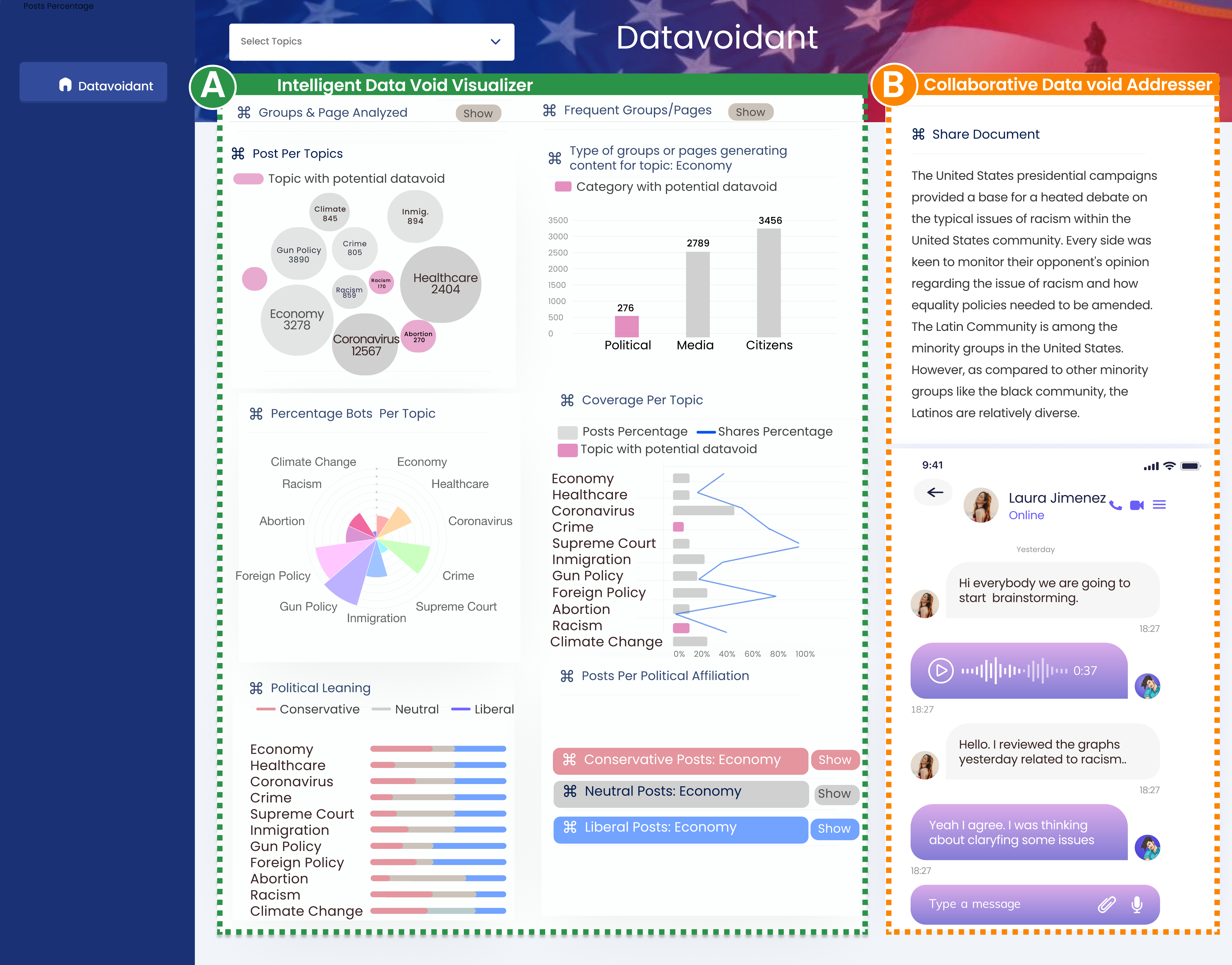}
\caption{{\small Datavoidant: A)   Intelligent Data Void Visualizer: Visualization module. B)  Collaborative Data void Addresser: composed of shared document (top) and chat module (bottom).}}~\label{fig:datavoidant}
\vspace{-0.3cm}
\end{wrapfigure}

\subsubsection{\bf Designing Datavoidant for Minorities.}
\textbf{}Our goal was to tailor our tool for independent journalists targeting underrepresented communities. For those journalists as opposed to more mainstream ones \cite{mesquita2021collaborative}, collaboration is key, as it allows them to broaden their reach in an already limited (minority) audience. Working with underrepresented populations also requires diverse expertise and knowledge \cite{Failuret73:online}, highlighting even more the importance of collaborations. We thus designed Datavoidant with features that enabled journalist collaborations. Additionally, these journalists are among the first to deliver content to underrepresented audiences (as they covered data voids). Consequently, they spent time strategizing how they would best present the content to resonate with their underrepresented audiences.  We then decided to integrate components for backstage planning. Based on prior work \cite{Failuret73:online}, we consider that journalists are experts on the underrepresented populations they target. We therefore designed our system to also allow journalists to use their expertise to drive the study of the data voids (e.g., by having journalists define what Facebook groups and pages to study). In the following section, we provide more information about Datavoidant's different components and further highlight how it is design to work with underrepresented groups.

\subsubsection{\bf Intelligent Data Void Visualizer} 
%\subsection{Collaborative Data Void }
This component of Datavoidant focuses on helping journalists to visualize and make sense of the data voids that exists in the information ecosystem of their desired underrepresented population. To accomplish this goal, the component has the following modules: 1) Data Collection module; 2) Smart Categorization module; and 3) Viz module. Each module integrates processes from the sensemaking loop of Pirolli et al. \cite{pirolli2005sensemaking}. Next, we explain each module in detail. 

{\it Data Collection Module (SG1, SG2).} Independent journalists provide Datavoidant with a list of Facebook pages and groups for which they want to identify possible data voids (notice that this corresponds to the {\it ``Step: Search and Filter''} in the sensemaking loop of Pirolli et al. \cite{pirolli2005sensemaking}).  Next, the system connects to the CrowdTangle API to read and extract all the posts, likes, number of comments and reshares from the public Facebook groups and pages that journalists initially provide (corresponding to {\it ``Step: Read and Extract''} in the sensemaking loop). 
%( Read and extract in the sensemaking loop)

{\it Smart Categorization Module (SG4).} %The previous module enables Datavoidant to access the data of the information ecosystems that journalists are interested in studying.
Given that the data collected by the \textit{Data Collection Module} can be massive and difficult for humans to interpret, this module focuses on structuring and categorizing the data to facilitate collective sensemaking.  For this purpose, Datavoidant uses state-of-the-art machine learning models to categorize social media content and then synthesize the results  ({\it ``Step: Schematize''} in the sensemaking loop).
This section provides an overview of how the module works (an in-depth explanation and evaluation can be found in the appendix \ref{appendixSmartCatModule}).

To categorize the content, Datavoidant uses basic NLP techniques to categorize the Facebook groups and pages into either ``content from political actors,'' ``content from citizen initiatives,'' or ``content from news sites.'' This type of categorization is important given that journalists expressed an interest in being able to bridge the data gap between these different online spaces. However it is also important for journalists to conduct a multi-level analysis where they can understand what topics were less covered than others across these different online spaces, which political actors were pushing certain topics, and whether automated methods were pushing certain topics (to understand manipulations around data voids). For this purpose, Datavoidant integrates state-of-the-art machine learning models to categorize the content on multiple levels and facilitate these types of data analysis.

{TOPIC LEVEL CATEGORIZATION. }
In the design of Datavoidant, we considered that journalists would likely not have the time or ability to interpret complex abstract topics without labels, like the ones that the topic modeling algorithm of LDA throws out \cite{blei2003latent}. We assume that most journalists will likely not know how to provide labeled data to train machine learning algorithms that can discern one topic from another. Therefore, we opted for automated methods that could remove the unnecessary burden and complexity to journalists, while still allowing them to automatically categorize their data at scale. Datavoidant simply asks journalists to provide the list of topics they are interested in exploring and a list of keywords associated with each topic. The system then uses these keywords and topics to automatically create a training and testing set to teach machine learning models how to classify posts into topics.

{POLITICAL LEANING CATEGORIZATION.}
In addition to topic-level data voids, Datavoidant also  helps journalists to identify {political-level data deficiencies}, where some topics might be less discussed by accounts from certain political or ideological perspectives. For example, climate change content might be rarely covered by liberals, while critical race theory could be less covered by conservatives, creating partisan echo chambers and political-level data voids. For this purpose, Datavoidant identifies each post's political leaning to facilitate visualization and understanding of political-level  data deficits. To conduct its automatic categorization of posts with respect to political leanings, Datavoidant resorts to external knowledge about the political leanings of websites~\cite{robertson2018partisan} and political actors~\cite{feng2021knowledge}. 

\begin{toexclude}
\begin{itemize}
    \item If the post comes from a Facebook page that represents a websites that is in the list (i.e., a website with a clear political leaning), the system:
    \begin{itemize}
    \item averages the mentioned websites' political leaning score based on ~\cite{robertson2018partisan} and through this obtains the post's final ``political leaning score'' as $b_w$. 
     \end{itemize}
    \item If the post mentions any website on the list or mentions any political actors then the system:
    \begin{itemize}
    \item calculates  the sentiment score $s$~\cite{devlin2018bert, wolf-etal-2020-transformers} for the post, with -1 as most negative and +1 as most positive.
   \item averages the political leaning score of the actors and websites mentioned based on~\cite{feng2021knowledge,robertson2018partisan} to obtain the ``political leaning score'' $b_a$. The final political leaning score of the post is then obtained by $b_a \times s$.
    \end{itemize}
  \item If the post mentions neither websites nor political actors, the system takes 0 for its political leaning score and regard the post as neutral.
\end{itemize}
\end{toexclude}

Notice that Datavoidant categorizes posts first based on the overall nature of the Facebook page from which the post is from. We consider that known conservative outlets will tend to always post conservative content and liberal outlets will tend to post liberal content. If the system cannot identify the nature of the Facebook page, it analyzes whether the post is discussing liberal or conservative actors in a positive or negative form, and uses this to calculate the political leaning score of the post. In all other cases, the system labels the post as neutral. In this way, Datavoidant  calculates political leaning scores for social media posts, which helps to illustrate political-level data deficiencies across topics.

{BOT CATEGORIZATION.} Automated social media users, also known as bots, widely exist on online social networks and induce undesirable social effects. In the past decade, malicious actors have launched bot campaigns to interfere with elections~\cite{ferrara2017disinformation,deb2019perils}, spread misinformation~\cite{feng2021satar} and propagate extreme ideology~\cite{berger2015isis}. To address these issues, Datavoidant includes a bot detection component that categorizes accounts into bots and none-bots. The aim is to help journalists identify biased information propagated by malicious actors. In Datavoidant, we focus on the textual content of posts to identify {Facebook bots and malicious actors}. Specifically, we follow the method in the state-of-the-art approach~\cite{feng2021botrgcn} to encode post content with pre-trained language models~\cite{liu2019roberta} and train a multi-layer perceptron for bot detection. We train our model with the comprehensive benchmark TwiBot-20~\cite{feng2021twibot}. \newline

{\it Viz Module (SG3, SG4)} Datavoidant employs diversified machine learning techniques to identify malicious actors, topic-level and political-level data deficiencies. However, these approaches can be highly technical and presenting the results as-is might that might confuse journalists. To address this issue, Datavoidant synthesizes results from different components to present an intuitive, easy-to-use and visualization-friendly summary of the system's findings ({\it ``Step: Schematize''} in the sensemaking loop). Datavoidant extracts the following information for the front-end visualization:

\begin{itemize}
    \item \textit{Number of posts per topic}: we use a bubble chart to show the number of posts per topic. Each bubble represent a specific topic. The bubbles then expand or shrink based on the number of posts that relate to each topic.
    \item \textit{Distribution of topical content by political leanings}: We use stacked bar charts to show the political leanings of each topic. Each stack bar represents a topic, and the segments in the bar indicate the percentage of posts that each political side (neutral, conservative, liberal) has generated for the topic (the total sum of the different perspectives is always 100\%).
    \item \textit{Percentage of comments and shares per topic}: we use grouped bar charts to allow users to compare the percentage of comments, and shares per topic. The topic with the greatest percentage of comments, likes, and shares will indicate that it has received the most engagement among all topics. This visualization also helps to highlight which topics are NOT receiving engagement and where there could be a possible void. It is important to note that in some cases a high number of comments and shares on a topic could come from very specific outlier posts. In the future, we aim to present the outliers in a separate graph to help journalists further understand the dynamic.
    \item \textit{Number of topical posts produced per type of Facebook Group/Page}: using separate bars for each type of page/group (news media, political, or citizens) indicates when certain actors are covering (or not covering)  specific topics.
    \item \textit{Percentage of bot content per topic}: we use bar charts to show the percentage of topical posts that were potentially were produced by automated accounts. It is important to note that there are news media outlets that utilize automated accounts to enhance their dissemination of news on social media \cite{lokot2016news,diakopoulos2019automating}, which might show in this graph. Journalists may conclude that all of these accounts are malicious. Our aim is to also educate journalists to realize that seeing automation does not necessarily equate to an account spreading manipulative content.
    \item \textit{Frequent groups and pages per topic}: we show in a table the names of the most frequent  groups/pages that cover each topic, along with the type of Facebook page to which they belong (news media, political groups or citizens). 
    \item \textit{Individual posts}: when a journalists selects a specific topic, Datavoids shows the individual posts of that topic separated by political leaning. This allows journalists to take deep dives and analyze the data on different fronts.
    \item \textit{Automatic translation: } if a journalist needs to translate the information on the platform, this feature allows them to instantly translate texts into more than one hundred languages.
\end{itemize}

Datavoidant presents these intuitive and easy-to-use visualizations to facilitate journalists' sensemaking efforts to counter data voids and prevent disinformation that could weaponize those voids (Fig. \ref{fig:datavoidant}). Notice that Datavoidant provides an interface that allows for deep-dive analysis of data voids on multiple levels.% (analysis on political leanings, topics, and automation).  

\subsubsection{\bf Collaborative Data Void Addresser} 
This piece is composed of two modules that help journalists to collaborate and make sense of the data voids. %information presented by the ``Data Void Visualizer'' and create content that addresses the data voids that they identify. %Notice that this piece addresses steps: {\it ``11. Build case in the sensemaking loop''}, and {\it ``14. Tell story''} in the sensemaking loop.

{\it Chat Module(SG5).}
This module allows journalists to communicate with each other to identify potential data voids based on the information presented in Datavoidant's {\it Intelligent Data Void Visualizer}. Notice that this corresponds to {\it “Step: Build Case”} in the sensemaking loop. For this, we integrated a chat room, in which participants can have conversations about the potential hypothesis they derive from the data presented in Datavoidant. This chat room can be seen as an ``investigation'' backspace where users can match their findings and discuss what hypotheses they are drawing. Through this chat room, users can discuss their findings and what they think the data might indicate. They can also start to devise strategies on how they will address the voids. To integrate the chat room we used RumbleTalk \cite{Onlinegr23:online}. The chatroom allows users to chat via text, voice, video, and have live video calls.

{\it Shared Document Module (SG6).}
When users understand what is going on (e.g., types of data voids that exist) and have decided how they will address the void, they can collaborate to create a final article or news report to fill the data void (\textit{``Step: Tell Story''} in the sensemaking loop). We implemented a shared document that appears directly within Datavoidant's. All users can use this document simultaneously to create a final document collaboratively. To integrate the shared document we used Pusher, an API service designed to facilitate adding real-time interactions \cite{PusherLe52:online}.

\section{Evaluation of Datavoidant}
To study Datavoidant we conduct an interface evaluation. Note that in our appendix, we also share an evaluation of the machine learning models used (See \ref{appendixML}).
For our interface evaluation we investigate the impact of our system on journalists and how our tools helps (or hinders) journalists in addressing data voids. We designed our evaluation based on standard usability measures of performance and satisfaction metrics \cite{nielsen1994usability, venkatagiri2019groundtruth}. Our aim was to understand how well Datavoidant allows journalists to identify data voids and collaborate to address them (performance). We were also interested in understanding journalists’ experiences when using Datavoidant (satisfaction).

% to understand the ability of our system to help independent journalists to identify and cover political data voids targeting an underrepresented population. 

%Our goal was to understand how independent journalists used our tool to identify data voids on multiple levels. We were also interested in determining how straightforward or difficult it was to identify data voids using our system; the challenges and opportunities that independent news journalists identified when using our tool, and which alternative methods they would use to complete the tasks if the interface were unavailable.

\subsection{Participant Recruitment.}
We recruited 22 independent journalists from Upwork to participate in our study. These individuals were all different than the journalists who took part in our initial interviews. To recruit participants, we posted a job on Upwork inviting people to our study. We set the Upwork job category to ``content writing'' and skills as: ``independent journalism writing,'' ``article writing,'' ``experience writing for minorities,'' ``social media monitoring,'', ``collaboration,'' ``experience exposing and debunking mis/disinformation''. We required that only U.S. based journalists apply for our study (to ensure they worked with underrepresented populations similar to the ones we studied previously). We also required people to show evidence that they were independent journalists who, as part of their day-to-day jobs, conducted social media monitoring of general political content for underrepresented groups. For this purpose, potential study participants had to share related articles they had authored as journalists with us. In our job description we told the participants they would be paid to use a new interface with another journalist to write an article together covering knowledge gaps in underrepresented populations. We paid participants  \$15 for taking part in a one-hour session. 12 of the participants were female; 9 male; 1 preferred not to disclose. 14 participants had a Bachelor’s Degree; 7 had a Master’s Degree; 1 had a Ph.D. 17  participants mentioned using social media four days a week or more for their journalist work; 5 used social media at least three days a week for their work. All primarily used Facebook for their work.  In the rest of the paper we refer to these participants with the identification of  ``P''.

\subsection{{Study Procedure.}}
To measure performance,  we conducted sessions over Zoom and had participants work together in pairs of two to complete a series of tasks. The tasks focused on identifying and addressing different types of data voids, and gathering information on a variety of usage scenarios. % and aimed to have participants exercise all the interface elements of Datavoidant,
Notice that we had all participants use Datavoidant with the exact same dataset (in particular, we used a dataset that journalists helped us to create to evaluate our machine learning algorithms. See our appendix for details \ref{dataset}). This helped us to better control our experiment and the data voids that participants were exposed to. During each session, participants first completed the IRB approved consent form and a pre-survey asking about their demographic information. The sessions were conducted in teams of two to allow for collaboration. For each participant pair, we presented a brief overview of the dataset, including the time frame, the Facebook pages and groups included.  We gave each participant a tour of Datavoidant and asked them to collaborate together on a series of different data void related tasks. In particular, participants were asked to work together to identify: (a) the topics with less content (measured in terms of number of Facebook posts), (b) the topics with missing or limited content for a specific political leaning, (c) groups or pages with limited content for specific topics, (d) groups or pages with limited content for specific political leanings and topics, and (e) a topic, political leaning, or group/page, with limited content. The goal was for participants to then create with their partner an article addressing that data void, especially for the underrepresented population in the dataset (i.e., Latinx). After journalists completed the tasks, we conducted short surveys to ask participants the level of difficulty they experienced in performing each task, using a five-point Likert scale. After that to measure satisfaction, we asked participants which aspects of the interface they liked and disliked, as well as any challenges and opportunities they experienced when using our system to complete the task. Furthermore, we asked participants to tell us about the alternative methods they would use to complete the task in question if Datavoidant were unavailable. Note that while participants completed the tasks in pairs, they responded survey questions individually. 

\subsection{Data Analysis of Journalists Usages of Datavoidant}
Our data analysis focuses on studying the performance of journalists using our tool and the perspectives (satisfaction) that journalists have about it, allowing for quantitative and qualitative ways of studying tool usages.

\subsubsection{Performance Data Analysis.} We were interested in studying how well our tool helped journalists to perform their work (performance). For this purpose, we quantitatively studied performance in terms of how long it took participants to complete all tasks using our tool, the number of participants who were able to use Datavoidant to identify data voids on multiple levels, the level of difficulty they had for performing the different tasks on Datavoidant, and average number of words that the journalists used for each article they created with our tool. 

\subsubsection{Satisfaction Data Analysis.} To analyze journalists' perspectives about Datavoidant (the challenges and opportunities they identified when using our system) we analyzed the open-ended responses that participants provided in the survey, where they shared their impressions of the tool. Based on prior work that characterized people's perspectives about different interfaces, we decided to use a hybrid approach  of inductive and deductive thematic analysis \cite{fereday2006demonstrating}. We first used the deductive approach to identify data patterns that were relevant to the usability themes of interest to our work. Deductive codes included interface learnability, efficiency, interface memorability, errors, and satisfaction \cite{nielsen1994usability}. We then used open coding to explore the qualitative data and allow for the discovery of emergent themes previously not identified (inductive analysis) \cite{mihas2019qualitative}. Two of the authors discussed the initial concepts (themes) as a group to iterate on them and created an initial codebook (the codebook included also the themes from the deductive process). We then had several iterations of the codebook and in-depth discussions among the research team to condense the codes into the final themes and created a finalized codebook. The finalized codebook with examples was shared with two coders who categorized the survey responses into the different themes. The coders agreed on 86.4\% of the responses they categorized (Cohen's kappa =0.82). We then asked a third coder to label the responses upon which the first two coders disagreed.

\subsection{Results User Interface Evaluation: Performance}

\begin{table}[h]
\begin{tabular}{|l|c|rr|cr|}
\hline
 & \textbf{\shortstack{\tiny Task a\\ \tiny (Topics)}} & \multicolumn{1}{c|}{ \textbf{\shortstack{\tiny Task b \\ \tiny (Political Leanings)}}} & \multicolumn{1}{c|}{\textbf{\shortstack{\tiny Task c \\ \tiny (Topics in Groups/Pages)}}} & \multicolumn{1}{c|}{ \textbf{\shortstack{ \tiny Task d \\ \tiny (Political Leanings \& \\ \tiny  Topics in Groups/Pages)}}} & \multicolumn{1}{c|}{\textbf{\tiny Average Tasks}} \\ \hline
\textbf{\small Correct} & \multicolumn{1}{r|}{{\color[HTML]{00009B} \textbf{\small 73\%}}} & \multicolumn{1}{r|}{{\color[HTML]{00009B} \textbf{ \small 77\%}}} & {\color[HTML]{00009B} \textbf{\small 68\%}} & \multicolumn{1}{r|}{{\color[HTML]{00009B} \textbf{\small 68\%}}} & {\color[HTML]{00009B} \textbf{\small 72\%}} \\ \hline
{\small \bf Incorrect} & \multicolumn{1}{r|}{ \small 27\%} & \multicolumn{1}{r|}{\small 23\%} & {\small 32\%} & \multicolumn{1}{r|}{\small 32\%} & \small 28\% \\ \hline
\end{tabular}
\caption{Overview of participants who identified data voids on multiple levels using Datavoidant. }
\label{tab:data-void-identification}
\vspace{-0.5cm}
\end{table}

\begin{figure}[h]
\centering
  \includegraphics[width=1\textwidth]{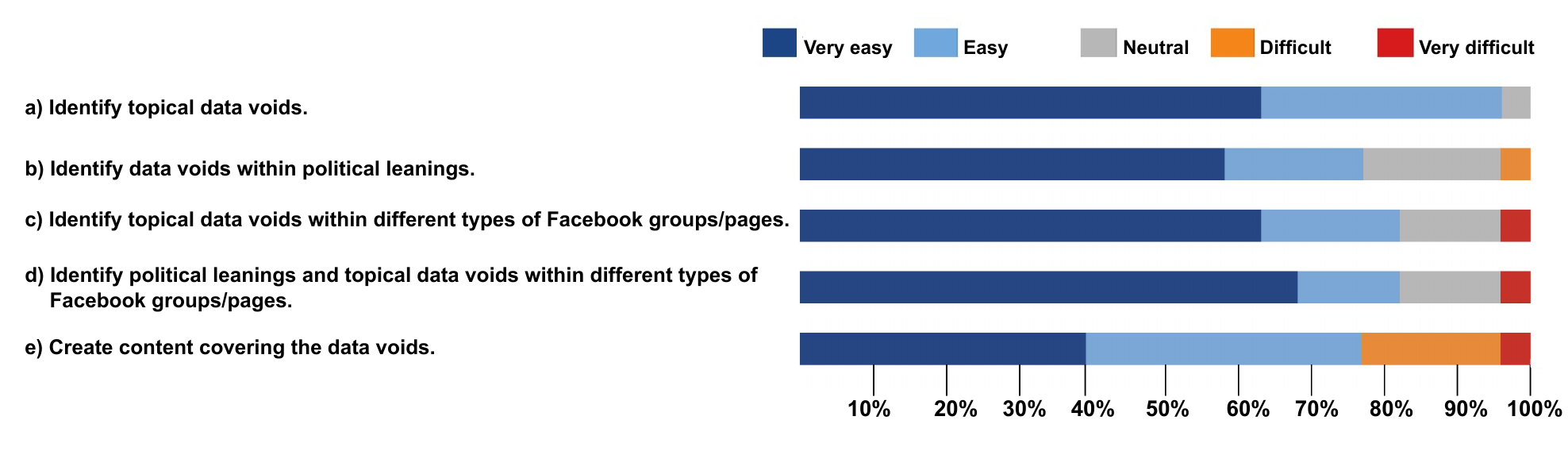}
  \caption{{\small Participants’ responses show that the majority found Datavoidant ``very easy'' to use for identifying and addressing data voids.}}~\label{fig:eval}
\end{figure}

Participants took an average of 36 minutes to complete the tasks in our study (SD=27.73 minutes). The articles they created to address data voids had an average of 144 words. All participants were able to complete all tasks in our study. Notice that for tasks a,b,c,d we can quantify the quality of how participants completed them, especially because we can measure whether participants indeed were able to identify the data voids that existed in the dataset that we used for our study. Table \ref{tab:data-void-identification} presents an overview of the percentage of participants who completed tasks a,b,c,d correctly. We were strict in our measurements and only considered that a participant completed a task correctly if they were able to find all the data voids related to the task at hand. In general, over half of the participants correctly identified the multiple level data voids (i.e., data voids in topics, political leanings, pages and groups). Overall the participants were better at identifying data voids about particular topics and political leanings than data voids within particular groups and pages. To better understand why this was happening, we analyzed details about participants' difficulties using Datavoidant. Participants evaluated the level of difficulty for performing the different tasks on Datavoidant using a five-point Likert scale, ranging from {\it ``very easy''} (+1) to
{\it ``very difficult''} (+5). Results are presented in Fig. \ref{fig:eval}. From Fig \ref{fig:eval} we observe that across tasks, the majority of participants considered that Datavoidant was {\it ``very easy''} or  {\it ``easy''} to use. Surprisingly, the task that most participants %(67\%)
(15) considered was the easiest to conduct, was the task of identifying data voids based on topic and type of Facebook groups/pages. We mention this is surprising as it was also one of the tasks that participants struggled with the most to complete correctly (See Table \ref{tab:data-void-identification}).  We believe that some participants likely mentioned only the first data voids they saw (note that when they did not provide the full list of data voids for a task, we marked the task as incorrect as we used strict measurements). In the future, we plan to explore interfaces that prompt end-users to explore data voids more and not just focus on the first results they see \cite{card1999readings}. Here it will be important to balance exploration with the tight deadlines in which journalists work. 

\subsection{Results User Interface Evaluation: Satisfaction}
Next, we analyze participants' open-ended responses to  their survey answers to shed light on the challenges they faced when identifying and addressing data voids, as well as any opportunities they saw with Datavoidant. 

%\subsubsection{\textbf{Intelligent Data Void Visualizer.\\ \newline}}

\textit{\textbf{Saving time: Without Datavoidant finding data voids is a time consuming process.}}
Most journalists in our study %(90\%)
(20) considered that Datavoidant helped them save time in identifying data voids, as normally, this process was much slower. Part of the reason was that they needed to analyze information manually: \textit{``To obtain these types of reports [lists of multi-level data voids], I would have to manually search all of Facebook and Google, convert the data, and then go through a lengthy process.''} P8. Similarly, participant P21 mentioned how without Datavoidant, the process of finding data voids could take them days: \textit{``...it [Datavoidant] is very useful, as from Facebook, we get everything manually, and it can take days...''} P21. 

\textbf{\textit{Helpful to have summarized information for countering data voids}}. Most (62\%) mentioned that the most helpful element of the interface was the ability to see the summarized information, such as the coverage per topic. These summaries helped them to understand what they should focus on to address the data voids: \textit{``I'm able to easily understand the graphs [summary graphs]  with ease after a few seconds of viewing them. On mouse hover it shows the percentages of the graphs as well, giving an even clearer picture of what things I should write about next.''} P18. Similarly,  P2 expressed: \textit{`` The political leaning graph [summary graph about what political topics are discussed by each side] is super helpful because you can see which issues matter to each political party; or what they want to push the most. In my community [the underrepresented population for which she writes] it's very common to hear the right-wing political party talk about crime with us and say that they are like Superman, coming to protect and save us from all the bad stuff. It seems to me that being able to know this is very helpful. I could say: ''well, how strange, they're talking a lot about that on this side''; then I could check in other sources, and I might realize the reality is totally different. When that happens, the topic for my next article will be clear to me.''} P2. Overall, participants saw an opportunity in using the summarized data that Datavoidant provided to identify what their forthcoming articles would cover. For instance, P13 expressed that part of a journalist's job was to help audiences make more informed decisions. He felt the summaries of Datavoidant helped him to find problematic content and identify what articles he would create for his audience to enable them to make more informed decisions: \textit{``The interface provides holistic, meta information [summaries] that gives an overview of all the information flowing on social media. Through the analysis of patterns of content across topics, I can determine if there is a large difference in coverage that might indicate that some topics have been artificially promoted. Then we can create notes that will make it easier for the audience to make educated decisions.''} P13.

\textbf{\textit{Deep dives allow journalists to understand data voids more easily}}.
Most participants %(81\%)
(18) expressed that one of their favorite aspects of Datavoidant was the ability to take deep dives and study data voids from different angles. In fact, this feature of the interface was the most used component in Datavoidant. P5 expressed how they enjoyed conducting deep dives to analyze topical data voids: \textit{``Selecting the topic from the drop down menu [deep dive interaction] was very useful. Like, a click is all it takes to learn everything you need to know about that topic.''} P5. Similarly, journalists %(29\%)
(6), expressed how they found the deep dive of the data voids within different groups/pages (media, citizen, political) to be useful. Some found it especially helpful for inspiring them on the interview questions that they could ask different actors to start addressing different voids:
\textit{``The graph illustrating how many Facebook pages are covering a particular topic and which types of pages they are [deep dive interaction], is really helpful. The graph can be used to determine, for example, when the media is trying to impose a particular issue and how that influences citizens. This interface lets us cross validate data quickly and easily.  For instance, by looking at the graph, I see that politicians aren't concerned about racism. Politicians don't seem to care about this issue. A very interesting interview scenario would be to meet with a politician, and ask him: ``racism has been on everybody's mouth, the media has been discussing it, and so have the citizens, but not you, why?...''} P22.

\textbf{\textit{Datavoidant gives journalists confidence on the content created for addressing data voids.}} Participants %(41\%)
(9) expressed that Datavoidant gave them confidence about the content they created to address voids as they had a better overview of what existed, what did not exist, and how people engaged with information: \textit{``[While using the tool] I realized that I'd do my job with more confidence. As I would be able to tell with certainty what information is needed or wanted by the people, and I could report on topics that are not covered.''} P16. Journalists also considered that our system could give them confidence in sharing more: \textit{``I think what journalism lacks today is journalists who dare to give their opinion; in general, I like to give my opinion about problems affecting my people [underrepresented population]. I know my perceptions are subjective, but if I knew the `exact count' of comments on a topic instead of randomly guessing, then I would realize that there are a lot of people who care about this. That way, I'd be more brave about the opinion pieces I publish.''} P5.

\textit{\textbf{Collaborative features were valued, but missing the richness of traditional tools.}} Participants who considered that it was {\it ``very easy''} or {\it ``easy''} to use Datavoidant to create content to address data voids, also said that the collaborative features needed improvements. Part of the reason was that these journalists were already well-versed and comfortable collaborating with other tools (e.g., Google Docs). The shared document that was used in Datavoidant, therefore, appeared to be ``under-featured'' for them  (especially when compared to the collaborative documents offered by Google):
\textit{``The writing interface [of Datavoidant] is impressive, and it would empower journalists to analyze and improve their content, but the application lacks certain features that boost collaboration. (Check Google Doc features for reference)'' P12}. This may indicate a unwillingness to deviate from the norm and utilize collaborative tools that are unfamiliar to them. Other participants suggested adding encryption features to the collaborative chat interface. They considered that there could be occasions where delicate topics are discussed on Datavoidant that could put journalists in danger. As a consequence, participants considered it was important to have encryption in place to keep journalists safe: \textit{``There are certain topics where it is better not to be known as the one who helped expose them to the world. The features that Signal [an encrypted messaging application] has for discussing sensitive topics might be useful to you. I think it would be great if the chat [on datavoidant] were encrypted. It would make life much safer for journalists if it had that capability.'' P7}. Similarly, participants also wanted features to easily share what they were doing with others on Datavoidant. The sharing feature that they requested resembled the ``Share'' button that several social media platforms offer: \textit{``It would be helpful if it had a `Share' button to share real-time data on other platforms such as Twitter, WhatsApp, Instagram, etc.''} P9. This suggests the possibility of piggybacking on existing software infrastructure to enable enhanced collaboration interactions among journalists\cite{Collabo83:online, noain2020collaborative}.

%\textit{\textbf{Location-Aware Data Voids.}}
%10\% of the journalists in our study also wanted to have ways to understand data voids per location, as P2 mentioned:
%\textit{``The first thing that came into my head is that I wish I could see information according to towns/states/regions. How many people in the Midwest are writing posts about ``racism''? ''} P2.

\section{Discussion}
We studied how journalists currently address data voids so we could enhance the process. We found that it was primarily the independent journalists who focused on underrepresented communities that addressed the data voids unlike journalists working with more general audiences \cite{Failuret73:online}. These journalists typically addressed the voids via a collaborative sensemaking process; however, the process was time-consuming and complex. Our system, Datavoidant, combines sensemaking theory, state-of-the-art machine learning models, and collaborative interfaces to empower journalists to understand data voids and create strategies for addressing the voids more easily. Through a user interface evaluation, we found that participants could use our tool to identify data voids on multiple levels, and were able to create content to cover the voids. 
Most journalists in our study found that our tool was easy to use, and appreciated the intelligent summaries and deep dives that Datavoidant offered. They felt these features allowed them to understand more rapidly what was happening in the information ecosystem in order to more effectively address the data voids. One benefit of this design is that by giving journalists a better sense of the information ecosystem, they felt more confident about the content they created and the unique perspectives they proposed. Datavoidant opens up a design space with potential impact on other domains, where people collaboratively make sense of their information ecosystem to proactively devise strategies for creating change and make unique contributions to their ecosystem.%and take action to prevent low quality information to be weaponized for disinformation.

%FeedMe opens up a design space
%with potential impact on other domains where AI is still
%error-prone, for example expert finding: “We think that
%your friend Sanjay can answer this question about Nikon
%cameras: [question]. Is he a good person to ask?”

%FeedMe opens up a design space
%with potential impact on other domains where AI is still
%error-prone, for example expert finding: “We think that
%your friend Sanjay can answer this question about Nikon
%cameras: [question]. Is he a good person to ask?”

%We designed Datavoidant to assist independent journalists in identifying and addressing data voids on Facebook targeting underrepresented populations. Datavoidant's design was grounded in interviews and in the theory of data voids \cite{golebiewski2018data}, theory of collective sensemaking \cite{pirolli2005sensemaking} and collaborative journalism \cite{noain2020collaborative,Collabo83:online,noain202013}. Datavoidant combines visualization and state-of-the-art machine learning models to help journalists identify and address data voids.  We discuss here the notion of Datavoidant as a novel design for addressing data voids and preventing disinformation.

%\subsection{Datavoidant's Most Useful Features}
%\subsection{Designing Systems for Understanding and Addressing Data Voids}
{\bf A Proactive Approach to Counter Disinformation.} Until now, journalists have primarily adopted a reactive approach to combating the problem of mis/disinformation where they use  fact-checking and content moderation to take down problematic content  \cite{wintersieck2017debating, graves2016journalists}. However, researchers and practitioners have recommended taking a more preventive approach to combating disinformation \cite{UndertheSurface:online,Talkinga39:online}, especially because ``reactive approaches'' are often not enough to persuade audience members to change their minds \cite{FaceCheckPersuade:online}. With Datavoidant, we aim to enable more system designs that proactively address disinformation. By helping journalists identify data voids, they can proactively create content to fill them and avoid disinformation campaigns weaponizing the voids.

\textbf{Designing to Address Disinformation Targeting Underrepresented Communities.} In our interviews, independent journalists were eager to address data voids. They considered they had fewer limitations on what articles they were able to produce, thus enabling them to fill the voids more easily than ``mainstream'' journalists. This was important when working with underrepresented communities as the dynamics of what mainstream media decides to cover (and NOT cover) within underrepresented groups leads to data voids. Unfortunately, unlike mainstream media, independent journalists also felt limited in their ability to analyze large amounts of data. These struggles are a recurring theme of independent journalism working with underrepresented populations \cite{Awarroom85:online,Facebook53:online}. It becomes even more problematic as the time-consuming process depletes them of valuable resources that could be used to advocate and provide services to their communities \cite{Facebook53:online}. In building Datavoidant, we aimed to address these struggles by automating parts of the operations that independent journalists conducted for identifying and addressing data voids in underrepresented communities. Some key design features that Datavoidant integrates to empower journalists working with underrepresented groups are:

{\it Visualizations of Data Voids on Multiple Levels}. Our interviews highlighted that within underrepresented communities, data voids appeared based on topic, political leaning, and the actors driving the conversation. It was thus crucial to understand the multiple types of information asymmetries that existed (a problem not always present when working with general audiences, who have the privilege of being able to access vast information from multiple perspectives about the topics they care about \cite{ji2014role}.) It was based on these points from our interviews that we decided to enable data visualizations in Datavoidant that would allow journalists to identify and study data voids on multiple levels.

{\it Collaborative Interface.} Independent journalists working with underrepresented populations are typically even more under resourced than mainstream media \cite{deane2016role}, and need more specialized knowledge in order to understand properly the information ecosystem of the underrepresented communities \cite{Failuret73:online}. Our interviews showed how these dynamics led independent journalists, in difference to mainstream journalists (who are more prone to compete for stories), to collaborate more. Collaborations also helped them to reach a wider network of underrepresented populations, which is crucial when working with these groups \cite{chung2021community}. Thus, we designed Datavoidant to be a collaborative tools for journalists. 

{\it Backstage Space.} Our interviews uncovered that journalists working with underrepresented communities had to strategize about what data voids they would cover and how they would address them. The strategies were important because they were heavily under resourced and hence, could not tackle all voids. It was also important to strategize about the content they would create to engage the communities, especially as they were the first to tailor the content for the underrepresented groups. (They did not have a reference for how the content should look like; it was important to collectively strategize on best ways to present the information). Thus, we enabled a backstage space to create strategies.

{\bf Datavoidant and Collaboratively Addressing Strategic Silences.}
The journalists in our study acknowledged the power of mainstream media and bad actors to silence certain voices, control all that is published, and set agendas, influencing public perceptions of reality. Donovan et al. call this a ``strategic silence'' \cite{donovan2021stop}. In our interview study, independent journalists described themselves as ``social justice warriors,'' willing to cover these strategic silences by providing quality, informative content. Nonetheless, the journalists reported a lack of tools to learn what mainstream media and other critical actors are covering or ignoring. To address this challenge, we proposed Datavoidant as a platform to allow journalists to strategically understand data voids. According to the journalists who evaluated our system, the process of locating data voids would be much slower without Datavoidant. Journalists also felt more confident  since they understood what information was necessary and how people engaged with particular types of information. Ideally, this will enable them to conduct strategic amplifications of content faster. Ultimately, Datavoidant enabled independent journalists to collaborate to fill critical voids in the information ecosystem and conduct ``strategic amplifications'' of content \cite{donovan2021stop}.  Datavoidant also provided journalists with the ability to identify unique angles for their news stories. During our evaluation, some journalists pointed out that Datavoidant had helped them identify novel interview questions for public officials. This brings several implications for designing new social computing systems that should help journalists to discover unique angles to stories. 

{\bf Mitigating Risks and Exploitation of Datavoidant by Bad Actors.}
Based on prior work, which has studied how to mitigate bad actors from exploiting tools intended for a collective good \cite{lilley2020using,li2018out}, an important next step in the development of Datavoidant is to define concrete mechanisms on who can access Datavoidant, and who is likely to be blacklisted. We can imagine that in order to use Datavoidant, journalists will need to share their reasons for wanting access. Journalists would be blacklisted and removed from access, if they are caught using the tool for other purposes. We envision connecting to the ``Ethical OS'' checklist to have an initial list of problematic usages that could be given to our tool. Journalists who express wanting to use our tool in problematic ways, or are caught engaging in such usages, would be banned from Datavoidant. We also imagine a group of trusted and experienced journalists helping to expand the checklist of problematic usages, based on their own experience, as well as motivated by the literature \cite{homoliak2019insight}. We believe it is critical to include the voices of trusted independent journalists working with underrepresented populations, as prior work might not understand in detail all of the problems and bad behaviors that can emerge when working with underrepresented communities. But journalists might have much deeper insight. it will also be important to understand how Datavoidant can create social and economic differences among independent journalists, specially those living in rural vs urban areas \cite{flores2020challenges}; and how Datavoidant could be used in collaborative settings in which paid senior journalists mentor aspiring journalists to create high-quality articles to fill data voids circulating within their communities \cite{saviaga2020understanding,flores2016leadwise}. Part of the solution is to release Datavoidant as open source, and hold workshops to ensure that a wide range of journalists can access our tool, which we plan to do. 

{\bf Limitations and Future Work.} Currently, Datavoidant works with Facebook information, which generates some challenges. For example, Facebook's algorithm may downrank or filter publications written by independent journalists. Secondly, independent journalists and news outlets may not have enough Facebook followers, thereby affecting their reach. We start to counter these challenges by helping journalists to collaborate to expand their network and visibility. Furthermore, CrowdTangle tracks interactions from popular public Facebook groups and pages (with at least 25k followers and 2K members, respectively). While this means that we cannot help journalists to engage with small private groups, we consider Datavoidant a step forward in enabling journalists to understand data voids targeting underrepresented populations. In future work, we plan to expand Datavoidant to include other social media sources and allow journalists to study data voids across platforms. Datavoidant also works with the groups and pages that journalists define. Despite doing their best to include pages and groups from across the political spectrum, there may be asymmetries in the political leanings of the pages and groups that Datavoidant is fed. (It can be unintentional biases generated from the groups and pages that journalists originally select.) This may result in an over or under representation of certain political viewpoints. If, for example, journalists feed Datavoidant with only left-leaning groups, Datavoidant will show them that no one from the right-leaning side of the political spectrum is discussing certain topics. Evidently, this may lead to a false impression of reality. In the future, Datavoidant could be modified to inform journalists that the number of groups and pages is unbalanced and encourage them to draw a more accurate picture of the ecosystem. Finally, our methods focused on breadth rather than depth. Future work could conduct an in-depth analysis of how journalists across the globe address data voids, and  how datavoidant is used long term by journalists.%In the future, Datavoidant will be made available as open-source.  

%Future work can also add geographical information to Datavoidant, this way Datavoidant can also be used to understand how the media portrays different groups in different contexts, for example, the media representation of underrepresented communities in urban areas vs rural areas, or how certain political topics are discussed within different geogrpahical regions. This could help to address problems related to the rural-urban divide.

\section{Conclusion}
In this study, we examined the practices of 22 independent journalists for covering political data voids targeted at underrepresented populations. Based on our findings, we created Datavoidant, an online collaborative tool that combines sensemaking theory, state-of-the-art machine learning models and data visualizations to help journalists on Facebook to collectively identify data voids in underrepresented communities at multiple levels.  Our evaluation revealed that journalists found that our tool was easy to use, and appreciated the collaborative features, intelligent summaries, deep dives, and multiple perspectives that Datavoidant offered to inspect and address data voids. 

\section{ACKNOWLEDGMENTS}
Special thanks to all the anonymous reviewers who helped us to strengthen the paper as well as the journalists who participated in the interviews and the user evaluation. This work was partially supported by NSF grant FW-HTF-19541.
\bibliographystyle{ACM-Reference-Format}
\bibliography{sample-base}


\begin{thebibliography}{156}

%%% the Web address.

\ifx \showCODEN    \undefined \def \showCODEN     #1{\unskip}     \fi
\ifx \showDOI      \undefined \def \showDOI       #1{#1}\fi
\ifx \showISBNx    \undefined \def \showISBNx     #1{\unskip}     \fi
\ifx \showISBNxiii \undefined \def \showISBNxiii  #1{\unskip}     \fi
\ifx \showISSN     \undefined \def \showISSN      #1{\unskip}     \fi
\ifx \showLCCN     \undefined \def \showLCCN      #1{\unskip}     \fi
\ifx \shownote     \undefined \def \shownote      #1{#1}          \fi
\ifx \showarticletitle \undefined \def \showarticletitle #1{#1}   \fi
\ifx \showURL      \undefined \def \showURL       {\relax}        \fi
% invisible to TeX
\providecommand\bibfield[2]{#2}
\providecommand\bibinfo[2]{#2}
\providecommand\natexlab[1]{#1}
\providecommand\showeprint[2][]{arXiv:#2}

\bibitem[\protect\citeauthoryear{??}{20T}{[n.\,d.]}]%
        {20TheNew25:online}
 \bibinfo{year}{[n.\,d.]}\natexlab{}.
\newblock \bibinfo{title}{(20+) The New York Times | Facebook}.
\newblock \bibinfo{howpublished}{\url{https://www.facebook.com/nytimes}}.
\newblock


\bibitem[\protect\citeauthoryear{??}{Che}{[n.\,d.]}]%
        {Checkche53:online}
 \bibinfo{year}{[n.\,d.]}\natexlab{}.
\newblock \bibinfo{title}{Check (@checkdesk) / Twitter}.
\newblock \bibinfo{howpublished}{\url{https://twitter.com/checkdesk}}.
\newblock


\bibitem[\protect\citeauthoryear{??}{Pen}{[n.\,d.]}]%
        {Pennsylv12:online}
 \bibinfo{year}{[n.\,d.]}\natexlab{}.
\newblock \bibinfo{title}{The Expanding News Desert}.
\newblock \bibinfo{howpublished}{\url{https://www.usnewsdeserts.com}}.
\newblock


\bibitem[\protect\citeauthoryear{??}{Joi}{[n.\,d.]}]%
        {Joinusin5:online}
 \bibinfo{year}{[n.\,d.]}\natexlab{}.
\newblock \bibinfo{title}{Join us in pushing back on misinformation.}
\newblock \bibinfo{howpublished}{\url{https://our.news/}}.
\newblock


\bibitem[\protect\citeauthoryear{??}{Pus}{[n.\,d.]}]%
        {PusherLe52:online}
 \bibinfo{year}{[n.\,d.]}\natexlab{}.
\newblock \bibinfo{title}{Pusher | Leader In Realtime Technologies}.
\newblock \bibinfo{howpublished}{\url{https://pusher.com/}}.
\newblock


\bibitem[\protect\citeauthoryear{??}{Dec}{2018}]%
        {Decadeso78:online}
 \bibinfo{year}{2018}\natexlab{}.
\newblock \bibinfo{title}{Decades of Failure - Columbia Journalism Review}.
\newblock
  \bibinfo{howpublished}{\url{https://www.cjr.org/special_report/race-ethnicity-newsrooms-data.php}}.
\newblock


\bibitem[\protect\citeauthoryear{??}{202}{2020a}]%
        {2020inUn45:online}
 \bibinfo{year}{2020}\natexlab{a}.
\newblock \bibinfo{title}{2020 in United States politics and government -
  Wikipedia}.
\newblock
  \bibinfo{howpublished}{\url{https://en.wikipedia.org/wiki/2020_in_United_States_politics_and_government}}.
\newblock


\bibitem[\protect\citeauthoryear{??}{202}{2020b}]%
        {2020Unit27:online}
 \bibinfo{year}{2020}\natexlab{b}.
\newblock \bibinfo{title}{2020 United States presidential election -
  Wikipedia}.
\newblock
  \bibinfo{howpublished}{\url{https://en.wikipedia.org/wiki/2020_United_States_presidential_election}}.
\newblock


\bibitem[\protect\citeauthoryear{??}{Lat}{2022}]%
        {LatinoMe80:online}
 \bibinfo{year}{2022}\natexlab{}.
\newblock \bibinfo{title}{Latino Media Report | The State of Latino News
  Media}.
\newblock
  \bibinfo{howpublished}{\url{http://thelatinomediareport.journalism.cuny.edu/latino-media-report/}}.
\newblock


\bibitem[\protect\citeauthoryear{??}{Onl}{2022}]%
        {Onlinegr23:online}
 \bibinfo{year}{2022}\natexlab{}.
\newblock \bibinfo{title}{Online group chat | Rumbletalk, chat for live events
  and websites}.
\newblock \bibinfo{howpublished}{\url{https://rumbletalk.com/}}.
\newblock


\bibitem[\protect\citeauthoryear{Ackoff}{Ackoff}{1989}]%
        {ackoff1989data}
\bibfield{author}{\bibinfo{person}{Russell~L Ackoff}.}
  \bibinfo{year}{1989}\natexlab{}.
\newblock \showarticletitle{From data to wisdom}.
\newblock \bibinfo{journal}{\emph{Journal of Applied Systems Analysis}}
  \bibinfo{volume}{16}, \bibinfo{number}{1} (\bibinfo{year}{1989}),
  \bibinfo{pages}{3--9}.
\newblock


\bibitem[\protect\citeauthoryear{Afrika~Check}{Afrika~Check}{2019}]%
        {FaceCheckPersuade:online}
\bibfield{author}{\bibinfo{person}{Full~Fact Afrika~Check, Chequeando}.}
  \bibinfo{year}{2019}\natexlab{}.
\newblock \bibinfo{title}{Fact checking doesn’t work (the way you think it
  does)}.
\newblock
  \bibinfo{howpublished}{\url{https://fullfact.org/blog/2019/jun/how-fact-checking-works/}}.
\newblock


\bibitem[\protect\citeauthoryear{Alam, Cresci, Chakraborty, Silvestri,
  Dimitrov, Da~San~Martino, Shaar, Firooz, and Nakov}{Alam
  et~al\mbox{.}}{2021}]%
        {alam2021survey}
\bibfield{author}{\bibinfo{person}{Firoj Alam}, \bibinfo{person}{Stefano
  Cresci}, \bibinfo{person}{Tanmoy Chakraborty}, \bibinfo{person}{Fabrizio
  Silvestri}, \bibinfo{person}{Dimiter Dimitrov}, \bibinfo{person}{Giovanni
  Da~San~Martino}, \bibinfo{person}{Shaden Shaar}, \bibinfo{person}{Hamed
  Firooz}, {and} \bibinfo{person}{Preslav Nakov}.}
  \bibinfo{year}{2021}\natexlab{}.
\newblock \showarticletitle{A Survey on Multimodal Disinformation Detection}.
\newblock \bibinfo{journal}{\emph{arXiv e-prints}} (\bibinfo{year}{2021}).
\newblock


\bibitem[\protect\citeauthoryear{Allen, Arechar, Pennycook, and Rand}{Allen
  et~al\mbox{.}}{2021}]%
        {allen2020scaling}
\bibfield{author}{\bibinfo{person}{Jennifer Allen}, \bibinfo{person}{Antonio~A
  Arechar}, \bibinfo{person}{Gordon Pennycook}, {and} \bibinfo{person}{David~G
  Rand}.} \bibinfo{year}{2021}\natexlab{}.
\newblock \showarticletitle{Scaling up fact-checking using the wisdom of
  crowds}.
\newblock \bibinfo{journal}{\emph{Science Advances}} \bibinfo{volume}{7},
  \bibinfo{number}{36} (\bibinfo{year}{2021}).
\newblock


\bibitem[\protect\citeauthoryear{Anderson and Rainie}{Anderson and
  Rainie}{2017}]%
        {anderson2017future}
\bibfield{author}{\bibinfo{person}{Janna Anderson} {and} \bibinfo{person}{Lee
  Rainie}.} \bibinfo{year}{2017}\natexlab{}.
\newblock \showarticletitle{The future of truth and misinformation online}.
\newblock \bibinfo{journal}{\emph{Pew Research Center}}  \bibinfo{volume}{19}
  (\bibinfo{year}{2017}).
\newblock


\bibitem[\protect\citeauthoryear{Arif, Stewart, and Starbird}{Arif
  et~al\mbox{.}}{2018}]%
        {arif2018acting}
\bibfield{author}{\bibinfo{person}{Ahmer Arif}, \bibinfo{person}{Leo~Graiden
  Stewart}, {and} \bibinfo{person}{Kate Starbird}.}
  \bibinfo{year}{2018}\natexlab{}.
\newblock \showarticletitle{Examining information operations within\#
  BlackLivesMatter discourse}.
\newblock \bibinfo{journal}{\emph{Proceedings of CSCW'18}}  \bibinfo{volume}{2}
  (\bibinfo{year}{2018}), \bibinfo{pages}{1--27}.
\newblock


\bibitem[\protect\citeauthoryear{Auxier}{Auxier}{2020}]%
        {Demograp95:online}
\bibfield{author}{\bibinfo{person}{Brooke Auxier}.}
  \bibinfo{year}{2020}\natexlab{}.
\newblock \bibinfo{title}{Social Media Use by Race | Pew Research Center}.
\newblock
  \bibinfo{howpublished}{\url{https://www.pewresearch.org/internet/chart/social-media-use-by-race/}}.
\newblock


\bibitem[\protect\citeauthoryear{Bayer, Bitiukova, Bard, Szak{\'a}cs, Alemanno,
  and Uszkiewicz}{Bayer et~al\mbox{.}}{2019}]%
        {bayer2019disinformation}
\bibfield{author}{\bibinfo{person}{Judit Bayer}, \bibinfo{person}{Natalija
  Bitiukova}, \bibinfo{person}{Petra Bard}, \bibinfo{person}{Judit
  Szak{\'a}cs}, \bibinfo{person}{Alberto Alemanno}, {and} \bibinfo{person}{Erik
  Uszkiewicz}.} \bibinfo{year}{2019}\natexlab{}.
\newblock \showarticletitle{Disinformation and propaganda--impact on the
  functioning of the rule of law in the EU and its Member States}.
\newblock \bibinfo{journal}{\emph{European Parliament, LIBE Committee, Policy
  Department for Citizens' Rights and Constitutional Affairs}}
  (\bibinfo{year}{2019}).
\newblock


\bibitem[\protect\citeauthoryear{Beers, Haughey, Arif, and Starbird}{Beers
  et~al\mbox{.}}{2020}]%
        {beers2020examining}
\bibfield{author}{\bibinfo{person}{Andrew Beers},
  \bibinfo{person}{Melinda~McClure Haughey}, \bibinfo{person}{Ahmer Arif},
  {and} \bibinfo{person}{Kate Starbird}.} \bibinfo{year}{2020}\natexlab{}.
\newblock \showarticletitle{Examining the digital toolsets of journalists
  reporting on disinformation}.
\newblock \bibinfo{journal}{\emph{Proceedings of Computation+ Journalism 2020
  (C+ J’20)}} (\bibinfo{year}{2020}).
\newblock


\bibitem[\protect\citeauthoryear{Berger and Morgan}{Berger and Morgan}{2015}]%
        {berger2015isis}
\bibfield{author}{\bibinfo{person}{Jonathon~M Berger} {and}
  \bibinfo{person}{Jonathon Morgan}.} \bibinfo{year}{2015}\natexlab{}.
\newblock \showarticletitle{Defining and describing the population of ISIS
  supporters on Twitter}.
\newblock \bibinfo{journal}{\emph{Center For Middle East Policy - Brookings}}
  (\bibinfo{year}{2015}).
\newblock


\bibitem[\protect\citeauthoryear{Bittman}{Bittman}{1985}]%
        {bittman1985kgb}
\bibfield{author}{\bibinfo{person}{Ladislav Bittman}.}
  \bibinfo{year}{1985}\natexlab{}.
\newblock \bibinfo{booktitle}{\emph{The KGB and Soviet disinformation: an
  insider's view}}.
\newblock \bibinfo{publisher}{Washington: Pergamon-Brassey's}.
\newblock


\bibitem[\protect\citeauthoryear{Blei, Ng, and Jordan}{Blei
  et~al\mbox{.}}{2003}]%
        {blei2003latent}
\bibfield{author}{\bibinfo{person}{David~M Blei}, \bibinfo{person}{Andrew~Y
  Ng}, {and} \bibinfo{person}{Michael~I Jordan}.}
  \bibinfo{year}{2003}\natexlab{}.
\newblock \showarticletitle{Latent dirichlet allocation}.
\newblock \bibinfo{journal}{\emph{the Journal of Machine Learning Research}}
  \bibinfo{volume}{3} (\bibinfo{year}{2003}), \bibinfo{pages}{993--1022}.
\newblock


\bibitem[\protect\citeauthoryear{Bradshaw and Howard}{Bradshaw and
  Howard}{2018}]%
        {bradshaw2018global}
\bibfield{author}{\bibinfo{person}{Samantha Bradshaw} {and}
  \bibinfo{person}{Philip~N Howard}.} \bibinfo{year}{2018}\natexlab{}.
\newblock \showarticletitle{The global organization of social media
  disinformation campaigns}.
\newblock \bibinfo{journal}{\emph{Journal of International Affairs}}
  \bibinfo{volume}{71}, \bibinfo{number}{1.5} (\bibinfo{year}{2018}),
  \bibinfo{pages}{23--32}.
\newblock


\bibitem[\protect\citeauthoryear{Brands, Graham, and Broersma}{Brands
  et~al\mbox{.}}{2018}]%
        {brands2018social}
\bibfield{author}{\bibinfo{person}{Bert~Jan Brands}, \bibinfo{person}{Todd
  Graham}, {and} \bibinfo{person}{Marcel Broersma}.}
  \bibinfo{year}{2018}\natexlab{}.
\newblock \showarticletitle{Social media sourcing practices: How Dutch
  newspapers use tweets in political news coverage}.
\newblock In \bibinfo{booktitle}{\emph{Managing Democracy in the Digital Age}}.
  \bibinfo{publisher}{Springer}, \bibinfo{pages}{159--178}.
\newblock


\bibitem[\protect\citeauthoryear{Brandtzaeg, L{\"u}ders, Spangenberg,
  Rath-Wiggins, and F{\o}lstad}{Brandtzaeg et~al\mbox{.}}{2016}]%
        {brandtzaeg2016emerging}
\bibfield{author}{\bibinfo{person}{Petter~Bae Brandtzaeg},
  \bibinfo{person}{Marika L{\"u}ders}, \bibinfo{person}{Jochen Spangenberg},
  \bibinfo{person}{Linda Rath-Wiggins}, {and} \bibinfo{person}{Asbj{\o}rn
  F{\o}lstad}.} \bibinfo{year}{2016}\natexlab{}.
\newblock \showarticletitle{Emerging journalistic verification practices
  concerning social media}.
\newblock \bibinfo{journal}{\emph{Journalism Practice}} \bibinfo{volume}{10},
  \bibinfo{number}{3} (\bibinfo{year}{2016}), \bibinfo{pages}{323--342}.
\newblock


\bibitem[\protect\citeauthoryear{Butcher and Sumner}{Butcher and
  Sumner}{2011}]%
        {butcher2011self}
\bibfield{author}{\bibinfo{person}{Kirsten~R Butcher} {and}
  \bibinfo{person}{Tamara Sumner}.} \bibinfo{year}{2011}\natexlab{}.
\newblock \showarticletitle{Self-directed learning and the sensemaking
  paradox}.
\newblock \bibinfo{journal}{\emph{Human--Computer Interaction}}
  \bibinfo{volume}{26}, \bibinfo{number}{1-2} (\bibinfo{year}{2011}).
\newblock


\bibitem[\protect\citeauthoryear{Caled and Silva}{Caled and Silva}{2021}]%
        {caled2021digital}
\bibfield{author}{\bibinfo{person}{Danielle Caled} {and}
  \bibinfo{person}{M{\'a}rio~J Silva}.} \bibinfo{year}{2021}\natexlab{}.
\newblock \showarticletitle{Digital media and misinformation: An outlook on
  multidisciplinary strategies against manipulation}.
\newblock \bibinfo{journal}{\emph{Journal of Computational Social Science}}
  (\bibinfo{year}{2021}), \bibinfo{pages}{1--37}.
\newblock


\bibitem[\protect\citeauthoryear{Card}{Card}{1999}]%
        {card1999readings}
\bibfield{author}{\bibinfo{person}{Mackinlay Card}.}
  \bibinfo{year}{1999}\natexlab{}.
\newblock \bibinfo{booktitle}{\emph{Readings in Information Visualization:
  Using Vision to Think}}.
\newblock \bibinfo{publisher}{Morgan Kaufmann}.
\newblock


\bibitem[\protect\citeauthoryear{Carlson, Robinson, and Lewis}{Carlson
  et~al\mbox{.}}{2021}]%
        {carlson2021digital}
\bibfield{author}{\bibinfo{person}{Matt Carlson}, \bibinfo{person}{Sue
  Robinson}, {and} \bibinfo{person}{Seth~C Lewis}.}
  \bibinfo{year}{2021}\natexlab{}.
\newblock \showarticletitle{Digital press criticism: The symbolic dimensions of
  Donald Trump’s assault on US journalists as the “enemy of the people”}.
\newblock \bibinfo{journal}{\emph{Digital Journalism}} \bibinfo{volume}{9},
  \bibinfo{number}{6} (\bibinfo{year}{2021}), \bibinfo{pages}{737--754}.
\newblock


\bibitem[\protect\citeauthoryear{Carneiro, Nacenta, Toniolo, Mendez, and
  Quigley}{Carneiro et~al\mbox{.}}{2019}]%
        {carneiro2019deb8}
\bibfield{author}{\bibinfo{person}{Guilherme Carneiro}, \bibinfo{person}{Miguel
  Nacenta}, \bibinfo{person}{Alice Toniolo}, \bibinfo{person}{Gonzalo Mendez},
  {and} \bibinfo{person}{Aaron~J Quigley}.} \bibinfo{year}{2019}\natexlab{}.
\newblock \showarticletitle{Deb8: A Tool for Collaborative Analysis of Video}.
  In \bibinfo{booktitle}{\emph{Proceedings of the 2019 ACM International
  Conference on Interactive Experiences for TV and Online Video}}.
  \bibinfo{pages}{47--58}.
\newblock


\bibitem[\protect\citeauthoryear{Castelo, Almeida, Elghafari, Santos, Pham,
  Nakamura, and Freire}{Castelo et~al\mbox{.}}{2019}]%
        {castelo2019topic}
\bibfield{author}{\bibinfo{person}{Sonia Castelo}, \bibinfo{person}{Thais
  Almeida}, \bibinfo{person}{Anas Elghafari}, \bibinfo{person}{A{\'e}cio
  Santos}, \bibinfo{person}{Kien Pham}, \bibinfo{person}{Eduardo Nakamura},
  {and} \bibinfo{person}{Juliana Freire}.} \bibinfo{year}{2019}\natexlab{}.
\newblock \showarticletitle{A topic-agnostic approach for identifying fake news
  pages}. In \bibinfo{booktitle}{\emph{WWW'19 Companion}}.
  \bibinfo{pages}{975--980}.
\newblock


\bibitem[\protect\citeauthoryear{Chou, Gaysynsky, and Vanderpool}{Chou
  et~al\mbox{.}}{2021}]%
        {chou2021covid}
\bibfield{author}{\bibinfo{person}{Wen-Ying~Sylvia Chou}, \bibinfo{person}{Anna
  Gaysynsky}, {and} \bibinfo{person}{Robin~C Vanderpool}.}
  \bibinfo{year}{2021}\natexlab{}.
\newblock \showarticletitle{The COVID-19 Misinfodemic: moving beyond
  fact-checking}.
\newblock \bibinfo{journal}{\emph{Health Education \& Behavior}}
  \bibinfo{volume}{48}, \bibinfo{number}{1} (\bibinfo{year}{2021}),
  \bibinfo{pages}{9--13}.
\newblock


\bibitem[\protect\citeauthoryear{Christopher~Bing}{Christopher~Bing}{2020}]%
        {Spanishl89:online}
\bibfield{author}{\bibinfo{person}{Paresh~Dave Christopher~Bing,
  Elizabeth~Culliford}.} \bibinfo{year}{2020}\natexlab{}.
\newblock \bibinfo{title}{Spanish-language misinformation dogged Democrats in
  U.S. election | Reuters}.
\newblock
  \bibinfo{howpublished}{\url{https://www.reuters.com/article/us-usa-election-disinformation-spanish/spanish-language-misinformation-dogged-democrats-in-u-s-election-idUSKBN27N0ED}}.
\newblock


\bibitem[\protect\citeauthoryear{Chung and Nah}{Chung and Nah}{2021}]%
        {chung2021community}
\bibfield{author}{\bibinfo{person}{Deborah~S Chung} {and}
  \bibinfo{person}{Seungahn Nah}.} \bibinfo{year}{2021}\natexlab{}.
\newblock \showarticletitle{Community Newspaper Editors’ Perspectives on News
  Collaboration: Participatory Opportunities and Ethical Considerations Toward
  Citizen News Engagement}.
\newblock \bibinfo{journal}{\emph{Journalism Practice}} (\bibinfo{year}{2021}),
  \bibinfo{pages}{1--21}.
\newblock


\bibitem[\protect\citeauthoryear{Cobian}{Cobian}{2019}]%
        {HowMisin70:online}
\bibfield{author}{\bibinfo{person}{Jessica Cobian}.}
  \bibinfo{year}{2019}\natexlab{}.
\newblock \bibinfo{title}{How Misinformation Fueled Anti-Immigrant Sentiment in
  the Tijuana Border Region - Center for American Progress}.
\newblock
  \bibinfo{howpublished}{\url{https://www.americanprogress.org/article/misinformation-fueled-anti-immigrant-sentiment-tijuana-border-region/}}.
\newblock


\bibitem[\protect\citeauthoryear{Dalgali and Crowston}{Dalgali and
  Crowston}{2020}]%
        {dalgali2020algorithmic}
\bibfield{author}{\bibinfo{person}{Ayse Dalgali} {and} \bibinfo{person}{Kevin
  Crowston}.} \bibinfo{year}{2020}\natexlab{}.
\newblock \showarticletitle{Algorithmic Journalism and Its Impacts on Work}.
\newblock  (\bibinfo{year}{2020}).
\newblock


\bibitem[\protect\citeauthoryear{Daniel and Jacquelyn}{Daniel and
  Jacquelyn}{2020}]%
        {Failuret73:online}
\bibfield{author}{\bibinfo{person}{Acosta-Ramos Daniel} {and}
  \bibinfo{person}{Mason Jacquelyn}.} \bibinfo{year}{2020}\natexlab{}.
\newblock \bibinfo{title}{Failure to understand Black and Latinx communities
  will result in a critical misunderstanding of the impact of disinformation}.
\newblock
  \bibinfo{howpublished}{\url{https://firstdraftnews.org/articles/black-latinx-disinformation-impact/}}.
\newblock


\bibitem[\protect\citeauthoryear{Deane}{Deane}{2016}]%
        {deane2016role}
\bibfield{author}{\bibinfo{person}{James Deane}.}
  \bibinfo{year}{2016}\natexlab{}.
\newblock \showarticletitle{The role of the independent media in curbing
  corruption in fragile settings}.
\newblock \bibinfo{journal}{\emph{Policy Brief}} (\bibinfo{year}{2016}).
\newblock


\bibitem[\protect\citeauthoryear{Deb, Luceri, Badaway, and Ferrara}{Deb
  et~al\mbox{.}}{2019}]%
        {deb2019perils}
\bibfield{author}{\bibinfo{person}{Ashok Deb}, \bibinfo{person}{Luca Luceri},
  \bibinfo{person}{Adam Badaway}, {and} \bibinfo{person}{Emilio Ferrara}.}
  \bibinfo{year}{2019}\natexlab{}.
\newblock \showarticletitle{Perils and challenges of social media and election
  manipulation analysis: The 2018 us midterms}. In
  \bibinfo{booktitle}{\emph{WWW'19 Companion}}. \bibinfo{pages}{237--247}.
\newblock


\bibitem[\protect\citeauthoryear{Devlin, Chang, Lee, and Toutanova}{Devlin
  et~al\mbox{.}}{2018}]%
        {devlin2018bert}
\bibfield{author}{\bibinfo{person}{Jacob Devlin}, \bibinfo{person}{Ming-Wei
  Chang}, \bibinfo{person}{Kenton Lee}, {and} \bibinfo{person}{Kristina
  Toutanova}.} \bibinfo{year}{2018}\natexlab{}.
\newblock \showarticletitle{Bert: Pre-training of deep bidirectional
  transformers for language understanding}.
\newblock  (\bibinfo{year}{2018}).
\newblock


\bibitem[\protect\citeauthoryear{Diakopoulos}{Diakopoulos}{2019}]%
        {diakopoulos2019automating}
\bibfield{author}{\bibinfo{person}{Nicholas Diakopoulos}.}
  \bibinfo{year}{2019}\natexlab{}.
\newblock \bibinfo{booktitle}{\emph{Automating the News}}.
\newblock \bibinfo{publisher}{Harvard University Press}.
\newblock


\bibitem[\protect\citeauthoryear{Diakopoulos, De~Choudhury, and
  Naaman}{Diakopoulos et~al\mbox{.}}{2012}]%
        {diakopoulos2012finding}
\bibfield{author}{\bibinfo{person}{Nicholas Diakopoulos},
  \bibinfo{person}{Munmun De~Choudhury}, {and} \bibinfo{person}{Mor Naaman}.}
  \bibinfo{year}{2012}\natexlab{}.
\newblock \showarticletitle{Finding and assessing social media information
  sources in the context of journalism}. In
  \bibinfo{booktitle}{\emph{Proceedings of the CHI'12}}.
  \bibinfo{pages}{2451--2460}.
\newblock


\bibitem[\protect\citeauthoryear{Direito-Rebollal, Negreira-Rey, and
  Rodr{\'\i}guez-V{\'a}zquez}{Direito-Rebollal et~al\mbox{.}}{2020}]%
        {direito2020social}
\bibfield{author}{\bibinfo{person}{Sabela Direito-Rebollal},
  \bibinfo{person}{Mar{\'\i}a-Cruz Negreira-Rey}, {and}
  \bibinfo{person}{Ana-Isabel Rodr{\'\i}guez-V{\'a}zquez}.}
  \bibinfo{year}{2020}\natexlab{}.
\newblock \showarticletitle{Social Media Guidelines for Journalists in European
  Public Service Media}.
\newblock In \bibinfo{booktitle}{\emph{Journalistic Metamorphosis}}.
  \bibinfo{publisher}{Springer}, \bibinfo{pages}{129--141}.
\newblock


\bibitem[\protect\citeauthoryear{Donovan and Boyd}{Donovan and Boyd}{2021}]%
        {donovan2021stop}
\bibfield{author}{\bibinfo{person}{Joan Donovan} {and} \bibinfo{person}{Danah
  Boyd}.} \bibinfo{year}{2021}\natexlab{}.
\newblock \showarticletitle{Stop the presses? Moving from strategic silence to
  strategic amplification in a networked media ecosystem}.
\newblock \bibinfo{journal}{\emph{American Behavioral Scientist}}
  \bibinfo{volume}{65}, \bibinfo{number}{2} (\bibinfo{year}{2021}),
  \bibinfo{pages}{333--350}.
\newblock


\bibitem[\protect\citeauthoryear{Dupuis, Chhor, and Ly}{Dupuis
  et~al\mbox{.}}{2021}]%
        {dupuis2021misinformation}
\bibfield{author}{\bibinfo{person}{Marc Dupuis}, \bibinfo{person}{Kelly Chhor},
  {and} \bibinfo{person}{Nhu Ly}.} \bibinfo{year}{2021}\natexlab{}.
\newblock \showarticletitle{Misinformation and Disinformation in the Era of
  COVID-19: The Role of Primary Information Sources and the Development of
  Attitudes Toward Vaccination}. In \bibinfo{booktitle}{\emph{Proceedings of
  the 22st Annual Conference on Information Technology Education}}.
\newblock


\bibitem[\protect\citeauthoryear{Entman}{Entman}{2010}]%
        {entman2010media}
\bibfield{author}{\bibinfo{person}{Robert~M Entman}.}
  \bibinfo{year}{2010}\natexlab{}.
\newblock \showarticletitle{Media framing biases and political power:
  Explaining slant in news of Campaign 2008}.
\newblock \bibinfo{journal}{\emph{Journalism}} \bibinfo{volume}{11},
  \bibinfo{number}{4} (\bibinfo{year}{2010}), \bibinfo{pages}{389--408}.
\newblock


\bibitem[\protect\citeauthoryear{Feng, Chen, Li, and Luo}{Feng
  et~al\mbox{.}}{2021a}]%
        {feng2021knowledge}
\bibfield{author}{\bibinfo{person}{Shangbin Feng}, \bibinfo{person}{Zilong
  Chen}, \bibinfo{person}{Qingyao Li}, {and} \bibinfo{person}{Minnan Luo}.}
  \bibinfo{year}{2021}\natexlab{a}.
\newblock \showarticletitle{Augmented Political Perspective Detection in News
  Media}.
\newblock \bibinfo{journal}{\emph{CoRR}} (\bibinfo{year}{2021}).
\newblock


\bibitem[\protect\citeauthoryear{Feng, Wan, Wang, Li, and Luo}{Feng
  et~al\mbox{.}}{2021c}]%
        {feng2021satar}
\bibfield{author}{\bibinfo{person}{Shangbin Feng}, \bibinfo{person}{Herun Wan},
  \bibinfo{person}{Ningnan Wang}, \bibinfo{person}{Jundong Li}, {and}
  \bibinfo{person}{Minnan Luo}.} \bibinfo{year}{2021}\natexlab{c}.
\newblock \showarticletitle{SATAR: A Self-supervised Approach to Twitter
  Account Representation Learning and its Application in Bot Detection}. In
  \bibinfo{booktitle}{\emph{Proceedings of CIKM'21}}.
  \bibinfo{pages}{3808--3817}.
\newblock


\bibitem[\protect\citeauthoryear{Feng, Wan, Wang, Li, and Luo}{Feng
  et~al\mbox{.}}{2021d}]%
        {feng2021twibot}
\bibfield{author}{\bibinfo{person}{Shangbin Feng}, \bibinfo{person}{Herun Wan},
  \bibinfo{person}{Ningnan Wang}, \bibinfo{person}{Jundong Li}, {and}
  \bibinfo{person}{Minnan Luo}.} \bibinfo{year}{2021}\natexlab{d}.
\newblock \showarticletitle{TwiBot-20: A Comprehensive Twitter Bot Detection
  Benchmark}. In \bibinfo{booktitle}{\emph{Proceedings of CIKM'21}}.
  \bibinfo{pages}{4485--4494}.
\newblock


\bibitem[\protect\citeauthoryear{Feng, Wan, Wang, and Luo}{Feng
  et~al\mbox{.}}{2021b}]%
        {feng2021botrgcn}
\bibfield{author}{\bibinfo{person}{Shangbin Feng}, \bibinfo{person}{Herun Wan},
  \bibinfo{person}{Ningnan Wang}, {and} \bibinfo{person}{Minnan Luo}.}
  \bibinfo{year}{2021}\natexlab{b}.
\newblock \showarticletitle{BotRGCN: Twitter bot detection with relational
  graph convolutional networks}. In \bibinfo{booktitle}{\emph{Proceedings of
  ASONAM'21}}. \bibinfo{pages}{236--239}.
\newblock


\bibitem[\protect\citeauthoryear{Fereday and Muir-Cochrane}{Fereday and
  Muir-Cochrane}{2006}]%
        {fereday2006demonstrating}
\bibfield{author}{\bibinfo{person}{Jennifer Fereday} {and}
  \bibinfo{person}{Eimear Muir-Cochrane}.} \bibinfo{year}{2006}\natexlab{}.
\newblock \showarticletitle{A hybrid approach of inductive and deductive coding
  and theme development}.
\newblock \bibinfo{journal}{\emph{International journal of qualitative
  methods}} \bibinfo{volume}{5}, \bibinfo{number}{1} (\bibinfo{year}{2006}),
  \bibinfo{pages}{80--92}.
\newblock


\bibitem[\protect\citeauthoryear{Ferrara}{Ferrara}{2017}]%
        {ferrara2017disinformation}
\bibfield{author}{\bibinfo{person}{Emilio Ferrara}.}
  \bibinfo{year}{2017}\natexlab{}.
\newblock \showarticletitle{Disinformation and social bot operations in the run
  up to the 2017 French presidential election}.
\newblock \bibinfo{journal}{\emph{First Monday}} (\bibinfo{year}{2017}).
\newblock


\bibitem[\protect\citeauthoryear{Flores-Saviaga, Granados, Savage, Escobedo,
  and Savage}{Flores-Saviaga et~al\mbox{.}}{2020a}]%
        {saviaga2020understanding}
\bibfield{author}{\bibinfo{person}{Claudia Flores-Saviaga},
  \bibinfo{person}{Ricardo Granados}, \bibinfo{person}{Liliana Savage},
  \bibinfo{person}{Lizbeth Escobedo}, {and} \bibinfo{person}{Saiph Savage}.}
  \bibinfo{year}{2020}\natexlab{a}.
\newblock \showarticletitle{Understanding the complementary nature of paid and
  volunteer crowds for content creation}.
\newblock \bibinfo{journal}{\emph{AIHC'20}} (\bibinfo{year}{2020}).
\newblock


\bibitem[\protect\citeauthoryear{Flores-Saviaga, Keegan, and
  Savage}{Flores-Saviaga et~al\mbox{.}}{2018}]%
        {flores2018mobilizing}
\bibfield{author}{\bibinfo{person}{Claudia Flores-Saviaga},
  \bibinfo{person}{Brian Keegan}, {and} \bibinfo{person}{Saiph Savage}.}
  \bibinfo{year}{2018}\natexlab{}.
\newblock \showarticletitle{Mobilizing the trump train: Understanding
  collective action in a political trolling community}. In
  \bibinfo{booktitle}{\emph{Proceedings of ICWSM'18}},
  Vol.~\bibinfo{volume}{12}.
\newblock


\bibitem[\protect\citeauthoryear{Flores-Saviaga, Li, Hanrahan, Bigham, and
  Savage}{Flores-Saviaga et~al\mbox{.}}{2020b}]%
        {flores2020challenges}
\bibfield{author}{\bibinfo{person}{Claudia Flores-Saviaga},
  \bibinfo{person}{Yuwen Li}, \bibinfo{person}{Benjamin Hanrahan},
  \bibinfo{person}{Jeffrey Bigham}, {and} \bibinfo{person}{Saiph Savage}.}
  \bibinfo{year}{2020}\natexlab{b}.
\newblock \showarticletitle{The challenges of crowd workers in rural and urban
  America}. In \bibinfo{booktitle}{\emph{Proceedings of HCOMP'2020}},
  Vol.~\bibinfo{volume}{8}. \bibinfo{pages}{159--162}.
\newblock


\bibitem[\protect\citeauthoryear{Flores-Saviaga and Savage}{Flores-Saviaga and
  Savage}{2019}]%
        {flores2019anti}
\bibfield{author}{\bibinfo{person}{Claudia Flores-Saviaga} {and}
  \bibinfo{person}{Saiph Savage}.} \bibinfo{year}{2019}\natexlab{}.
\newblock \showarticletitle{Anti-Latinx Computational Propaganda in the United
  States}.
\newblock \bibinfo{journal}{\emph{Institute For the Future}}
  (\bibinfo{year}{2019}).
\newblock


\bibitem[\protect\citeauthoryear{Flores-Saviaga and Savage}{Flores-Saviaga and
  Savage}{2021}]%
        {flores2021fighting}
\bibfield{author}{\bibinfo{person}{Claudia Flores-Saviaga} {and}
  \bibinfo{person}{Saiph Savage}.} \bibinfo{year}{2021}\natexlab{}.
\newblock \showarticletitle{Fighting disaster misinformation in Latin America:
  the\# 19S Mexican earthquake case study}.
\newblock \bibinfo{journal}{\emph{Personal and Ubiquitous Computing}}
  \bibinfo{volume}{25}, \bibinfo{number}{2} (\bibinfo{year}{2021}),
  \bibinfo{pages}{353--373}.
\newblock


\bibitem[\protect\citeauthoryear{Flores-Saviaga, Savage, and
  Taraborelli}{Flores-Saviaga et~al\mbox{.}}{2016}]%
        {flores2016leadwise}
\bibfield{author}{\bibinfo{person}{Claudia Flores-Saviaga},
  \bibinfo{person}{Saiph Savage}, {and} \bibinfo{person}{Dario Taraborelli}.}
  \bibinfo{year}{2016}\natexlab{}.
\newblock \showarticletitle{Leadwise: using online bots to recruite and guide
  expert volunteers}. In \bibinfo{booktitle}{\emph{CSCW'16 Companion}}.
  \bibinfo{pages}{257--260}.
\newblock


\bibitem[\protect\citeauthoryear{Forelle, Howard, Monroy-Hernandez, and
  Savage}{Forelle et~al\mbox{.}}{2015}]%
        {forelle2015political}
\bibfield{author}{\bibinfo{person}{Michelle~C Forelle},
  \bibinfo{person}{Philip~N Howard}, \bibinfo{person}{Andres Monroy-Hernandez},
  {and} \bibinfo{person}{Saiph Savage}.} \bibinfo{year}{2015}\natexlab{}.
\newblock \showarticletitle{Political Bots and the Manipulation of Public
  Opinion in Venezuela}.
\newblock \bibinfo{journal}{\emph{SSRN Electronic Journal}}
  (\bibinfo{year}{2015}).
\newblock


\bibitem[\protect\citeauthoryear{Gamboa}{Gamboa}{2020}]%
        {RepsMuca18:online}
\bibfield{author}{\bibinfo{person}{Suzanne Gamboa}.}
  \bibinfo{year}{2020}\natexlab{}.
\newblock \bibinfo{title}{Reps. Mucarsel-Powell, Castro call for FBI
  investigation of disinformation aimed at Latinos}.
\newblock
  \bibinfo{howpublished}{\url{https://www.nbcnews.com/news/latino/reps-mucarsel-powell-castro-call-fbi-investigation-disinformation-aimed-latinos-n1240853}}.
\newblock


\bibitem[\protect\citeauthoryear{Ghaffary}{Ghaffary}{2020}]%
        {Howfaken58:online}
\bibfield{author}{\bibinfo{person}{Shirin Ghaffary}.}
  \bibinfo{year}{2020}\natexlab{}.
\newblock \bibinfo{title}{How fake news aimed at Latinos thrives on social
  media - Vox}.
\newblock
  \bibinfo{howpublished}{\url{https://www.vox.com/recode/21574293/social-media-latino-voters-2020-election}}.
\newblock


\bibitem[\protect\citeauthoryear{Godel, Sanderson, Aslett, Nagler, Bonneau,
  Persily, and Tucker}{Godel et~al\mbox{.}}{2021}]%
        {godel2021moderating}
\bibfield{author}{\bibinfo{person}{William Godel}, \bibinfo{person}{Zeve
  Sanderson}, \bibinfo{person}{Kevin Aslett}, \bibinfo{person}{Jonathan
  Nagler}, \bibinfo{person}{Richard Bonneau}, \bibinfo{person}{Nathaniel
  Persily}, {and} \bibinfo{person}{Joshua Tucker}.}
  \bibinfo{year}{2021}\natexlab{}.
\newblock \showarticletitle{Moderating with the Mob: Evaluating the Efficacy of
  Real-Time Crowdsourced Fact-Checking}.
\newblock \bibinfo{journal}{\emph{Journal of Online Trust and Safety}}
  \bibinfo{volume}{1}, \bibinfo{number}{1} (\bibinfo{year}{2021}).
\newblock


\bibitem[\protect\citeauthoryear{Golebiewski and boyd}{Golebiewski and
  boyd}{2019}]%
        {golebiewski2018data}
\bibfield{author}{\bibinfo{person}{Michael Golebiewski} {and}
  \bibinfo{person}{danah boyd}.} \bibinfo{year}{2019}\natexlab{}.
\newblock \showarticletitle{Data Voids: Where Missing Data Can Easily Be
  Exploited}.
\newblock \bibinfo{journal}{\emph{Data \& Society}} (\bibinfo{date}{October}
  \bibinfo{year}{2019}).
\newblock
\urldef\tempurl%
\url{https://www.microsoft.com/en-us/research/publication/data-voids-where-missing-data-can-easily-be-exploited/}
\showURL{%
\tempurl}


\bibitem[\protect\citeauthoryear{Goodman}{Goodman}{1961}]%
        {goodman1961snowball}
\bibfield{author}{\bibinfo{person}{Leo~A Goodman}.}
  \bibinfo{year}{1961}\natexlab{}.
\newblock \showarticletitle{Snowball sampling}.
\newblock \bibinfo{journal}{\emph{The annals of mathematical statistics}}
  (\bibinfo{year}{1961}), \bibinfo{pages}{148--170}.
\newblock


\bibitem[\protect\citeauthoryear{Graves, Nyhan, and Reifler}{Graves
  et~al\mbox{.}}{2016}]%
        {graves2016journalists}
\bibfield{author}{\bibinfo{person}{Lucas Graves}, \bibinfo{person}{Brendan
  Nyhan}, {and} \bibinfo{person}{Jason Reifler}.}
  \bibinfo{year}{2016}\natexlab{}.
\newblock \showarticletitle{Why do journalists fact-check}.
\newblock \bibinfo{journal}{\emph{https://www. dartmouth. edu/\~{}
  nyhan/journalist-fact-checking. pdf}} (\bibinfo{year}{2016}).
\newblock


\bibitem[\protect\citeauthoryear{Guest, Bunce, and Johnson}{Guest
  et~al\mbox{.}}{2006}]%
        {guest2006many}
\bibfield{author}{\bibinfo{person}{Greg Guest}, \bibinfo{person}{Arwen Bunce},
  {and} \bibinfo{person}{Laura Johnson}.} \bibinfo{year}{2006}\natexlab{}.
\newblock \showarticletitle{How many interviews are enough? An experiment with
  data saturation and variability}.
\newblock \bibinfo{journal}{\emph{Field methods}} \bibinfo{volume}{18},
  \bibinfo{number}{1} (\bibinfo{year}{2006}), \bibinfo{pages}{59--82}.
\newblock


\bibitem[\protect\citeauthoryear{Hagen, Falling, Lisnichenko, Elmadany, Mehta,
  Abdul-Mageed, Costakis, and Keller}{Hagen et~al\mbox{.}}{2019}]%
        {hagen2019emoji}
\bibfield{author}{\bibinfo{person}{Loni Hagen}, \bibinfo{person}{Mary Falling},
  \bibinfo{person}{Oleksandr Lisnichenko}, \bibinfo{person}{AbdelRahim~A
  Elmadany}, \bibinfo{person}{Pankti Mehta}, \bibinfo{person}{Muhammad
  Abdul-Mageed}, \bibinfo{person}{Justin Costakis}, {and}
  \bibinfo{person}{Thomas~E Keller}.} \bibinfo{year}{2019}\natexlab{}.
\newblock \showarticletitle{Emoji use in Twitter white nationalism
  communication}. In \bibinfo{booktitle}{\emph{Proceedings of CSCW'19}}.
  \bibinfo{pages}{201--205}.
\newblock


\bibitem[\protect\citeauthoryear{Halper}{Halper}{2020}]%
        {Awarroom85:online}
\bibfield{author}{\bibinfo{person}{Evan Halper}.}
  \bibinfo{year}{2020}\natexlab{}.
\newblock \bibinfo{title}{A 'war room' arms Black, Latino voters against
  disinformation - Los Angeles Times}.
\newblock
  \bibinfo{howpublished}{\url{https://www.latimes.com/politics/story/2020-08-06/war-room-arms-black-latino-voters-against-disinformation}}.
\newblock


\bibitem[\protect\citeauthoryear{Haque, Yousuf, Alam, Saha, Ahmed, and
  Hassan}{Haque et~al\mbox{.}}{2020}]%
        {haque2020combating}
\bibfield{author}{\bibinfo{person}{Md~Mahfuzul Haque},
  \bibinfo{person}{Mohammad Yousuf}, \bibinfo{person}{Ahmed~Shatil Alam},
  \bibinfo{person}{Pratyasha Saha}, \bibinfo{person}{Syed~Ishtiaque Ahmed},
  {and} \bibinfo{person}{Naeemul Hassan}.} \bibinfo{year}{2020}\natexlab{}.
\newblock \showarticletitle{Combating Misinformation in Bangladesh: Roles and
  Responsibilities as Perceived by Journalists, Fact-checkers, and Users}.
\newblock \bibinfo{journal}{\emph{Proceedings of the ACM on Human-Computer
  Interaction}} \bibinfo{volume}{4}, \bibinfo{number}{CSCW2}
  (\bibinfo{year}{2020}), \bibinfo{pages}{1--32}.
\newblock


\bibitem[\protect\citeauthoryear{Haughey, Povolo, and Starbird}{Haughey
  et~al\mbox{.}}{2022}]%
        {haughey2022bridging}
\bibfield{author}{\bibinfo{person}{Melinda~McClure Haughey},
  \bibinfo{person}{Martina Povolo}, {and} \bibinfo{person}{Kate Starbird}.}
  \bibinfo{year}{2022}\natexlab{}.
\newblock \showarticletitle{Bridging Contextual and Methodological Gaps on the
  “Misinformation Beat”: Insights from Journalist-Researcher Collaborations
  at Speed}.
\newblock  (\bibinfo{year}{2022}).
\newblock


\bibitem[\protect\citeauthoryear{Hazard}{Hazard}{2020}]%
        {Spanishl58:online}
\bibfield{author}{\bibinfo{person}{Laura Hazard}.}
  \bibinfo{year}{2020}\natexlab{}.
\newblock \bibinfo{title}{Spanish-language misinformation is flourishing —
  and often hidden. Is help on the way? | Nieman Journalism Lab}.
\newblock
  \bibinfo{howpublished}{\url{https://www.niemanlab.org/2020/09/spanish-language-misinformation-is-flourishing-and-often-hidden-is-help-on-the-way/}}.
\newblock


\bibitem[\protect\citeauthoryear{Herman and Chomsky}{Herman and
  Chomsky}{2010}]%
        {herman2010manufacturing}
\bibfield{author}{\bibinfo{person}{Edward~S Herman} {and} \bibinfo{person}{Noam
  Chomsky}.} \bibinfo{year}{2010}\natexlab{}.
\newblock \bibinfo{booktitle}{\emph{Manufacturing consent: The political
  economy of the mass media}}.
\newblock \bibinfo{publisher}{Random House}.
\newblock


\bibitem[\protect\citeauthoryear{Hermida}{Hermida}{2012}]%
        {hermida2012social}
\bibfield{author}{\bibinfo{person}{Alfred Hermida}.}
  \bibinfo{year}{2012}\natexlab{}.
\newblock \showarticletitle{Social journalism: Exploring how social media is
  shaping journalism}.
\newblock \bibinfo{journal}{\emph{The Handbook of Global Online Journalism}}
  \bibinfo{volume}{12} (\bibinfo{year}{2012}), \bibinfo{pages}{309--328}.
\newblock


\bibitem[\protect\citeauthoryear{Hernandez}{Hernandez}{2020}]%
        {Talkinga39:online}
\bibfield{author}{\bibinfo{person}{John Hernandez}.}
  \bibinfo{year}{2020}\natexlab{}.
\newblock \bibinfo{title}{Talking about misinformation with First Draft -
  American Press Institute}.
\newblock
  \bibinfo{howpublished}{\url{https://www.americanpressinstitute.org/publications/elections/trusted-elections-network/talking-about-misinformation-with-first-draft/}}.
\newblock


\bibitem[\protect\citeauthoryear{Homoliak, Toffalini, Guarnizo, Elovici, and
  Ochoa}{Homoliak et~al\mbox{.}}{2019}]%
        {homoliak2019insight}
\bibfield{author}{\bibinfo{person}{Ivan Homoliak}, \bibinfo{person}{Flavio
  Toffalini}, \bibinfo{person}{Juan Guarnizo}, \bibinfo{person}{Yuval Elovici},
  {and} \bibinfo{person}{Mart{\'\i}n Ochoa}.} \bibinfo{year}{2019}\natexlab{}.
\newblock \showarticletitle{Insight into insiders and it: A survey of insider
  threat taxonomies, analysis, modeling, and countermeasures}.
\newblock \bibinfo{journal}{\emph{ACM Computing Surveys (CSUR)}}
  \bibinfo{volume}{52}, \bibinfo{number}{2} (\bibinfo{year}{2019}),
  \bibinfo{pages}{1--40}.
\newblock


\bibitem[\protect\citeauthoryear{Hyde}{Hyde}{2002}]%
        {hyde2002independent}
\bibfield{author}{\bibinfo{person}{Gene Hyde}.}
  \bibinfo{year}{2002}\natexlab{}.
\newblock \showarticletitle{Independent Media Centers: Cyber-subversion and the
  alternative press}.
\newblock \bibinfo{journal}{\emph{First Monday}} (\bibinfo{year}{2002}).
\newblock


\bibitem[\protect\citeauthoryear{Ireton and Posetti}{Ireton and
  Posetti}{2018}]%
        {ireton2018journalism}
\bibfield{author}{\bibinfo{person}{Cherilyn Ireton} {and}
  \bibinfo{person}{Julie Posetti}.} \bibinfo{year}{2018}\natexlab{}.
\newblock \bibinfo{booktitle}{\emph{Journalism, Fake News \& Disinformation:
  Handbook for Journalism Education and Training}}.
\newblock \bibinfo{publisher}{Unesco}.
\newblock


\bibitem[\protect\citeauthoryear{Ismail}{Ismail}{2018}]%
        {ismail2018strengthening}
\bibfield{author}{\bibinfo{person}{Zenobia Ismail}.}
  \bibinfo{year}{2018}\natexlab{}.
\newblock \showarticletitle{Strengthening the financial independence of
  independent media organisations}.
\newblock  (\bibinfo{year}{2018}).
\newblock


\bibitem[\protect\citeauthoryear{Jack}{Jack}{2017}]%
        {jack2017lexicon}
\bibfield{author}{\bibinfo{person}{Caroline Jack}.}
  \bibinfo{year}{2017}\natexlab{}.
\newblock \showarticletitle{Lexicon of lies: Terms for problematic
  information}.
\newblock \bibinfo{journal}{\emph{Data \& Society}} \bibinfo{volume}{3},
  \bibinfo{number}{22} (\bibinfo{year}{2017}), \bibinfo{pages}{1094--1096}.
\newblock


\bibitem[\protect\citeauthoryear{Ji, Ha, and Sypher}{Ji et~al\mbox{.}}{2014}]%
        {ji2014role}
\bibfield{author}{\bibinfo{person}{Qihao Ji}, \bibinfo{person}{Louisa Ha},
  {and} \bibinfo{person}{Ulla Sypher}.} \bibinfo{year}{2014}\natexlab{}.
\newblock \showarticletitle{The role of news media use and demographic
  characteristics in the possibility of information overload prediction}.
\newblock \bibinfo{journal}{\emph{International Journal of Communication}}
  \bibinfo{volume}{8} (\bibinfo{year}{2014}), \bibinfo{pages}{16}.
\newblock


\bibitem[\protect\citeauthoryear{Jiang and Wilson}{Jiang and Wilson}{2021}]%
        {jiang2021structurizing}
\bibfield{author}{\bibinfo{person}{Shan Jiang} {and} \bibinfo{person}{Christo
  Wilson}.} \bibinfo{year}{2021}\natexlab{}.
\newblock \showarticletitle{Structurizing misinformation stories via
  rationalizing fact-checks}. In \bibinfo{booktitle}{\emph{Proceedings of ACL |
  IJCNLP'21}}. \bibinfo{pages}{617--631}.
\newblock


\bibitem[\protect\citeauthoryear{Judit and Bognar}{Judit and Bognar}{2021}]%
        {judit2021}
\bibfield{author}{\bibinfo{person}{Szakacs Judit} {and} \bibinfo{person}{Eva
  Bognar}.} \bibinfo{year}{2021}\natexlab{}.
\newblock \showarticletitle{The impact of disinformation campaigns about
  migrants and minority groups in the EU}.
\newblock  (\bibinfo{year}{2021}).
\newblock


\bibitem[\protect\citeauthoryear{Karmakharm, Aletras, and Bontcheva}{Karmakharm
  et~al\mbox{.}}{2019}]%
        {karmakharm2019journalist}
\bibfield{author}{\bibinfo{person}{Twin Karmakharm}, \bibinfo{person}{Nikolaos
  Aletras}, {and} \bibinfo{person}{Kalina Bontcheva}.}
  \bibinfo{year}{2019}\natexlab{}.
\newblock \showarticletitle{Journalist-in-the-loop: Continuous learning as a
  service for rumour analysis}. In \bibinfo{booktitle}{\emph{Proceedings of
  EMNLP-IJCNLP'19}}. \bibinfo{pages}{115--120}.
\newblock


\bibitem[\protect\citeauthoryear{Komatsu, Gutierrez~Lopez, Makri, Porlezza,
  Cooper, MacFarlane, and Missaoui}{Komatsu et~al\mbox{.}}{2020}]%
        {komatsu2020ai}
\bibfield{author}{\bibinfo{person}{Tomoko Komatsu}, \bibinfo{person}{Marisela
  Gutierrez~Lopez}, \bibinfo{person}{Stephann Makri}, \bibinfo{person}{Colin
  Porlezza}, \bibinfo{person}{Glenda Cooper}, \bibinfo{person}{Andrew
  MacFarlane}, {and} \bibinfo{person}{Sondess Missaoui}.}
  \bibinfo{year}{2020}\natexlab{}.
\newblock \showarticletitle{AI should embody our values: Investigating
  journalistic values to inform AI technology design}. In
  \bibinfo{booktitle}{\emph{Proceedings of NordiCHI'20}}.
  \bibinfo{pages}{1--13}.
\newblock


\bibitem[\protect\citeauthoryear{Kou, Gui, Chen, and Pine}{Kou
  et~al\mbox{.}}{2017}]%
        {kou2017conspiracy}
\bibfield{author}{\bibinfo{person}{Yubo Kou}, \bibinfo{person}{Xinning Gui},
  \bibinfo{person}{Yunan Chen}, {and} \bibinfo{person}{Kathleen Pine}.}
  \bibinfo{year}{2017}\natexlab{}.
\newblock \showarticletitle{Conspiracy talk on social media: collective
  sensemaking during a public health crisis}.
\newblock \bibinfo{journal}{\emph{Proceedings of the ACM on Human-Computer
  Interaction}} \bibinfo{volume}{1}, \bibinfo{number}{CSCW}
  (\bibinfo{year}{2017}), \bibinfo{pages}{1--21}.
\newblock


\bibitem[\protect\citeauthoryear{Krogstad and Lopez}{Krogstad and
  Lopez}{2020}]%
        {Topelect86:online}
\bibfield{author}{\bibinfo{person}{Jens Krogstad} {and} \bibinfo{person}{Mark
  Lopez}.} \bibinfo{year}{2020}\natexlab{}.
\newblock \bibinfo{title}{Top election issues for Latino voters: Economy,
  health care, COVID-19 | Pew Research Center}.
\newblock
  \bibinfo{howpublished}{\url{https://www.pewresearch.org/fact-tank/2020/09/11/hispanic-voters-say-economy-health-care-and-covid-19-are-top-issues-in-2020-presidential-election/}}.
\newblock


\bibitem[\protect\citeauthoryear{Lampinen, Lehtinen, Lehmuskallio, and
  Tamminen}{Lampinen et~al\mbox{.}}{2011}]%
        {lampinen2011we}
\bibfield{author}{\bibinfo{person}{Airi Lampinen}, \bibinfo{person}{Vilma
  Lehtinen}, \bibinfo{person}{Asko Lehmuskallio}, {and} \bibinfo{person}{Sakari
  Tamminen}.} \bibinfo{year}{2011}\natexlab{}.
\newblock \showarticletitle{We're in it together: interpersonal management of
  disclosure in social network services}. In
  \bibinfo{booktitle}{\emph{Proceedings of CHI'11}}.
  \bibinfo{pages}{3217--3226}.
\newblock


\bibitem[\protect\citeauthoryear{Lee, Yang, Inchoco, Jones, and
  Satyanarayan}{Lee et~al\mbox{.}}{2021}]%
        {lee2021viral}
\bibfield{author}{\bibinfo{person}{Crystal Lee}, \bibinfo{person}{Tanya Yang},
  \bibinfo{person}{Gabrielle~D Inchoco}, \bibinfo{person}{Graham~M Jones},
  {and} \bibinfo{person}{Arvind Satyanarayan}.}
  \bibinfo{year}{2021}\natexlab{}.
\newblock \showarticletitle{Viral Visualizations: How Coronavirus Skeptics Use
  Orthodox Data Practices to Promote Unorthodox Science Online}. In
  \bibinfo{booktitle}{\emph{Proceedings of CHI'21}}. \bibinfo{pages}{1--18}.
\newblock


\bibitem[\protect\citeauthoryear{Lewandowsky, Ecker, Seifert, Schwarz, and
  Cook}{Lewandowsky et~al\mbox{.}}{2012}]%
        {lewandowsky2012misinformation}
\bibfield{author}{\bibinfo{person}{Stephan Lewandowsky},
  \bibinfo{person}{Ullrich~KH Ecker}, \bibinfo{person}{Colleen~M Seifert},
  \bibinfo{person}{Norbert Schwarz}, {and} \bibinfo{person}{John Cook}.}
  \bibinfo{year}{2012}\natexlab{}.
\newblock \showarticletitle{Misinformation and its correction: Continued
  influence and successful debiasing}.
\newblock \bibinfo{journal}{\emph{Psychological science in the public
  interest}} \bibinfo{volume}{13}, \bibinfo{number}{3} (\bibinfo{year}{2012}),
  \bibinfo{pages}{106--131}.
\newblock


\bibitem[\protect\citeauthoryear{Li, Alarcon, Milkes~Espinosa, and Hecht}{Li
  et~al\mbox{.}}{2018a}]%
        {li2018out}
\bibfield{author}{\bibinfo{person}{Hanlin Li}, \bibinfo{person}{Bodhi Alarcon},
  \bibinfo{person}{Sara Milkes~Espinosa}, {and} \bibinfo{person}{Brent Hecht}.}
  \bibinfo{year}{2018}\natexlab{a}.
\newblock \showarticletitle{Out of site: Empowering a new approach to online
  boycotts}.
\newblock \bibinfo{journal}{\emph{Proceedings of the ACM on Human-Computer
  Interaction}} \bibinfo{volume}{2}, \bibinfo{number}{CSCW}
  (\bibinfo{year}{2018}), \bibinfo{pages}{1--28}.
\newblock


\bibitem[\protect\citeauthoryear{Li, Luther, and North}{Li
  et~al\mbox{.}}{2018b}]%
        {li2018crowdia}
\bibfield{author}{\bibinfo{person}{Tianyi Li}, \bibinfo{person}{Kurt Luther},
  {and} \bibinfo{person}{Chris North}.} \bibinfo{year}{2018}\natexlab{b}.
\newblock \showarticletitle{Crowdia: Solving mysteries with crowdsourced
  sensemaking}.
\newblock \bibinfo{journal}{\emph{Proceedings of the ACM on Human-Computer
  Interaction}} \bibinfo{volume}{2}, \bibinfo{number}{CSCW}
  (\bibinfo{year}{2018}), \bibinfo{pages}{1--29}.
\newblock


\bibitem[\protect\citeauthoryear{Lilley, Currie, Pyper, and Attwood}{Lilley
  et~al\mbox{.}}{2020}]%
        {lilley2020using}
\bibfield{author}{\bibinfo{person}{Mariana Lilley}, \bibinfo{person}{Anne
  Currie}, \bibinfo{person}{Andrew Pyper}, {and} \bibinfo{person}{Sue
  Attwood}.} \bibinfo{year}{2020}\natexlab{}.
\newblock \showarticletitle{Using the Ethical OS Toolkit to Mitigate the Risk
  of Unintended Consequences}. In \bibinfo{booktitle}{\emph{International
  Conference on Human-Computer Interaction}}. Springer,
  \bibinfo{pages}{77--82}.
\newblock


\bibitem[\protect\citeauthoryear{Lin, Hsu, Yao, Lu, Kuo, Lin, and Chang}{Lin
  et~al\mbox{.}}{2021}]%
        {lin2021sync}
\bibfield{author}{\bibinfo{person}{Jin-An Lin}, \bibinfo{person}{Feng-Yi Hsu},
  \bibinfo{person}{Hsin-Yu Yao}, \bibinfo{person}{Shang-Hsun Lu},
  \bibinfo{person}{Tsai-Yu Kuo}, \bibinfo{person}{Chieh-Kai Lin}, {and}
  \bibinfo{person}{Yung-Ju Chang}.} \bibinfo{year}{2021}\natexlab{}.
\newblock \showarticletitle{SYNC: A Crowdsourcing Platform for News
  Co-editing}. In \bibinfo{booktitle}{\emph{CSCW'21}}.
  \bibinfo{pages}{126--129}.
\newblock


\bibitem[\protect\citeauthoryear{Liu, Ott, Goyal, Du, Joshi, Chen, Levy, Lewis,
  Zettlemoyer, and Stoyanov}{Liu et~al\mbox{.}}{2019}]%
        {liu2019roberta}
\bibfield{author}{\bibinfo{person}{Yinhan Liu}, \bibinfo{person}{Myle Ott},
  \bibinfo{person}{Naman Goyal}, \bibinfo{person}{Jingfei Du},
  \bibinfo{person}{Mandar Joshi}, \bibinfo{person}{Danqi Chen},
  \bibinfo{person}{Omer Levy}, \bibinfo{person}{Mike Lewis},
  \bibinfo{person}{Luke Zettlemoyer}, {and} \bibinfo{person}{Veselin
  Stoyanov}.} \bibinfo{year}{2019}\natexlab{}.
\newblock \showarticletitle{Roberta: A robustly optimized bert pretraining
  approach}.
\newblock  (\bibinfo{year}{2019}).
\newblock


\bibitem[\protect\citeauthoryear{Lokot and Diakopoulos}{Lokot and
  Diakopoulos}{2016}]%
        {lokot2016news}
\bibfield{author}{\bibinfo{person}{Tetyana Lokot} {and}
  \bibinfo{person}{Nicholas Diakopoulos}.} \bibinfo{year}{2016}\natexlab{}.
\newblock \showarticletitle{Automating news and information dissemination on
  Twitter}.
\newblock \bibinfo{journal}{\emph{Digital Journalism}} (\bibinfo{year}{2016}).
\newblock


\bibitem[\protect\citeauthoryear{Lu, Jiang, Shen, Jack, Wigdor, and Naaman}{Lu
  et~al\mbox{.}}{2021}]%
        {lu2021positive}
\bibfield{author}{\bibinfo{person}{Zhicong Lu}, \bibinfo{person}{Yue Jiang},
  \bibinfo{person}{Chenxinran Shen}, \bibinfo{person}{Margaret~C Jack},
  \bibinfo{person}{Daniel Wigdor}, {and} \bibinfo{person}{Mor Naaman}.}
  \bibinfo{year}{2021}\natexlab{}.
\newblock \showarticletitle{" Positive Energy" Perceptions and Attitudes
  Towards COVID-19 Information on Social Media in China}.
\newblock \bibinfo{journal}{\emph{Proceedings of the ACM on Human-Computer
  Interaction}} \bibinfo{volume}{5}, \bibinfo{number}{CSCW1}
  (\bibinfo{year}{2021}), \bibinfo{pages}{1--25}.
\newblock


\bibitem[\protect\citeauthoryear{Maly}{Maly}{2021}]%
        {PoliticalDataVoid2:online}
\bibfield{author}{\bibinfo{person}{Ico Maly}.} \bibinfo{year}{2021}\natexlab{}.
\newblock \bibinfo{title}{Data voids: What are they and do they threaten
  democracy?}
\newblock
  \bibinfo{howpublished}{\url{https://www.diggitmagazine.com/articles/data-voids-what-are-they-and-do-they-threaten-democracy}}.
\newblock


\bibitem[\protect\citeauthoryear{Matatov, Bechhofer, Aroyo, Amir, and
  Naaman}{Matatov et~al\mbox{.}}{2018}]%
        {matatov2018dejavu}
\bibfield{author}{\bibinfo{person}{Hana Matatov}, \bibinfo{person}{Adina
  Bechhofer}, \bibinfo{person}{Lora Aroyo}, \bibinfo{person}{Ofra Amir}, {and}
  \bibinfo{person}{Mor Naaman}.} \bibinfo{year}{2018}\natexlab{}.
\newblock \showarticletitle{DejaVu: a system for journalists to collaboratively
  address visual misinformation}. In \bibinfo{booktitle}{\emph{Computation+
  Journalism Symposium. Miami}}.
\newblock


\bibitem[\protect\citeauthoryear{McClure~Haughey, Muralikumar, Wood, and
  Starbird}{McClure~Haughey et~al\mbox{.}}{2020}]%
        {mcclure2020misinformation}
\bibfield{author}{\bibinfo{person}{Melinda McClure~Haughey},
  \bibinfo{person}{Meena~Devii Muralikumar}, \bibinfo{person}{Cameron~A Wood},
  {and} \bibinfo{person}{Kate Starbird}.} \bibinfo{year}{2020}\natexlab{}.
\newblock \showarticletitle{On the Misinformation Beat: Understanding the Work
  of Investigative Journalists Reporting on Problematic Information Online}.
\newblock \bibinfo{journal}{\emph{Proceedings of CSCW'20}}  \bibinfo{volume}{4}
  (\bibinfo{year}{2020}), \bibinfo{pages}{1--22}.
\newblock


\bibitem[\protect\citeauthoryear{Mesquita and de~Lima-Santos}{Mesquita and
  de~Lima-Santos}{2021}]%
        {mesquita2021collaborative}
\bibfield{author}{\bibinfo{person}{Lucia Mesquita} {and}
  \bibinfo{person}{Mathias-Felipe de Lima-Santos}.}
  \bibinfo{year}{2021}\natexlab{}.
\newblock \showarticletitle{Collaborative Journalism from a Latin American
  Perspective: An Empirical Analysis}.
\newblock \bibinfo{journal}{\emph{Journalism and Media}} \bibinfo{volume}{2},
  \bibinfo{number}{4} (\bibinfo{year}{2021}), \bibinfo{pages}{545--571}.
\newblock


\bibitem[\protect\citeauthoryear{Mihas}{Mihas}{2019}]%
        {mihas2019qualitative}
\bibfield{author}{\bibinfo{person}{Paul Mihas}.}
  \bibinfo{year}{2019}\natexlab{}.
\newblock \showarticletitle{Qualitative data analysis}.
\newblock In \bibinfo{booktitle}{\emph{Oxford research encyclopedia of
  education}}.
\newblock


\bibitem[\protect\citeauthoryear{Newman}{Newman}{2009}]%
        {newman2009rise}
\bibfield{author}{\bibinfo{person}{Nic Newman}.}
  \bibinfo{year}{2009}\natexlab{}.
\newblock \showarticletitle{The rise of social media and its impact on
  mainstream journalism}.
\newblock \bibinfo{journal}{\emph{Reuters Institute for the Study of
  Journalism}} \bibinfo{volume}{8}, \bibinfo{number}{2} (\bibinfo{year}{2009}).
\newblock


\bibitem[\protect\citeauthoryear{Newman}{Newman}{2011}]%
        {newman2011mainstream}
\bibfield{author}{\bibinfo{person}{Nic Newman}.}
  \bibinfo{year}{2011}\natexlab{}.
\newblock \showarticletitle{Mainstream media and the distribution of news in
  the age of social discovery}.
\newblock \bibinfo{journal}{\emph{Reuters Institute for the Study of
  Journalism, University of Oxford}} (\bibinfo{year}{2011}),
  \bibinfo{pages}{6}.
\newblock


\bibitem[\protect\citeauthoryear{Newman, Richard~Fletcher, Simge~Andı, and
  Nielsen}{Newman et~al\mbox{.}}{2021}]%
        {DigitalN17:online}
\bibfield{author}{\bibinfo{person}{Nic Newman}, \bibinfo{person}{Anne~Schulz
  Richard~Fletcher}, \bibinfo{person}{Craig T.~Robertson Simge~Andı}, {and}
  \bibinfo{person}{Rasmus~Kleis Nielsen}.} \bibinfo{year}{2021}\natexlab{}.
\newblock \bibinfo{title}{Digital News Report 2021 | Reuters Institute for the
  Study of Journalism}.
\newblock
  \bibinfo{howpublished}{\url{https://reutersinstitute.politics.ox.ac.uk/digital-news-report/2021}}.
\newblock


\bibitem[\protect\citeauthoryear{nhmc}{nhmc}{2020}]%
        {Facebook53:online}
\bibfield{author}{\bibinfo{person}{nhmc}.} \bibinfo{year}{2020}\natexlab{}.
\newblock \bibinfo{title}{Facebook’s Spanish Language Disinformation Gap –
  NHMC National Hispanic Media Coalition}.
\newblock
  \bibinfo{howpublished}{\url{https://www.nhmc.org/facebooks-spanish-language-disinformation-gap/}}.
\newblock


\bibitem[\protect\citeauthoryear{Nielsen}{Nielsen}{1994}]%
        {nielsen1994usability}
\bibfield{author}{\bibinfo{person}{Jakob Nielsen}.}
  \bibinfo{year}{1994}\natexlab{}.
\newblock \bibinfo{booktitle}{\emph{Usability Engineering}}.
\newblock \bibinfo{publisher}{Morgan Kaufmann}.
\newblock


\bibitem[\protect\citeauthoryear{Noain-S{\'a}nchez}{Noain-S{\'a}nchez}{2020a}]%
        {noain202013}
\bibfield{author}{\bibinfo{person}{Amaya Noain-S{\'a}nchez}.}
  \bibinfo{year}{2020}\natexlab{a}.
\newblock \showarticletitle{13 Collaborative journalism versus disinformation}.
\newblock \bibinfo{journal}{\emph{The Politics of Technology in Latin America
  (Volume 2): Digital Media, Daily Life and Public Engagement}}
  (\bibinfo{year}{2020}), \bibinfo{pages}{194}.
\newblock


\bibitem[\protect\citeauthoryear{Noain-S{\'a}nchez}{Noain-S{\'a}nchez}{2020b}]%
        {noain2020collaborative}
\bibfield{author}{\bibinfo{person}{Amaya Noain-S{\'a}nchez}.}
  \bibinfo{year}{2020}\natexlab{b}.
\newblock \showarticletitle{Collaborative journalism versus disinformation: An
  approach to fact-checking projects in Mexico, Argentina, Colombia, Brazil,
  and Spain}.
\newblock \bibinfo{journal}{\emph{The Politics of Technology in Latin America
  (Volume 2)}} (\bibinfo{year}{2020}), \bibinfo{pages}{194--211}.
\newblock


\bibitem[\protect\citeauthoryear{Norman}{Norman}{2013}]%
        {norman2013design}
\bibfield{author}{\bibinfo{person}{Don Norman}.}
  \bibinfo{year}{2013}\natexlab{}.
\newblock \bibinfo{booktitle}{\emph{The Design of Everyday Things: Revised and
  Expanded Edition}}.
\newblock \bibinfo{publisher}{Basic books}.
\newblock


\bibitem[\protect\citeauthoryear{page}{page}{2020}]%
        {sbee61:online}
\bibfield{author}{\bibinfo{person}{Tate Ryan-Mosleyarchive page}.}
  \bibinfo{year}{2020}\natexlab{}.
\newblock \bibinfo{title}{It’s been really, really bad: How Hispanic voters
  are being targeted by disinformation | MIT Technology Review}.
\newblock
  \bibinfo{howpublished}{\url{https://www.technologyreview.com/2020/10/12/1010061/hispanic-voter-political-targeting-facebook-whatsapp/}}.
\newblock


\bibitem[\protect\citeauthoryear{Pasquetto, Olivieri, Tacchetti, Riotta, and
  Spada}{Pasquetto et~al\mbox{.}}{2022}]%
        {pasquetto2022disinformation}
\bibfield{author}{\bibinfo{person}{Irene~V Pasquetto},
  \bibinfo{person}{Alberto~F Olivieri}, \bibinfo{person}{Luca Tacchetti},
  \bibinfo{person}{Gianni Riotta}, {and} \bibinfo{person}{Alessandra Spada}.}
  \bibinfo{year}{2022}\natexlab{}.
\newblock \showarticletitle{Disinformation as Infrastructure: Making and
  maintaining the QAnon conspiracy on Italian digital media}.
\newblock \bibinfo{journal}{\emph{Proceedings of the ACM on Human-Computer
  Interaction}} \bibinfo{volume}{6}, \bibinfo{number}{CSCW1}
  (\bibinfo{year}{2022}), \bibinfo{pages}{1--31}.
\newblock


\bibitem[\protect\citeauthoryear{Paul and Morris}{Paul and Morris}{2009}]%
        {paul2009cosense}
\bibfield{author}{\bibinfo{person}{Sharoda~A Paul} {and}
  \bibinfo{person}{Meredith~Ringel Morris}.} \bibinfo{year}{2009}\natexlab{}.
\newblock \showarticletitle{CoSense: enhancing sensemaking for collaborative
  web search}. In \bibinfo{booktitle}{\emph{Proceedings of CHI'09}}.
  \bibinfo{pages}{1771--1780}.
\newblock


\bibitem[\protect\citeauthoryear{PENAmerica}{PENAmerica}{2022}]%
        {HardNews11:online}
\bibfield{author}{\bibinfo{person}{PENAmerica}.}
  \bibinfo{year}{2022}\natexlab{}.
\newblock \bibinfo{title}{Hard News: Journalists and the Threat of
  Disinformation - PEN America}.
\newblock
  \bibinfo{howpublished}{\url{https://pen.org/report/hard-news-journalists-and-the-threat-of-disinformation/}}.
\newblock


\bibitem[\protect\citeauthoryear{Persily and Tucker}{Persily and
  Tucker}{2020}]%
        {persily2020social}
\bibfield{author}{\bibinfo{person}{Nathaniel Persily} {and}
  \bibinfo{person}{Joshua~A Tucker}.} \bibinfo{year}{2020}\natexlab{}.
\newblock \bibinfo{booktitle}{\emph{Social Media and Democracy: The State of
  the Field, Prospects for Reform}}.
\newblock \bibinfo{publisher}{Cambridge U. Press}.
\newblock


\bibitem[\protect\citeauthoryear{Phillips}{Phillips}{2018}]%
        {phillips2018oxygen}
\bibfield{author}{\bibinfo{person}{Whitney Phillips}.}
  \bibinfo{year}{2018}\natexlab{}.
\newblock \showarticletitle{The oxygen of amplification}.
\newblock \bibinfo{journal}{\emph{Data \& Society}}  \bibinfo{volume}{22}
  (\bibinfo{year}{2018}), \bibinfo{pages}{1--128}.
\newblock


\bibitem[\protect\citeauthoryear{Pirolli and Card}{Pirolli and Card}{2005}]%
        {pirolli2005sensemaking}
\bibfield{author}{\bibinfo{person}{Peter Pirolli} {and} \bibinfo{person}{Stuart
  Card}.} \bibinfo{year}{2005}\natexlab{}.
\newblock \showarticletitle{The sensemaking process and leverage points for
  analyst technology as identified through cognitive task analysis}. In
  \bibinfo{booktitle}{\emph{Proceedings of international conference on
  intelligence analysis}}, Vol.~\bibinfo{volume}{5}. McLean, VA, USA,
  \bibinfo{pages}{2--4}.
\newblock


\bibitem[\protect\citeauthoryear{Pirolli and Russell}{Pirolli and
  Russell}{2011}]%
        {pirolli2011introduction}
\bibfield{author}{\bibinfo{person}{Peter Pirolli} {and}
  \bibinfo{person}{Daniel~M Russell}.} \bibinfo{year}{2011}\natexlab{}.
\newblock \bibinfo{title}{Introduction to this special issue on sensemaking}.
\newblock
\newblock


\bibitem[\protect\citeauthoryear{Price and Krug}{Price and Krug}{2000}]%
        {price2000enabling}
\bibfield{author}{\bibinfo{person}{Monroe~E Price} {and} \bibinfo{person}{Peter
  Krug}.} \bibinfo{year}{2000}\natexlab{}.
\newblock \showarticletitle{The enabling environment for free and independent
  media}.
\newblock \bibinfo{journal}{\emph{Available at SSRN 245494}}
  (\bibinfo{year}{2000}).
\newblock


\bibitem[\protect\citeauthoryear{Qu and Hansen}{Qu and Hansen}{2008}]%
        {qu2008building}
\bibfield{author}{\bibinfo{person}{Yan Qu} {and} \bibinfo{person}{Derek~L
  Hansen}.} \bibinfo{year}{2008}\natexlab{}.
\newblock \showarticletitle{Building shared understanding in collaborative
  sensemaking}. In \bibinfo{booktitle}{\emph{Proceedings of CHI'08}}.
\newblock


\bibitem[\protect\citeauthoryear{Quackenbush}{Quackenbush}{2020}]%
        {Collabo83:online}
\bibfield{author}{\bibinfo{person}{Casey Quackenbush}.}
  \bibinfo{year}{2020}\natexlab{}.
\newblock \bibinfo{title}{“Collaboration is the Future of Journalism” -
  Nieman Reports}.
\newblock
  \bibinfo{howpublished}{\url{https://niemanreports.org/articles/collaboration-is-the-future-of-journalism/}}.
\newblock


\bibitem[\protect\citeauthoryear{Recuero, Soares, and Gruzd}{Recuero
  et~al\mbox{.}}{2020}]%
        {recuero2020hyperpartisanship}
\bibfield{author}{\bibinfo{person}{Raquel Recuero},
  \bibinfo{person}{Felipe~Bonow Soares}, {and} \bibinfo{person}{Anatoliy
  Gruzd}.} \bibinfo{year}{2020}\natexlab{}.
\newblock \showarticletitle{Hyperpartisanship, disinformation and political
  conversations on Twitter: The Brazilian presidential election of 2018}. In
  \bibinfo{booktitle}{\emph{Proceedings of ICWSM'20}},
  Vol.~\bibinfo{volume}{14}. \bibinfo{pages}{569--578}.
\newblock


\bibitem[\protect\citeauthoryear{Retis and Chacon}{Retis and Chacon}{2021}]%
        {Mappingd79:online}
\bibfield{author}{\bibinfo{person}{Jessica Retis} {and}
  \bibinfo{person}{Lourdes M.~Cueva Chacon}.} \bibinfo{year}{2021}\natexlab{}.
\newblock \showarticletitle{Auditing Partisan Audience Bias within Google
  Search}.
\newblock \bibinfo{journal}{\emph{ISOJ Journal}} \bibinfo{volume}{11},
  \bibinfo{number}{1} (\bibinfo{year}{2021}), \bibinfo{pages}{35--63}.
\newblock


\bibitem[\protect\citeauthoryear{Rid}{Rid}{2020}]%
        {rid2020active}
\bibfield{author}{\bibinfo{person}{Thomas Rid}.}
  \bibinfo{year}{2020}\natexlab{}.
\newblock \bibinfo{booktitle}{\emph{Active Measures: The Secret History of
  Disinformation and Political Warfare}}.
\newblock \bibinfo{publisher}{Farrar, Straus and Giroux}.
\newblock


\bibitem[\protect\citeauthoryear{Rob{\'e} and Wolfson}{Rob{\'e} and
  Wolfson}{2020}]%
        {robe2020reflections}
\bibfield{author}{\bibinfo{person}{Chris Rob{\'e}} {and} \bibinfo{person}{Todd
  Wolfson}.} \bibinfo{year}{2020}\natexlab{}.
\newblock \showarticletitle{Reflections on the inheritances of Indymedia in the
  age of surveillance and social media}.
\newblock \bibinfo{journal}{\emph{Media, Culture \& Society}}
  \bibinfo{volume}{42}, \bibinfo{number}{6} (\bibinfo{year}{2020}),
  \bibinfo{pages}{1024--1030}.
\newblock


\bibitem[\protect\citeauthoryear{Robertson}{Robertson}{2018}]%
        {robertson2018partisan}
\bibfield{author}{\bibinfo{person}{Ronald Robertson}.}
  \bibinfo{year}{2018}\natexlab{}.
\newblock \bibinfo{title}{{Partisan Bias Scores for Web Domains}}.
\newblock
\newblock
\urldef\tempurl%
\url{https://doi.org/10.7910/DVN/QAN5VX}
\showDOI{\tempurl}


\bibitem[\protect\citeauthoryear{Rodriguez and Caputo}{Rodriguez and
  Caputo}{2020}]%
        {Thisisf55:online}
\bibfield{author}{\bibinfo{person}{Sabrina Rodriguez} {and}
  \bibinfo{person}{Marc Caputo}.} \bibinfo{year}{2020}\natexlab{}.
\newblock \bibinfo{title}{‘This is f---ing crazy’: Florida Latinos swamped
  by wild conspiracy theories - POLITICO}.
\newblock
  \bibinfo{howpublished}{\url{https://www.politico.com/news/2020/09/14/florida-latinos-disinformation-413923}}.
\newblock


\bibitem[\protect\citeauthoryear{Rory~Smith}{Rory~Smith}{2021}]%
        {UndertheSurface:online}
\bibfield{author}{\bibinfo{person}{Claire~Wardle Rory~Smith, Seb~Cubbon}.}
  \bibinfo{year}{2021}\natexlab{}.
\newblock \bibinfo{title}{Under the surface: Covid-19 vaccine narratives,
  misinformation and data deficits on social media}.
\newblock
  \bibinfo{howpublished}{\url{https://firstdraftnews.org/long-form-article/under-the-surface-covid-19-vaccine-narratives-misinformation-and-data-deficits-on-social-media/}}.
\newblock


\bibitem[\protect\citeauthoryear{Schneider}{Schneider}{2021}]%
        {Howtroll78:online}
\bibfield{author}{\bibinfo{person}{Ari Schneider}.}
  \bibinfo{year}{2021}\natexlab{}.
\newblock \bibinfo{title}{How trolls are weaponizing data voids online.}
\newblock
  \bibinfo{howpublished}{\url{https://slate.com/technology/2020/11/data-voids-election-misinformation.html}}.
\newblock


\bibitem[\protect\citeauthoryear{Schwartz and Overdorf}{Schwartz and
  Overdorf}{2020}]%
        {schwartz2020disinformation}
\bibfield{author}{\bibinfo{person}{Christopher Schwartz} {and}
  \bibinfo{person}{Rebekah Overdorf}.} \bibinfo{year}{2020}\natexlab{}.
\newblock \showarticletitle{Disinformation from the Inside: Combining Machine
  Learning and Journalism to Investigate Sockpuppet Campaigns}. In
  \bibinfo{booktitle}{\emph{WWW'20 Companion}}. \bibinfo{pages}{623--628}.
\newblock


\bibitem[\protect\citeauthoryear{Sesin}{Sesin}{2020}]%
        {Spanishl94}
\bibfield{author}{\bibinfo{person}{Carmen Sesin}.}
  \bibinfo{year}{2020}\natexlab{}.
\newblock \bibinfo{title}{Spanish-language disinformation intensifies among
  Florida Latinos, worrying Democrats}.
\newblock
  \bibinfo{howpublished}{\url{https://www.nbcnews.com/news/latino/spanish-language-disinformation-intensifies-among-florida-latinos-worrying-democrats-n1240361}}.
\newblock


\bibitem[\protect\citeauthoryear{Sethi and Rangaraju}{Sethi and
  Rangaraju}{2018}]%
        {sethi2018extinguishing}
\bibfield{author}{\bibinfo{person}{Ricky Sethi} {and} \bibinfo{person}{Raghuram
  Rangaraju}.} \bibinfo{year}{2018}\natexlab{}.
\newblock \showarticletitle{Extinguishing the backfire effect: using emotions
  in online social collaborative argumentation for fact checking}. In
  \bibinfo{booktitle}{\emph{IEEE ICWS'18}}. IEEE, \bibinfo{pages}{363--366}.
\newblock


\bibitem[\protect\citeauthoryear{Shane}{Shane}{2021}]%
        {NeimanLab:online}
\bibfield{author}{\bibinfo{person}{Tommy Shane}.}
  \bibinfo{year}{2021}\natexlab{}.
\newblock \bibinfo{title}{People are using Facebook and Instagram as search
  engines. During a pandemic, that’s dangerous}.
\newblock
  \bibinfo{howpublished}{\url{https://www.niemanlab.org/2020/08/people-are-using-facebook-and-instagram-as-search-engines-during-a-pandemic-thats-dangerous/}}.
\newblock


\bibitem[\protect\citeauthoryear{Shane and Noel}{Shane and Noel}{2020}]%
        {Datadefi93:online}
\bibfield{author}{\bibinfo{person}{T Shane} {and} \bibinfo{person}{P Noel}.}
  \bibinfo{year}{2020}\natexlab{}.
\newblock \bibinfo{title}{Data deficits: why we need to monitor the demand and
  supply of information in real time - First Draft}.
\newblock
  \bibinfo{howpublished}{\url{https://firstdraftnews.org/long-form-article/data-deficits/}}.
\newblock


\bibitem[\protect\citeauthoryear{Shearer}{Shearer}{2018}]%
        {shearer2018social}
\bibfield{author}{\bibinfo{person}{Elisa Shearer}.}
  \bibinfo{year}{2018}\natexlab{}.
\newblock \bibinfo{title}{Social media outpaces print newspapers in the US as a
  news source | Pew Research Center}.
\newblock
  \bibinfo{howpublished}{\url{https://www.pewresearch.org/fact-tank/2018/12/10/social-media-outpaces-print-newspapers-in-the-u-s-as-a-news-source/}}.
\newblock


\bibitem[\protect\citeauthoryear{Shu, Cui, Wang, Lee, and Liu}{Shu
  et~al\mbox{.}}{2019}]%
        {shu2019defend}
\bibfield{author}{\bibinfo{person}{Kai Shu}, \bibinfo{person}{Limeng Cui},
  \bibinfo{person}{Suhang Wang}, \bibinfo{person}{Dongwon Lee}, {and}
  \bibinfo{person}{Huan Liu}.} \bibinfo{year}{2019}\natexlab{}.
\newblock \showarticletitle{defend: Explainable fake news detection}. In
  \bibinfo{booktitle}{\emph{Proceedings of SIGKDD '19}}.
  \bibinfo{pages}{395--405}.
\newblock


\bibitem[\protect\citeauthoryear{Smith and Cubbon}{Smith and Cubbon}{2020}]%
        {TheCovid22:online}
\bibfield{author}{\bibinfo{person}{Rory Smith} {and} \bibinfo{person}{Seb
  Cubbon}.} \bibinfo{year}{2020}\natexlab{}.
\newblock \bibinfo{title}{The Covid-19 and other vaccines: Where we’re
  failing to provide the right information - First Draft}.
\newblock
  \bibinfo{howpublished}{\url{https://firstdraftnews.org/long-form-article/the-covid-19-and-other-vaccines-where-were-failing-to-provide-the-right-information/}}.
\newblock


\bibitem[\protect\citeauthoryear{Spangenberg and Heise}{Spangenberg and
  Heise}{2014}]%
        {spangenberg2014news}
\bibfield{author}{\bibinfo{person}{Jochen Spangenberg} {and}
  \bibinfo{person}{Nicolaus Heise}.} \bibinfo{year}{2014}\natexlab{}.
\newblock \showarticletitle{News from the crowd: Grassroots and collaborative
  journalism in the digital age}. In \bibinfo{booktitle}{\emph{Proceedings of
  WWW'14}}. \bibinfo{pages}{765--768}.
\newblock


\bibitem[\protect\citeauthoryear{Starbird, Arif, and Wilson}{Starbird
  et~al\mbox{.}}{2019}]%
        {starbird2019disinformation}
\bibfield{author}{\bibinfo{person}{Kate Starbird}, \bibinfo{person}{Ahmer
  Arif}, {and} \bibinfo{person}{Tom Wilson}.} \bibinfo{year}{2019}\natexlab{}.
\newblock \showarticletitle{Disinformation as collaborative work: Surfacing the
  participatory nature of strategic information operations}.
\newblock \bibinfo{journal}{\emph{Proceedings of CSCW '19}}
  \bibinfo{volume}{3} (\bibinfo{year}{2019}), \bibinfo{pages}{1--26}.
\newblock


\bibitem[\protect\citeauthoryear{Starbird, Maddock, Orand, Achterman, and
  Mason}{Starbird et~al\mbox{.}}{2014}]%
        {starbird2014rumors}
\bibfield{author}{\bibinfo{person}{Kate Starbird}, \bibinfo{person}{Jim
  Maddock}, \bibinfo{person}{Mania Orand}, \bibinfo{person}{Peg Achterman},
  {and} \bibinfo{person}{Robert~M Mason}.} \bibinfo{year}{2014}\natexlab{}.
\newblock \showarticletitle{Rumors, false flags, and digital vigilantes:
  Misinformation on twitter after the 2013 boston marathon bombing}.
\newblock \bibinfo{journal}{\emph{IConference Proceedings}}
  (\bibinfo{year}{2014}).
\newblock


\bibitem[\protect\citeauthoryear{Stray}{Stray}{2019}]%
        {stray2019institutional}
\bibfield{author}{\bibinfo{person}{Jonathan Stray}.}
  \bibinfo{year}{2019}\natexlab{}.
\newblock \showarticletitle{Institutional Counter-disinformation Strategies in
  a Networked Democracy}. In \bibinfo{booktitle}{\emph{Proceedings of WWW'19}}.
  \bibinfo{pages}{1020--1025}.
\newblock


\bibitem[\protect\citeauthoryear{Thakur and Hankerson}{Thakur and
  Hankerson}{2021}]%
        {Factsand79:online}
\bibfield{author}{\bibinfo{person}{Dhanaraj Thakur} {and}
  \bibinfo{person}{DeVan Hankerson}.} \bibinfo{year}{2021}\natexlab{}.
\newblock \bibinfo{title}{Facts and their Discontents: A Research Agenda for
  Online Disinformation, Race, and Gender - CDT}.
\newblock
  \bibinfo{howpublished}{\url{https://cdt.org/insights/facts-and-their-discontents-a-research-agenda-for-online-disinformation-race-and-gender/}}.
\newblock


\bibitem[\protect\citeauthoryear{Tolmie, Procter, Randall, Rouncefield, Burger,
  Wong Sak~Hoi, Zubiaga, and Liakata}{Tolmie et~al\mbox{.}}{2017}]%
        {tolmie2017supporting}
\bibfield{author}{\bibinfo{person}{Peter Tolmie}, \bibinfo{person}{Rob
  Procter}, \bibinfo{person}{David~William Randall}, \bibinfo{person}{Mark
  Rouncefield}, \bibinfo{person}{Christian Burger}, \bibinfo{person}{Geraldine
  Wong Sak~Hoi}, \bibinfo{person}{Arkaitz Zubiaga}, {and}
  \bibinfo{person}{Maria Liakata}.} \bibinfo{year}{2017}\natexlab{}.
\newblock \showarticletitle{Supporting the use of user generated content in
  journalistic practice}. In \bibinfo{booktitle}{\emph{Proceedings of CHI'17}}.
  \bibinfo{pages}{3632--3644}.
\newblock


\bibitem[\protect\citeauthoryear{Venkatagiri, Thebault-Spieker, Kohler,
  Purviance, Mansur, and Luther}{Venkatagiri et~al\mbox{.}}{2019}]%
        {venkatagiri2019groundtruth}
\bibfield{author}{\bibinfo{person}{Sukrit Venkatagiri}, \bibinfo{person}{Jacob
  Thebault-Spieker}, \bibinfo{person}{Rachel Kohler}, \bibinfo{person}{John
  Purviance}, \bibinfo{person}{Rifat~Sabbir Mansur}, {and}
  \bibinfo{person}{Kurt Luther}.} \bibinfo{year}{2019}\natexlab{}.
\newblock \showarticletitle{GroundTruth: Augmenting expert image geolocation
  with crowdsourcing and shared representations}.
\newblock \bibinfo{journal}{\emph{Proceedings of CSCW'19}}  \bibinfo{volume}{3}
  (\bibinfo{year}{2019}), \bibinfo{pages}{1--30}.
\newblock


\bibitem[\protect\citeauthoryear{Waisbord}{Waisbord}{2020}]%
        {waisbord2020mob}
\bibfield{author}{\bibinfo{person}{Silvio Waisbord}.}
  \bibinfo{year}{2020}\natexlab{}.
\newblock \showarticletitle{Mob censorship: Online harassment of US journalists
  in times of digital hate and populism}.
\newblock \bibinfo{journal}{\emph{Digital Journalism}} \bibinfo{volume}{8},
  \bibinfo{number}{8} (\bibinfo{year}{2020}).
\newblock


\bibitem[\protect\citeauthoryear{Weller, Vickers, Bernard, Blackburn, Borgatti,
  Gravlee, and Johnson}{Weller et~al\mbox{.}}{2018}]%
        {weller2018open}
\bibfield{author}{\bibinfo{person}{Susan~C Weller}, \bibinfo{person}{Ben
  Vickers}, \bibinfo{person}{H~Russell Bernard}, \bibinfo{person}{Alyssa~M
  Blackburn}, \bibinfo{person}{Stephen Borgatti}, \bibinfo{person}{Clarence~C
  Gravlee}, {and} \bibinfo{person}{Jeffrey~C Johnson}.}
  \bibinfo{year}{2018}\natexlab{}.
\newblock \showarticletitle{Open-ended interview questions and saturation}.
\newblock \bibinfo{journal}{\emph{PloS one}} \bibinfo{volume}{13},
  \bibinfo{number}{6} (\bibinfo{year}{2018}), \bibinfo{pages}{e0198606}.
\newblock


\bibitem[\protect\citeauthoryear{Wild, Ciortea, and Mayer}{Wild
  et~al\mbox{.}}{2020}]%
        {wild2020designing}
\bibfield{author}{\bibinfo{person}{Antonia Wild}, \bibinfo{person}{Andrei
  Ciortea}, {and} \bibinfo{person}{Simon Mayer}.}
  \bibinfo{year}{2020}\natexlab{}.
\newblock \showarticletitle{Designing social machines for tackling online
  disinformation}. In \bibinfo{booktitle}{\emph{WWW'20 Companion}}.
\newblock


\bibitem[\protect\citeauthoryear{Wilson and Starbird}{Wilson and
  Starbird}{2021}]%
        {wilson2021cross}
\bibfield{author}{\bibinfo{person}{Tom Wilson} {and} \bibinfo{person}{Kate
  Starbird}.} \bibinfo{year}{2021}\natexlab{}.
\newblock \showarticletitle{Cross-platform Information Operations: Mobilizing
  Narratives \& Building Resilience through both'Big'\&'Alt'Tech}.
\newblock \bibinfo{journal}{\emph{Proceedings of CSCW'21}}  \bibinfo{volume}{5}
  (\bibinfo{year}{2021}), \bibinfo{pages}{1--32}.
\newblock


\bibitem[\protect\citeauthoryear{Wintersieck}{Wintersieck}{2017}]%
        {wintersieck2017debating}
\bibfield{author}{\bibinfo{person}{Amanda~L Wintersieck}.}
  \bibinfo{year}{2017}\natexlab{}.
\newblock \showarticletitle{Debating the truth: The impact of fact-checking
  during electoral debates}.
\newblock \bibinfo{journal}{\emph{American politics research}}
  \bibinfo{volume}{45}, \bibinfo{number}{2} (\bibinfo{year}{2017}).
\newblock


\bibitem[\protect\citeauthoryear{Wolf, Debut, Sanh, Chaumond, Delangue, Moi,
  Cistac, Rault, Louf, Funtowicz, Davison, Shleifer, von Platen, Ma, Jernite,
  Plu, Xu, Scao, Gugger, Drame, Lhoest, and Rush}{Wolf et~al\mbox{.}}{2020}]%
        {wolf-etal-2020-transformers}
\bibfield{author}{\bibinfo{person}{Thomas Wolf}, \bibinfo{person}{Lysandre
  Debut}, \bibinfo{person}{Victor Sanh}, \bibinfo{person}{Julien Chaumond},
  \bibinfo{person}{Clement Delangue}, \bibinfo{person}{Anthony Moi},
  \bibinfo{person}{Pierric Cistac}, \bibinfo{person}{Tim Rault},
  \bibinfo{person}{Rémi Louf}, \bibinfo{person}{Morgan Funtowicz},
  \bibinfo{person}{Joe Davison}, \bibinfo{person}{Sam Shleifer},
  \bibinfo{person}{Patrick von Platen}, \bibinfo{person}{Clara Ma},
  \bibinfo{person}{Yacine Jernite}, \bibinfo{person}{Julien Plu},
  \bibinfo{person}{Canwen Xu}, \bibinfo{person}{Teven~Le Scao},
  \bibinfo{person}{Sylvain Gugger}, \bibinfo{person}{Mariama Drame},
  \bibinfo{person}{Quentin Lhoest}, {and} \bibinfo{person}{Alexander~M. Rush}.}
  \bibinfo{year}{2020}\natexlab{}.
\newblock \showarticletitle{Transformers: State-of-the-Art Natural Language
  Processing}. In \bibinfo{booktitle}{\emph{Proceedings of EMNLP'20}}.
  \bibinfo{pages}{38--45}.
\newblock


\bibitem[\protect\citeauthoryear{Woolley and Howard}{Woolley and
  Howard}{2017}]%
        {woolley2017computational}
\bibfield{author}{\bibinfo{person}{Samuel~C Woolley} {and}
  \bibinfo{person}{Philip Howard}.} \bibinfo{year}{2017}\natexlab{}.
\newblock \showarticletitle{Computational propaganda worldwide: Executive
  summary}.
\newblock  (\bibinfo{year}{2017}).
\newblock


\bibitem[\protect\citeauthoryear{Xu, Sheng, and Wang}{Xu et~al\mbox{.}}{2021}]%
        {xu2021unified}
\bibfield{author}{\bibinfo{person}{Fan Xu}, \bibinfo{person}{Victor~S Sheng},
  {and} \bibinfo{person}{Mingwen Wang}.} \bibinfo{year}{2021}\natexlab{}.
\newblock \showarticletitle{A Unified Perspective for Disinformation Detection
  and Truth Discovery in Social Sensing: A Survey}.
\newblock \bibinfo{journal}{\emph{ACM Computing Surveys (CSUR)}}
  \bibinfo{volume}{55}, \bibinfo{number}{1} (\bibinfo{year}{2021}),
  \bibinfo{pages}{1--33}.
\newblock


\bibitem[\protect\citeauthoryear{Zeng, Abumansour, and Zubiaga}{Zeng
  et~al\mbox{.}}{2021}]%
        {zeng2021automated}
\bibfield{author}{\bibinfo{person}{Xia Zeng}, \bibinfo{person}{Amani~S
  Abumansour}, {and} \bibinfo{person}{Arkaitz Zubiaga}.}
  \bibinfo{year}{2021}\natexlab{}.
\newblock \showarticletitle{Automated fact-checking: A survey}.
\newblock \bibinfo{journal}{\emph{Language and Linguistics Compass}}
  \bibinfo{volume}{15}, \bibinfo{number}{10} (\bibinfo{year}{2021}).
\newblock


\bibitem[\protect\citeauthoryear{Zhang, Ranganathan, Metz, Appling, Sehat,
  Gilmore, Adams, Vincent, Lee, Robbins, et~al\mbox{.}}{Zhang
  et~al\mbox{.}}{2018}]%
        {zhang2018structured}
\bibfield{author}{\bibinfo{person}{Amy~X Zhang}, \bibinfo{person}{Aditya
  Ranganathan}, \bibinfo{person}{Sarah~Emlen Metz}, \bibinfo{person}{Scott
  Appling}, \bibinfo{person}{Connie~Moon Sehat}, \bibinfo{person}{Norman
  Gilmore}, \bibinfo{person}{Nick~B Adams}, \bibinfo{person}{Emmanuel Vincent},
  \bibinfo{person}{Jennifer Lee}, \bibinfo{person}{Martin Robbins},
  {et~al\mbox{.}}} \bibinfo{year}{2018}\natexlab{}.
\newblock \showarticletitle{A structured response to misinformation}. In
  \bibinfo{booktitle}{\emph{WWW'18 Companion}}. \bibinfo{pages}{603--612}.
\newblock


\bibitem[\protect\citeauthoryear{Zhou and Zafarani}{Zhou and Zafarani}{2019}]%
        {zhou2019fake}
\bibfield{author}{\bibinfo{person}{Xinyi Zhou} {and} \bibinfo{person}{Reza
  Zafarani}.} \bibinfo{year}{2019}\natexlab{}.
\newblock \showarticletitle{Fake news detection: An interdisciplinary
  research}. In \bibinfo{booktitle}{\emph{WWW'19 Companion}}.
  \bibinfo{pages}{1292--1292}.
\newblock


\bibitem[\protect\citeauthoryear{Ziff}{Ziff}{2016}]%
        {ziff2016countering}
\bibfield{author}{\bibinfo{person}{Benjamin Ziff}.}
  \bibinfo{year}{2016}\natexlab{}.
\newblock \showarticletitle{Countering Disinformation Through a Free Press}.
\newblock \bibinfo{journal}{\emph{Hampton Roads International Security
  Quarterly}} (\bibinfo{year}{2016}), \bibinfo{pages}{35}.
\newblock


\bibitem[\protect\citeauthoryear{Zubiaga, Aker, Bontcheva, Liakata, and
  Procter}{Zubiaga et~al\mbox{.}}{2018}]%
        {zubiaga2018detection}
\bibfield{author}{\bibinfo{person}{Arkaitz Zubiaga}, \bibinfo{person}{Ahmet
  Aker}, \bibinfo{person}{Kalina Bontcheva}, \bibinfo{person}{Maria Liakata},
  {and} \bibinfo{person}{Rob Procter}.} \bibinfo{year}{2018}\natexlab{}.
\newblock \showarticletitle{Detection and resolution of rumours in social
  media: A survey}.
\newblock \bibinfo{journal}{\emph{ACM Computing Surveys (CSUR)}}
  \bibinfo{volume}{51}, \bibinfo{number}{2} (\bibinfo{year}{2018}),
  \bibinfo{pages}{1--36}.
\newblock


\end{thebibliography}
\newpage
\textbf{A APPENDIX - EVALUATION OF DATAVOIDANT} \\ \\
\textbf{A.1 Evaluation of the Machine Learning Algorithms in Datavoidant.}\\ \label{appendixML}

Our system uses state-of-the-art machine learning models that learn how to categorize social media data to help end-users identify data voids. We study whether the automated approaches that we utilize match human intuition, specifically how humans themselves would categorize the social media data. \\

\textit{A.1.1 Dataset for Studying Machine Learning Models.} \label{dataset}
To support the evaluation of our machine learning algorithms, we created a dataset. We use the dataset to help us have a way for comparing the categorization conducted by our machine learning algorithms to how humans would categorize the same data. For this purpose, we asked 10 journalists to first provide a list of Facebook groups and pages from underrepresented communities for which they would like to potentially study and address data voids. The 10 journalists provided a list of 1,150 Facebook groups and pages. Next, we collected a month's worth of data from the groups and pages, collecting a total of 271,717 Facebook posts. \\

\textit{A.1.2 Evaluation of the Topic Categorization Machine Algorithms.}
We believe that our system will be more intuitive and better for independent journalists to use, the more the system's automation matches human decision-making. We therefore, compared the topic categorization of our system to the topic categorization done by humans. For this purpose, we first asked the 10 journalists who helped us to create our initial dataset (See \ref{dataset}) to provide the number of topics that they wanted to consider for studying data voids (recall this is one of the minimal inputs that our system needs for the categorization of content). The journalists agreed on having 11 topics to study the data voids. They defined the number of topics based on the number of issues that the Pew Research Center reported as mattering the most to underrepresented communities in the 2020 US presidential election \cite{Topelect86:online}. The journalists considered they wanted to cover data voids within these 11 key issues. Next, we asked three of the journalists to categorize a subset of posts in our dataset independently into the 11 topics. For this purpose, we used stratified sampling to collect 5\% of the Facebook posts from our dataset ensuring the posts covered all 11 topics. Next, we asked two of the coders to manually categorize each of the 13,585 posts using one of the 11 topics. We asked the workers to pick the ``most relevant'' topic for each post. The two workers agreed on  81.82\% posts (Cohen’s kappa: .80). We then asked the third coder to label the remaining posts upon which the first two coders had disagreed. We then used a ``majority rule'' approach to determine the topic for those posts. At the end of this step, we had all the posts of our dataset categorized into one of the 11 topics. We considered this human categorized dataset to be our ``Gold Standard''. Next, we separated this Gold Standard dataset into 80\% and 20\% for the training and validation of our system's data categorization. We evaluate our approach on the validation set. We had the following \textit{results}: 208,869 / 271,717 = 76.87\% accuracy (with 11 topics). These results suggest that the machine learning algorithm of our system can successfully categorize posts into topics that are similar to how humans would do the categorization.  \\

\textit{A.1.3 Evaluation of the Political Leaning Categorization.}
Our system has a module that automatically categorizes content into its political leaning (primarily ``conservative'' or ''liberal''). We are interested in studying how accurate this automated categorization is, especially in comparison to how humans would categorize the political leaning of the same content. For this purpose, we used stratified sampling to collect 5\% of the Facebook posts from our dataset, ensuring the posts covered all 11 topics and also had a balance of Facebook groups and pages from citizens, political actors, and news media outlets. We then asked the three journalists who had done the topic categorization to help us again to conduct the categorization of the political leaning of the posts. We asked two of the coders to categorize each of the 13,585 posts into whether they were ``liberal'' or ``conservative''. We asked them to take into account if the post mentioned a political actor, the website political leaning score from Robertson et al. \cite{robertson2018partisan} dataset, and  the tone (sentiment) of the posts  pick the “most relevant” political leaning for each post. If the post mentioned neither websites nor political actors we asked them to classify it as neutral. The two coders agreed on 80\% of posts (Cohen’s kappa: 0.70). We then asked the third coder to label the  posts upon which the first two coders had disagreed. We again used a “majority rule” approach to determine the political leaning of those posts. After this step, we had a dataset with ``gold-standard'' labels of the political leanings. Armed with our dataset, we tested how much our algorithm could accurately classify posts into their political leanings according to the gold standard dataset. Our algorithm achieved a precision of 74.42\%; recall of 94.12\%; and accuracy of 81.43\%. Details are in Table \ref{tab:pol_leaning}. This result suggests that our political leaning identification module can successfully categorize liberal and conservative posts. This helps the system to identify data voids that relate to political leanings.

% \usepackage{multirow}
\begin{table}[h]
\begin{tabular}{|l|c|c|c|l|l|l|}
\hline
\textbf{} &
  \multicolumn{1}{l|}{Real: liberal} &
  \multicolumn{1}{l|}{Real: conservative} &
  \multicolumn{1}{l|}{Precision} &
  Recall &
  Accuracy &
  F1-Score \\ \hline
Pred: liberal &
  32 &
  11 &
  \multirow{2}{*}{74.42\%} &
  \multirow{2}{*}{94.12\%} &
  \multirow{2}{*}{81.43\%} &
  \multirow{2}{*}{83.12\%} \\ \cline{1-3}
Pred: conservative &
  2 &
  25 &
   &
   &
   &
   \\ \hline
\end{tabular}
\caption{Results of the classification of political leaning of posts.}
\label{tab:pol_leaning}
\end{table}
%table1 

\textit{A.1.4 Evaluation of the Bot Detection Machine Learning Algorithm.}
We trained our machine learning algorithms that detect bots on the comprehensive bot detection benchmark of TwiBot-20 \cite{feng2021twibot}. The benchmark provides a dataset that has manually categorized social media accounts into ``bots'' and ''humans'' (i.e., they provide a gold standard).  We evaluate the machine learning algorithm that we use for detecting bots  on the test set of \cite{feng2021twibot}. Our algorithm achieved a precision of  76.09\%; a recall of 86.19\% and accuracy of 80.47\%, which is comparable to other state-of-the-art bot detection algorithms. See details on Table \ref{tab:bot}. Given these results, we argue that our bot detection module enables our system to identify and present the online political narratives that automated accounts could push.

% \usepackage{multirow}
\begin{table}[h]
\begin{tabular}{|l|c|c|c|l|l|l|}
\hline
\textbf{} &
  \multicolumn{1}{l|}{Gold: bot} &
  \multicolumn{1}{l|}{Gold: human} &
  \multicolumn{1}{l|}{Precision} &
  Recall &
  Accuracy &
  F1-Score \\ \hline
Pred: bot &
  487 &
  153 &
  \multirow{2}{*}{80.47\%} &
  \multirow{2}{*}{76.09\%} &
  \multirow{2}{*}{86.19\%} &
  \multirow{2}{*}{80.83\%} \\ \cline{1-3}
Pred: human &
  78 &
  465 &
   &
   &
   &
   \\ \hline
\end{tabular}
\caption{Results of our Bot Detection Machine Learning Algorithm.}
\label{tab:bot}
\end{table}
%table2

\textbf{A2 Smart Categorization Module.}\\ \label{appendixSmartCatModule}%The previous module enables Datavoidant to access the data of the information ecosystems that journalists are interested in studying.

Given that the data collected by the \textit{Data Collection Module} can be massive and difficult for humans to interpret, this module focuses on structuring and categorizing the data to facilitate collective sensemaking.  For this purpose, Datavoidant uses state-of-the-art machine learning models to categorize social media content and then synthesize results  ({\it ``Step: Schematize''} in the sensemaking loop). First, Datavoidant uses basic NLP techniques to categorize the Facebook groups and pages into either ``content from political actors,'' ``content from citizen initiatives,'' or ``content from news sites.'' In particular, the system uses public datasets that list different news sites \cite{Pennsylv12:online}, especially datasets of news sites targeting underrepresented populations \cite{LatinoMe80:online}, to analyze whether the name of a given Facebook group or page matches any of the news sites in the datasets. If it finds a match, the system labels that Facebook group or page as ``content from news sites.'' For example, if a journalist inputs the Facebook page of {\it “The New York Times''} \cite{20TheNew25:online}, the system will label that page as being ``content from news sites'' because it found a match in the dataset. Similarly, to identify whether a Facebook page or group is ``content from political actors,'' Datavoidant takes  the name and description of the  page, and  analyzes whether it directly mentions the term {\it “political”} (or related synonyms). Datavoidant also analyzes whether the page mentions a political actor or political party in its name. For this purpose, Datavoidant crawls Wikipedia to obtain lists of political actors and political parties to consider\cite{2020Unit27:online,2020inUn45:online}. All other Facebook groups and pages are labeled as ``content from citizen initiatives''. Notice that we allow journalists to correct the system’s categorization of Facebook groups and pages and re-categorize the content as they consider more appropriate. The system also allows journalists to input the sources of data they want Datavoidant to use for this first categorization (e.g., Wikipedia articles about political actors, lists of newspapers etc). After this step, we have all the Facebook groups and pages categorized into three main types: news media, political spaces, and citizen groups. This type of categorization was important given that journalists expressed an interest in being able to bridge the data gap between these different online spaces. However, it was also important for journalists to be able to conduct a multi-level analysis where they could understand what topics were covered less than others across these different online spaces, what political actors were pushing certain content, as well as identify whether automated methods were pushing certain topics (to understand manipulations around the data voids). For this purpose, Datavoidant integrates state-of-the-art machine learning models to categorize the content on multiple levels and facilitate these types of data analysis. %In the following we provide details of how Datavoidant conducts each type of data categorization.

{TOPIC LEVEL CATEGORIZATION. }
In the design of Datavoidant, we considered that journalists would likely not have the time or ability to interpret complex abstract topics without labels, like the ones that the topic modeling algorithm of LDA throws out \cite{blei2003latent}. We assume that most journalists will likely not know how to provide labeled data to train machine learning algorithms that can discern one topic from another. Therefore, we opted for automated methods that could remove the unnecessary burden and complexity to journalists, while still allowing them to automatically categorize their data at scale. Datavoidant simply asks journalists to provide the list of topics they are interested in exploring and a list of keywords associated with each topic. The system then uses these keywords and topics to automatically create a training and testing set to teach machine learning models how to classify posts into the different topics. In specific, for each topic, the system uses its associated keywords to randomly sample Facebook posts that mention the keywords. The posts are taken from the lists of Facebook groups and pages that the end-user provided initially. The system then labels each post with one topic, selecting the topic with the greatest number of keywords in the post. The system will aim to have the same number of posts for each topic, but allows the end user to know when this is is not the case. Through this, the end user can easily modify the topics and facilitate creating a more balanced dataset. Datavoidant then trains a  pre-trained language model RoBERTa~\cite{liu2019roberta} and uses fully connected layers for topic classification. This model is trained on the collected, labeled dataset of Facebook posts and their topics with a 8:2 split for training and validation set. 

{POLITICAL LEANING CATEGORIZATION.}
In addition to topic-level data voids, Datavoidant also  helps journalists to identify {political-level data deficiencies}, where some topics might be less discussed by accounts from certain political or ideological perspectives. For example, climate change content might be rarely covered by liberals, while critical race theory could be less covered by conservatives, creating partisan echo chambers and political-level data voids. For this purpose, Datavoidant identifies each post's political leaning to facilitate visualization and understanding of political-level  data deficits. To conduct its automatic categorization of posts with respect to political leanings, Datavoidant resorts to external knowledge about the political leanings of websites~\cite{robertson2018partisan} and political actors~\cite{feng2021knowledge}. Datavoidant conducts the following approach to calculate the political leaning score of a given Facebook post:
\begin{itemize}
    \item If the post comes from a Facebook page that represents a websites that is in the list (i.e., a website with a clear political leaning), the system:
    \begin{itemize}
    \item averages the mentioned websites' political leaning score based on ~\cite{robertson2018partisan} and through this obtains the post's final ``political leaning score'' as $b_w$. 
     \end{itemize}
    \item If the post mentions any website on the list or mentions any political actors then the system:
    \begin{itemize}
    \item calculates  the sentiment score $s$~\cite{devlin2018bert, wolf-etal-2020-transformers} for the post, with -1 as most negative and +1 as most positive.
    \item averages the political leaning score of the actors and websites mentioned based on~\cite{feng2021knowledge,robertson2018partisan}  to obtain the ``political leaning score'' $b_a$. The final political leaning score of the post is then obtained by $b_a \times s$.
    \end{itemize}
  \item If the post mentions neither websites nor political actors, the system takes 0 for its political leaning score and regard the post as neutral.
\end{itemize}

Notice that Datavoidant categorizes posts first based on the overall nature of the Facebook page from which the post is from. We consider that known conservative outlets will tend to always post conservative content and liberal outlets will tend to post liberal content. If the system cannot identify the nature of the Facebook page, it analyzes whether the post is discussing liberal or conservative actors in a positive or negative form, and uses this to calculate the political leaning score of the post. In all other cases, the system labels the post as neutral. In this way, Datavoidant  calculates political leaning scores for social media posts, which helps to illustrate political-level data deficiencies across topics.

{BOT CATEGORIZATION.} Automated social media users, also known as bots, widely exist on online social networks and induce undesirable social effects. In the past decade, malicious actors have launched bot campaigns to interfere with elections~\cite{ferrara2017disinformation,deb2019perils}, spread misinformation~\cite{feng2021satar} and propagate extreme ideology~\cite{berger2015isis}. To these issues, Datavoidant includes a bot detection component that categorizes accounts into bots and none-bots. The aim is to help journalists identify biased information propagated by malicious actors. In Datavoidant, we focus on the textual content of posts to identify {Facebook bots and malicious actors}. Specifically, we follow the method in the state-of-the-art approach~\cite{feng2021botrgcn} to encode post content with pre-trained language models~\cite{liu2019roberta} and train a multi-layer perceptron for bot detection. We train our model with the comprehensive benchmark TwiBot-20~\cite{feng2021twibot}. %We then use the trained model to enable datavoidant to identify automated accounts.

\newpage

\textbf{B PARTICIPANTS INTERVIEW STUDY}\\
For the purpose of protecting the anonymity of our interviewees, we have anonymized the data from the journalists we recruited for our interview study. We followed guidelines used in prior work for disclosing information about journalists who take part in interview studies  \cite{mcclure2020misinformation,haughey2022bridging}.

\begin{table}[h]
\begin{tabular}{|l|l|l|}
\hline
\textbf{\begin{tabular}[c]{@{}l@{}}Independent\\ Journalist\end{tabular}} & \textbf{Organization Type} & \textbf{Language} \\ \hline
J1 & Niche Newspaper & Monolingual \\ \hline
J2 & Niche Newspaper & Bilingual \\ \hline
J3 & Niche Newspaper & Monolingual \\ \hline
J4 & Niche Newspaper & Monolingual \\ \hline
J5 & Niche Newspaper & Monolingual \\ \hline
J6 & Niche Newspaper & Monolingual \\ \hline
J7 & U.S. Local Radio & Bilingual \\ \hline
J8 & Niche Newspaper & Monolingual \\ \hline
J9 & Niche Newspaper & Monolingual \\ \hline
J10 & Non-Profit Newsroom \& Civic Engagement Organization & Monolingual \\ \hline
J11 & Niche Newspaper & Monolingual \\ \hline
J12 & Digital First Outlet & Monolingual \\ \hline
J13 & Digital First Outlet & Monolingual \\ \hline
J14 &  Non-Profit Newsroom \& Civic Engagement Organization & Bilingual \\ \hline
J15 & Digital First Outlet & Monolingual \\ \hline
J16 & Digital First Outlet & Monolingual \\ \hline
J17 & Digital First Outlet & Bilingual \\ \hline
J18 & Digital First Outlet & Monolingual \\ \hline
J19 & Digital First Outlet & Monolingual \\ \hline
J20 & Digital First Outlet & Monolingual \\ \hline
J21 & Digital First Outlet & Monolingual \\ \hline
J22 & Digital First Outlet & Monolingual \\ \hline
\end{tabular}
\caption{Overview of participants in our interview study}
\label{tab:participants}
\end{table}
\end{document}